\documentclass[aps,prd,twocolumn,showpacs,eqsecnum,amsmath,amssymb,amsfonts]{revtex4}

\usepackage{graphicx}

\begin{document}


\title{Radiation from accelerated black holes in a de~Sitter universe}

\author{Pavel Krtou\v{s}}
\email{Pavel.Krtous@mff.cuni.cz}

\author{Ji\v{r}\'{\i} Podolsk\'y}
\email{Jiri.Podolsky@mff.cuni.cz}

\affiliation{
  Institute of Theoretical Physics,
  Faculty of Mathematics and Physics, Charles University in Prague,\\
  V Hole\v{s}ovi\v{c}k\'{a}ch 2, 180 00 Prague 8, Czech Republic
  }

\date{December 1, 2003}     

\begin{abstract}
Radiative properties of gravitational and electromagnetic
fields generated by uniformly accelerated charged black
holes in asymptotically de~Sitter spacetime are studied
by analyzing the $C$-metric exact solution of the Einstein-Maxwell
equations with a positive cosmological constant $\Lambda$.
Its global structure and physical properties are
thoroughly discussed. We explicitly find and describe
the specific pattern of radiation which exhibits the dependence
of the fields on a null direction along which the (spacelike) conformal
infinity is approached. This directional characteristic of
radiation supplements the peeling behavior of the
fields near infinity. The interpretation of the
solution is achieved by means of various coordinate systems,
and suitable tetrads.
The relation to the Robinson-Trautman framework is also  presented.
\end{abstract}


\vspace*{39pt}

\maketitle


\section{Introduction}
\label{sc:intro}

There has been great effort in general relativity devoted to investigation
of gravitational radiation in asymptotically flat spacetimes. Some of the now
classical works, which date back to the 1960s, set up rigorous frameworks within
which a general asymptotic character of radiative fields near infinity could
be elucidated
\cite{Bondi:1960,BondiBurgMetzner:1962,Sachs:1962,Burg:1969,NewmanPenrose:1962,%
NewmanUnti:1962,Sachs:1961,Penrose:1965,Penrose:1964,Penrose:1967,Penrose:1963}.
Also,   particular   examples of explicit exact
radiative spacetimes have been found and analyzed,  e.g.,
Refs.~\cite{BondiPiraniRobinson:1959,EhlersKundt:1962,RobinsonTrautman:1962,BonnorSwaminarayan:1964},
for a review of these important contributions to the theory of radiation see, for example,
Refs.~\cite{Pirani:1965,BonnorGriffithsMacCallum:1994,Bicak:Bonnor,Bicak:1997,Bicak:Ehlers}.

One of the fundamental approaches to investigate the radiative properties of
a gravitational field at large distances from a bounded source is based on introducing  a
suitable Bondi-Sachs coordinate system adapted to outgoing null hypersurfaces,
and expanding the metric functions in negative powers of the luminosity distance
\cite{Bondi:1960,BondiBurgMetzner:1962,Sachs:1962,Burg:1969}.
In the case of asymptotically flat spacetimes this
framework enables one to define the Bondi mass (total mass of the system as measured
at future null infinity $\scri^+$), and characterize the time evolution including
radiation in terms of the news functions. Using these concepts it is possible
to formulate a balance between the amount of energy radiated by gravitational waves
and the decrease of the Bondi mass of an isolated system. Unfortunately, this
standard explicit approach is not directly applicable to spacetimes whose conformal
infinity $\scri^+$ has a spacelike character as is the case of an asymptotically
de~Sitter universe which we wish to study here.

Alternatively, in accordance with the Newman-Penrose formalism \cite{NewmanPenrose:1962,NewmanUnti:1962},
information about the character of radiation in asymptotically flat spacetimes can
be extracted from the tetrad components of fields measured along a family of
null geodesics approaching $\scri^+$. The gravitational field is radiative if the dominant
components of the Weyl tensor $\WT_{\alpha\beta\gamma\delta}$ (or of the Maxwell tensor
$\EMF_{\alpha\beta}$ in the electromagnetic case)
fall off as ${1/\afp}$, where $\afp$ is an affine parameter along the null geodesics.
The rate of approach to zero of the Weyl and electromagnetic tensor is generally given
by the \vague{peeling off} theorem of Sachs \cite{Sachs:1961,Sachs:1962,Penrose:1965}.
In analogy to this well-known behavior it is natural to expect that those components of the
fields in parallelly transported tetrad which
are proportional to ${1/\afp}$  characterize gravitational
and electromagnetic radiation also in more general cases of spacetimes not asymptotically flat.
We shall adopt such a definition of radiation below.

In the presence of a positive cosmological constant $\Lambda$, however,
the conformal infinity $\scri^+$  has a \emph{spacelike} character,
and for  principal reasons the rigorous concept of gravitational and electromagnetic
radiation is much less clear. As Penrose noted
in the 1960s \cite{Penrose:1964,Penrose:1967} already, following his
geometrical formalization of the idea of asymptotical
flatness based of the conformal technique \cite{Penrose:1963,Penrose:1965}, radiation
is defined \vague{less invariantly}  when $\scri$ is spacelike than when it
has a null character.

One of the difficulties related to the spacelike character of the infinity is that
initial data on $\scri^-$ (or final data on $\scri^+$) for, e.g.,
electromagnetic field with sources cannot be
prescribed freely because the Gauss constraint has to be satisfied at $\scri^-$ (or
$\scri^+$). This results in the insufficiency of purely retarded solutions
in case of a spacelike~$\scri^-$ --- advanced effects must also be presented.
This phenomenon has been
demonstrated explicitly recently \cite{BicakKrtous:ASDS} by analyzing
test electromagnetic fields of uniformly accelerated charges in de~Sitter background.

We will concentrate on another crucial difference in behavior of radiative fields near
null versus spacelike infinity.
In the case of asymptotically flat spacetimes,
any point $N_+$ at null infinity $\scri^+$ can be approached
essentially only along \emph{one} null direction.
However, if future infinity $\scri^+$ has a spacelike character,
one can approach the point~$N_\fix$ from \emph{infinitely many different} null directions.
It is not a~priori clear how the radiation components of the fields
depend on a direction along which $N_\fix$ is approached. In this
paper such dependence will be thoroughly investigated.

In fact, radiative properties of a test electromagnetic
field of two uniformly  accelerated point-like
charges in the de~Sitter background has recently been studied
\cite{BicakKrtous:FUACS,BicakKrtous:BIS}. Within this context, the above mentioned
directional dependence has been explicitly found. In particular,
it has been demonstrated that there are always
exactly two special directions ---
those opposite  to the direction from the sources ---
along which the radiation vanishes. For all other
directions the radiation is nonvanishing and it is described by an explicit
formula which completely characterizes its angular dependence.

In the present paper, these results will  be considerably generalized
to both gravitational and electromagnetic field which  are not just test fields
in the de~Sitter background. Interestingly, it will be demonstrated
that the gravitational and electromagnetic fields of the $C$-metric with ${\Lambda>0}$,
which is an \emph{exact} solution  representing a pair of
uniformly accelerated possibly charged black holes in the de~Sitter-like
universe, exhibits exactly the \emph{same} asymptotic radiative behavior
as the test fields \cite{BicakKrtous:FUACS,BicakKrtous:BIS}.
We are thus able to supplement the information about the peeling behavior of the fields
near $\scri^+$ with an additional general property of  radiation, namely with the
specific \defterm{directional pattern of the radiation} at conformal infinity.

The $C$-metric with ${\Lambda=0}$ is a well-known solution of the Einstein(-Maxwell)
equations which, together with the famous Bonnor-Swaminarayan
solutions \cite{BonnorSwaminarayan:1964}, belongs to a large  class of asymptotically flat spacetimes with
boost and rotational symmetry \cite{BicakSchmidt:1989} representing
accelerated sources. It was discovered already
in 1917 by Levi-Civita \cite{LeviCivita:1917} and Weyl \cite{Weyl:1919}, and named by Ehlers and
Kundt \cite{EhlersKundt:1962}. Physical interpretation and understanding of the global structure
of the $C$-metric as a spacetime with radiation generated by a pair of accelerated
black holes came with the fundamental papers by Kinnersley and Walker
\cite{KinnersleyWalker:1970} and Bonnor \cite{Bonnor:1982}.
Consequently, a great number of works analyzed various aspects and
properties of this solution, including its generalization which admits a rotation
of the black holes. References and summary of the results
can be found, e.g., in Refs.~\cite{BicakSchmidt:1989,BicakPravda:1999,LetelierOliveira:2001,Pravdovi:2000}.
Another  possible generalization of the standard $C$-metric exists, namely, that to a
nonvanishing value of the cosmological constant $\Lambda$
\cite{PlebanskiDemianski:1976}, cf.\  \cite{Carter:1968,Debever:1971}.
However, in this case a complete understanding of global properties, mainly a character of
radiation, is still missing despite a successful application of this solution to the
problem of cosmological production of black holes \cite{MannRoss:1995},
and its recent analysis and interpretation
\cite{PodolskyGriffiths:2001,DiasLemos:2003a,DiasLemos:2003b}.

There exists  a strong motivation to investigate the \mbox{$C$-metric} solution
with ${\Lambda>0}$. As will be demonstrated below, it may serve as
an interesting exact model of gravitational and electromagnetic
radiation of bounded sources in the asymptotically de~Sitter
universe (in contrast to ${\Lambda=0}$, in which case the system is not permanently bounded).
The character of radiation, in particular the above
mentioned  dependence of the asymptotic fields on directions,
along which points on the de~Sitter-like infinity $\scri^+$ are
approached, can explicitly be found and studied. These results may
provide an important clue to formulation of a general theory of
radiation in spacetimes which are not asymptotically flat.
In addition to this purely theoretical motivation, understanding
the behavior of accelerated black holes in the universe with a
positive value of the cosmological constant can also be interesting from
perspective of contemporary cosmology.

The paper is organized as follows. First, in Section~\ref{sc:Cmetric} we
present the $C$-metric solution with a positive cosmological constant
in various coordinates which will be necessary for the subsequent
analysis. The global structure of the spacetime is described in detail in
Section~\ref{sc:GlobStr}. Next, in Section~\ref{sc:tetrads} we
introduce and discuss various privileged orthonormal and null
tetrads near the de~Sitter-like infinity $\scri^+$ together with
their mutual relations, and we give corresponding components of the
gravitational and electromagnetic $C$-metric fields.
Section~\ref{sc:RadChar} contains  the core of our analysis. We
carefully define interpretation tetrad
parallelly transported along all null geodesics approaching asymptotically
a given point on spacelike $\scri^+$ from different spatial directions.
The magnitude of the leading terms of gravitational and electromagnetic
fields in such a tetrad then provides us with a specific directional
pattern of radiation which is described and analyzed. This result
is subsequently rederived in Section~\ref{sc:RoTr} using the
Robinson-Trautman framework which also reveals some other aspects
of the radiative properties. Particular behavior of
radiation along the algebraically special null directions is
studied in Section~\ref{sc:AlgSpecDir}. For these
privileged geodesics the results are obtained explicitly without
performing asymptotic expansions of the physical quantities near ${\scri^+}$.
In this case we also study a specific dependence of the field components on
a choice of initial conditions on horizons.

The paper contains four appendixes. Appendix~\ref{apx:Coor} summarizes
known and also several new coordinates  for the $C$-metric with
${\Lambda>0}$. The properties of the specific metric functions
are described in Appendix~\ref{apx:FGprop}. In Appendix~\ref{apx:CoorFrames}
useful relations between the various coordinate 1-form
and vector frames are presented, together
with the relations between the different privileged null tetrads.
Appendix~\ref{apx:Transformations} contains general Lorentz
transformations of the null tetrad components of the
gravitational and electromagnetic fields.

\section{The $C$-metric with a cosmological constant in suitable coordinates}
\label{sc:Cmetric}

The generalization of the $C$-metric which admits a
nonvanishing cosmological constant ${\Lambda>0}$,
representing a pair of uniformly accelerated black holes
in a \vague{de~Sitter background}, has the form
\begin{equation}\label{KWmetric}
  \mtrc =
  \frac1{A^2(\xKW+\yKW)^2}\Bigl(
    -\FKW \,\grad\tKW^2
    +\frac1{\FKW} \,\grad\yKW^2
    +\frac1{\GKW} \,\grad\xKW^2
    +\GKW \,\grad\ph^2
    \Bigr)\commae
\end{equation}
where
\begin{equation}\label{KWFG}
\begin{aligned}
  \FKW &= -\frac1{\DSr^2\accl^2}-1+\yKW^2
  -2\mass\accl\,\yKW^3+\charge^2\accl^2\,\yKW^4\commae\\
  \GKW &= \mspace{81mu} 1-\xKW^2
  -2\mass\accl\,\xKW^3-\charge^2\accl^2\,\xKW^4\commae
\end{aligned}
\end{equation}
see Eqs.~\eqref{ae:KWmetric}, \eqref{ae:KWFG}.
Here ${\tKW\in\realn}$, ${\ph\in(-\pi\conpar,\pi\conpar)}$,
$\mass$, $\charge$, $\accl$, $\conpar$ are constants,
and ranges of the coordinates ${\xKW,\,\yKW}$
(or, more precisely, of the related coordinates ${\x,\,\y}$ defined
below by Eq. \eqref{KWtoxy})
will be discussed in detail in the next section. For convenience,
we have parametrized the cosmological constant $\Lambda$
by the \vague{de~Sitter radius} as
\begin{equation}\label{DSrLambda}
  \DSr= \sqrt{\frac{3}{\Lambda}}\period
\end{equation}
The metric \eqref{KWmetric} is a solution of the Einstein-Maxwell
equations with the electromagnetic field given by (\noteref{nt:Units})
\begin{equation}\label{KWEMF}
  \EMF = \charge\, \grad\yKW\wedge\grad\tKW\period
\end{equation}

The constants $\mass$, $\charge$, $\accl$, and $\conpar$ parametrize
\defterm{mass}, \defterm{charge}, \defterm{acceleration}, and \defterm{conicity}
of the black holes, although their relation to physical quantities is not, in
general, direct. For example, the total charge $\totcharge$ on a timelike hypersurface
${\tKW=\text{constant}}$ localized inside a surface ${\yKW=\text{constant}}$, defined
using the Gauss law, is given by
${\totcharge=\frac12(\x_\paxis-\x_\maxis)\,\conpar\charge}$,
where the constants ${\x_\maxis,\,\x_\paxis}$ are introduced at the beginning of the next section.
Obviously, $\totcharge$ depends not only on the charge parameter $\charge$. Similarly,
physical conicity
is proportional to the parameter $\conpar$, but it also depends
on other parameters, see Eq.~\eqref{conicity} below. The concept of mass
(outside the context of asymptotically flat spacetimes)
and of physical acceleration of black holes is even more complicated.
We will return to this point at the end of the next section.
For satisfactory interpretation of the parameters $\mass$, $\charge$, and $\accl$
in the limit of their small values see, e.g., Ref.~\cite{PodolskyGriffiths:2001}.

In the following we will always assume
\begin{equation}\label{assumtions}
  \mass>0\comma\charge^2<\mass^2\comma\accl>0\commae
\end{equation}
and $\FKW$, as a polynomial in $\yKW$, to have only distinct real roots.
Also, instead of the acceleration constant $\accl$ we will conveniently use the
dimensionless \defterm{acceleration parameter} $\acp$ defined as
\begin{equation}\label{acpdef}
  \sinh\acp=\DSr\accl\comma
  \cosh\acp=\sqrt{1+\DSr^2\accl^2}\period
\end{equation}

We will also use other suitable coordinates which are
introduced and discussed in more detail in Appendix~\ref{apx:Coor}.
Here we list only the basic definitions and the corresponding forms of metric.

The rescaled coordinates ${\tau,\,\y,\,\x,\,\ph}$ are defined
\begin{equation}\label{KWtoxy}
\begin{aligned}
  \tau &= \,\tKW\,\coth\acp \commae&
  \ph &= \ph\comma\\
  \y &= \yKW \,\tanh\acp \comma&
  \x &= -\xKW\commae
\end{aligned}
\end{equation}
cf.\  Eq.~\eqref{ae:KWtoxy}, in which
the metric takes the form \eqref{ae:xymetric},
\begin{equation}\label{xymetric}
  \mtrc =
  \rRT^2\Bigl(
    -\F \,\grad\tau^2 +\frac1{\F} \,\grad\y^2
    +\frac1{\G} \,\grad\x^2 +\G \,\grad\ph^2
    \Bigr)\commae
\end{equation}
where
\begin{equation}\label{rRTdef}
  \rRT = \frac1{\accl(\xKW+\yKW)}
  = \frac\DSr{\y\cosh\acp-\x\sinh\acp}\commae
\end{equation}
and
\begin{equation}\label{xyFG}
\begin{aligned}
  -\F &= 1-\y^2+\cosh\acp\,\frac{2\mass}{\DSr}\;\y^3
        -\cosh^2\acp\,\frac{\charge^2}{\DSr^2}\;\y^4\commae\\
  \G &= 1-\x^2+\sinh\acp\,\frac{2\mass}{\DSr}\;\x^3
        -\sinh^2\acp\,\frac{\charge^2}{\DSr^2}\;\x^4\period
\end{aligned}
\end{equation}

The coordinates ${\om,\,\tau,\,\sg,\,\ph}$ adapted to the Killing
vectors $\cvil{\tau}$, $\cvil{\ph}$
and the conformal infinity $\scri$ (${\om=0}$)
are defined by
\begin{equation}\label{omsgdef}
\begin{gathered}
  \om = - \y\cosh\acp + \x\sinh\acp\commae\\
  \grad\sg = \frac{\sinh\acp}{\F}\,\grad\y +
  \frac{\cosh\acp}{\G}\,\grad\x\commae
\end{gathered}
\end{equation}
cf.\ Eqs.~\eqref{ae:omdef}, \eqref{ae:sgdef},
and the metric is \eqref{ae:omsgmetric},
\begin{equation}\label{omsgmetric}
  \mtrc =
  \frac{\DSr^2}{\om^2}\Bigl(
    -\F \,\grad\tau^2
    +\frac1\E \,\grad\om^2
    +\frac{\F\G}{\E} \,\grad\sg^2
    +\G \,\grad\ph^2
    \Bigr)\commae
\end{equation}
where
\begin{equation}\label{Edef}
  \E = \F \cosh^2\acp + \G \sinh^2\acp\period
\end{equation}

Finally, we will also use the $C$-metric expressed in the Robinson-Trautman
coordinates ${\zRT,\,\bRT,\,\uRT,\,\rRT}$
which has the form \eqref{ae:RTmetric} (see \noteref{nt:SymTensProd} for a definition of the
symmetric product $\stp$)
\begin{equation}\label{RTmetric}
  \mtrc = \frac{\rRT^2}{\PRT^2}\,
    \grad\zRT\stp\grad\bRT
    -\grad\uRT\stp\grad\rRT
    -\HRT \,\grad\uRT^2
    \commae
\end{equation}
with
\begin{equation}\label{PHRTdef}
  \frac1{\PRT^2}=\G\comma
  \HRT = \frac{\rRT^2}{\DSr^2}\,\E\period
\end{equation}
It follows immediately from Eqs.~\eqref{rRTdef} and \eqref{omsgdef} that
\begin{equation}\label{rRTomRel}
  \rRT=-\frac\DSr\om\period
\end{equation}
For explicit definitions of the coordinates $\uRT$, $\zRT$
and further details see Eqs.~\eqref{ae:rRTKW}, \eqref{ae:zetaRTKW}, and
related text in Appendix~\ref{apx:Coor}.

\section{The global structure}
\label{sc:GlobStr}

\begin{figure}
\includegraphics{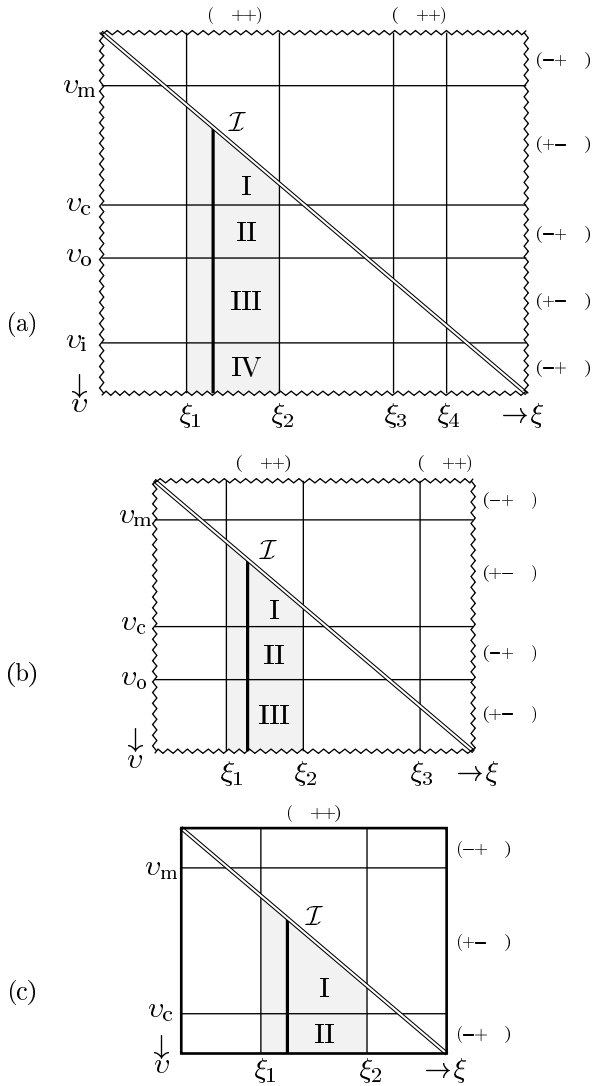}
\caption{\label{fig:xyDiag}
A qualitative diagram of the ${\x\textdash\y}$ section
(${\tau,\,\ph}=\text{constant}$) of the studied spacetime.
The three cases correspond to
(a) charged accelerated black holes in asymptotically
de~Sitter universe (${\charge\neq0}$, ${\mass\neq0}$),
(b) uncharged black holes (${\charge=0}$, ${\mass\neq0}$), and
(c) de~Sitter universe (${\charge=0}$, ${\mass=0}$).
Horizontal lines indicate the horizons, vertical lines
are axes of $\ph$~symmetry.
The diagonal double line ${\y=\x\tanh\acp}$ corresponds to
infinity $\scri$. Singularities are depicted by \vague{zig-zag} lines.
The boundary of each diagram corresponds to ${\x,\,\y=\pm\infty}$.
Mutual intersections of different lines are governed by  relations \eqref{zerosord}.
Different columns and rows correspond to different
signs of the functions $\G$ and $\F$, respectively, and thus to different signatures of the
metric, which are indicated on the sides of the diagrams.
The metric \eqref{xymetric} describes, in general, four distinct spacetimes ---
the domains in columns ${\x_\maxis<\x<\x_\paxis}$ and ${\x_3<\x<\x_4}$,
separated in addition by the infinity (the diagonal line).
In this paper we discuss only the physically most
reasonable spacetime with the coordinates ${\x,\,\y}$ in the ranges
${\x_\maxis<\x<\x_\paxis}$ and ${\y>\x\tanh\acp}$ (the shaded areas).
Sections ${\x=\text{constant}}$ which
correspond to the conformal diagrams in Fig.~\ref{fig:ConfDiag}
are indicated by thick lines.}%
\end{figure}

In this section we shall describe the global structure of the $C$-metric with
${\Lambda>0}$. In particular, we shall analyze the character of
infinity, singularities, and possible horizons.
(Cf. recent work \cite{DiasLemos:2003b} for similar discussion
that covers also cases not studied here.)
From the form \eqref{xymetric} of the metric we observe that it is
necessary to investigate zeros of the metric functions $\F$ and
$\G$ given by Eq.~\eqref{xyFG}. We will only
discuss the particular case when the function $\F$ has $n$ distinct real
roots, where $n$ is the degree of polynomial dependence of $\F$ on $\y$
(${n=4}$ for ${\charge\ne0}$).
Let us denote these roots as $\;\y_\ihor$, $\y_\ohor$, $\y_\chor$, and
$\;\y_\mhor\;$ in a descending order (the meaning of the subscripts will be explained below).
In the case ${\charge=0}$, the value of
$\y_\ihor$ is not defined, etc. Analogously, we denote the roots of $\G$
as $\x_\maxis$, $\x_\paxis$, $\x_3$, and $\x_4$ in an ascending order.
Similarly to discussion of the $C$-metric with vanishing
$\Lambda$ \cite{KinnersleyWalker:1970,BicakPravda:1999,LetelierOliveira:2001},
the zeros of the function
$\F$ correspond to \emph{horizons}, and the zeros of $\G$ to \emph{axes} of
$\ph$~symmetry. Following these works, and Ref.~\cite{PodolskyGriffiths:2001} for
${\Lambda>0}$ in particular, the qualitative diagrams of the ${\x\textdash\y}$ slice
(i.e.,  ${\tau,\,\ph=\text{constant}}$) are drawn in Fig.~\ref{fig:xyDiag}.
In this diagram we use  relations
\begin{equation}\label{zerosord}
\y_\mhor\coth\acp<\x_\maxis<0
<\x_\paxis<\y_\chor\coth\acp\commae
\end{equation}
which are obvious from Fig.~\ref{fig:PolS} in Appendix~\ref{apx:FGprop}.
Different columns and rows in the diagrams in Fig.~\ref{fig:xyDiag}
correspond to different signs of the functions $\G$ and $\F$.
The metric has a physical signature ${({-}{+}{+}{+})}$ for
${\x_\maxis<\x<\x_\paxis}$ and ${\x_3<\x<\x_4}$. We will be interested
only in the first region.

Infinity $\scri$ of the spacetime corresponds to
${\rRT=\infty}$, or equivalently to
\begin{equation}\label{xyscricond}
  \om=0\comma\text{i.e.,}\quad\y = \x\tanh\acp\commae
\end{equation}
(double line in Fig.~\ref{fig:xyDiag}).
We will restrict to the region ${\y>\x\tanh\acp}$ (i.e., ${\rRT>0}$) which
describes both interior and exterior of accelerated
black holes in de~Sitter-like spacetime
(the shaded areas in Fig.~\ref{fig:xyDiag}).
The metric has an unbounded
curvature for ${\rRT=0}$ which corresponds to
a physical singularity inside the black holes, a zig-zag line on
the boundary of the diagrams in Fig.~\ref{fig:xyDiag},
in particular of the column ${\x_\maxis<\x<\x_\paxis}$.

\begin{figure}
\includegraphics{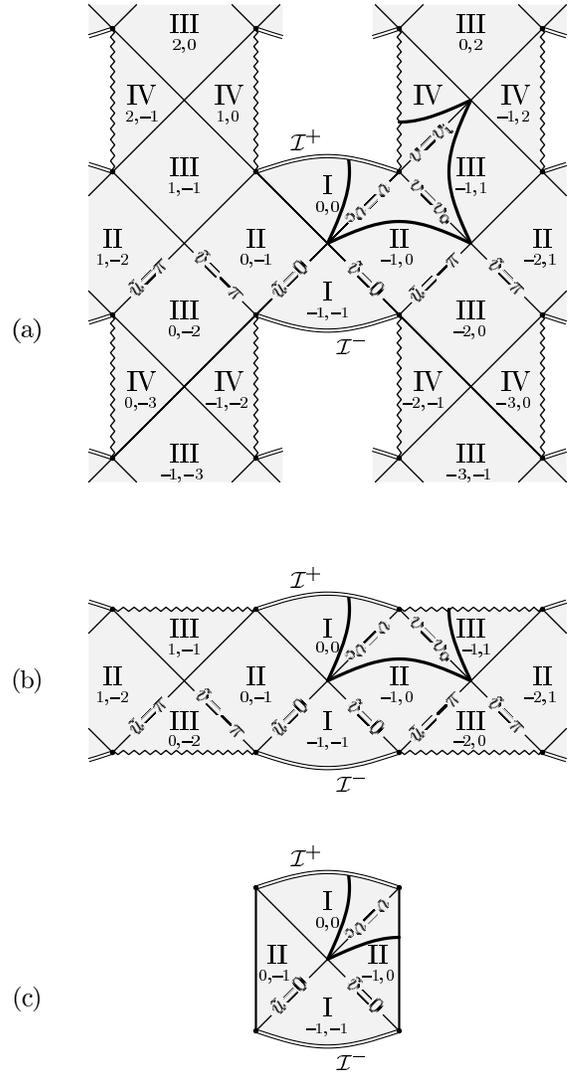}
\caption{\label{fig:ConfDiag}
The conformal diagrams of the ${\tau\textdash\y}$ section (${\x,\ph}=\text{constant}$).
Similarly to Fig.~\ref{fig:xyDiag}, the three
diagrams correspond to the cases of (a) charged accelerated black holes,
(b) uncharged black holes, and (c) de~Sitter
universe (in accelerated coordinates). The conformal infinities $\scri$ are
indicated by double lines, the singularities are drawn by \vague{zig-zag} lines,
and horizons by thin lines.
The horizons ${\y=\y_\chor,\,\y_\ohor,\,\y_\ihor}$ correspond to
the values ${\uGN=m\pi}$ or ${\vGN=n\pi}$, ${m,n\in\integern}$.
Thus, the integers ${(m,\,n)}$, indicated in the figure, label different blocks
${\uGN\in( m\pi,(m+1)\pi)}$,
${\vGN\in( n\pi,(n+1)\pi)}$
of the conformal diagrams. There are four types of these blocks, labeled by I--IV,
which correspond to the regions I--IV in Fig.~\ref{fig:xyDiag}.
The sections ${\tau=\text{constant}}$ (drawn in Fig.~\ref{fig:xyDiag}) are indicated by thick lines.
Similar lines could, of course, be drawn also in other blocks.
Only a part of the complete conformal diagram is shown in the cases (a) and
(b), however, the rest of the diagram would have a similar structure as the part
shown. The complete diagram depends on a freedom in the choice of a global topology of the spacetime
given by identifications of different blocks of the conformal
diagram. In the case (c), the diagram does not contain any black hole ---
it is \vague{closed on its sides} by poles of a spacelike section $S^3$
of de~Sitter universe (see the discussion at the end of Section~\ref{sc:GlobStr}).}%
\end{figure}

For a further discussion of the global structure
we employ the double null
coordinates ${\uGN,\,\vGN,\,\x,\,\ph}$ defined  by Eqs.~\eqref{ae:GNcoor},
\eqref{ae:DNcoor}, \eqref{ae:ysder} in which the metric
is \eqref{ae:GNmetric},
\begin{equation}\label{GNmetric}
  \mtrc = \rRT^2\Bigl(\frac{2\,\GNcoef^2\F}{\sin\uGN\sin\vGN}\,
    \grad\uGN\stp\grad\vGN
    +\frac1\G\,\grad\x^2+\G\,\grad\ph^2\Bigr)\period
\end{equation}
Using these coordinates we can  draw the conformal diagram
of the spacetime section ${\tau\textdash\y}$, i.e., for
${\x,\,\ph=\text{constant}}$ --- see Fig.~\ref{fig:ConfDiag}.
The domains I--IV in this figure correspond to the regions I--IV in
Fig.~\ref{fig:xyDiag}.

The region~I describes the domain of spacetime
above the cosmological horizons given by ${\y=\y_\chor}$, which has a
similar structure as an analogous domain in the de~Sitter spacetime.
The region~II corresponds to a static
spacetime domain between the cosmological
horizon and the (outer) horizon of the black hole.
If the black hole is charged (${\charge\ne0}$), region~III
corresponds to a spacetime domain between the outer horizon
${\y=\y_\ohor}$ and the inner horizon ${\y=\y_\ihor}$,
and region~IV to a domain below the inner horizon of the
black hole (similar to the analogous domains of the Reissner-Nordstr\"om
spacetime). The domain~IV contains a timelike singularity at ${\y=\infty}$ (${\rRT=0}$).
In the uncharged case (${\charge=0}$, ${\mass\ne0}$) there is
only region~III which corresponds to a domain below
the single black hole horizon ${\y=\y_\ohor}$. In this case the singularity at ${\y=\infty}$
has a spacelike character, similarly as for the Schwarzschild black hole.
If both ${\charge=0}$, ${\mass=0}$, we obtain de~Sitter spacetime expressed in accelerated
coordinates. In this case there is no black hole horizon,
and region~II (the domain below the cosmological horizon ${\y_\chor}$)
is \vague{cut off} by nonsingular poles ${\y=\infty}$. We will return to this
particular case at the end of this section.

\begin{figure*}
\includegraphics{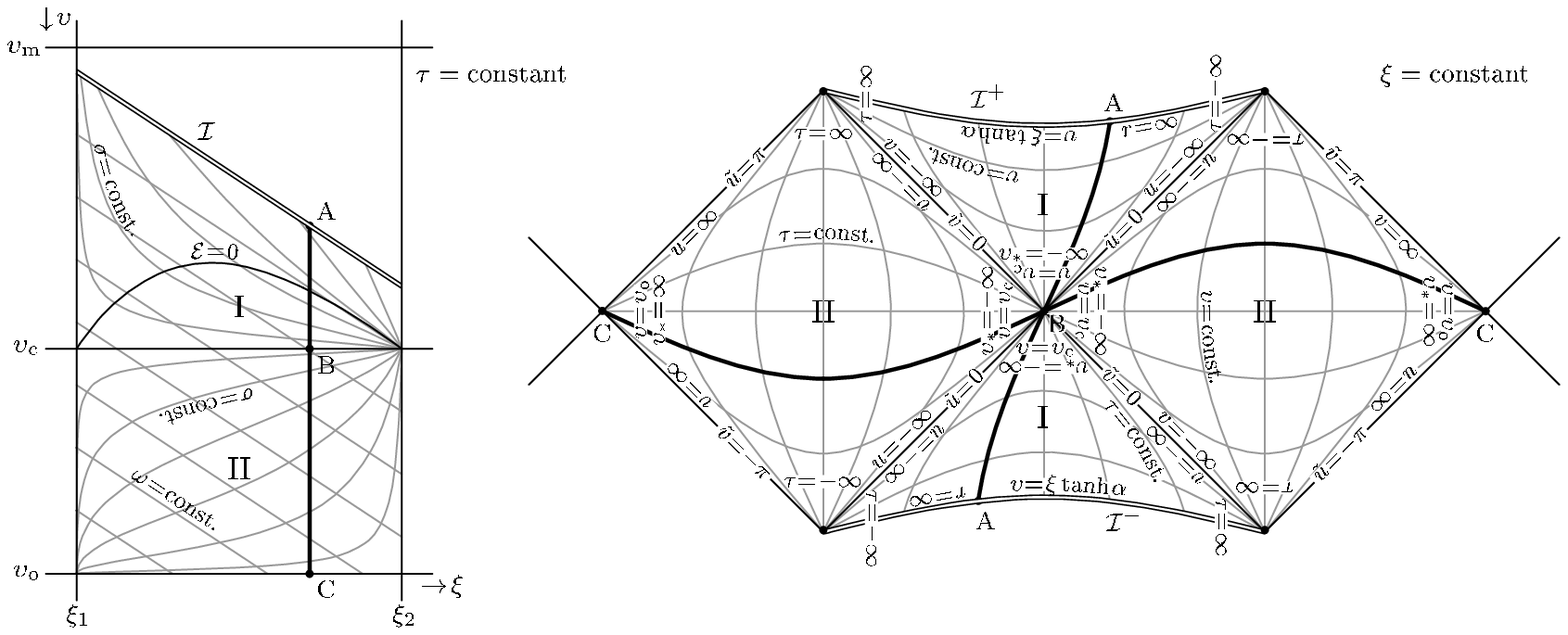}
\caption{\label{fig:DiagDet}
The ${\x\textdash\y}$ (left) and the conformal (right) diagrams near  infinity $\scri$.
Only one asymptotically de~Sitter-like region of the spacetime
(domains I and II of Figs.~\ref{fig:xyDiag} and \ref{fig:ConfDiag})
is shown. Ranges of various coordinates introduced in the paper are indicated
(orientation of the coordinate labels suggests a direction in which the coordinates increase).
The thick line in the conformal diagram corresponds to the
${\x\textdash\y}$ diagram and vice versa. In the ${\x\textdash\y}$ diagram
the lines of constant $\om$ and $\sg$ are also drawn. The coordinates ${\om,\,\sg}$ are not
unambiguous in the full domain I, however, they are invertible near $\scri$, in the domain ${\E<0}$.
On the boundary ${\E=0}$ (shown in diagram) the coordinates ${\om,\,\sg}$ change their timelike/spacelike character.
(Notice the difference between the null coordinate $\vRT$
--- diagonal straight lines --- and the \vague{radial}
coordinate $\y$ --- curved lines --- in the ${\tau\textdash\y}$ conformal diagram.)}%
\end{figure*}

Before we proceed to discuss further properties in detail let us note
that (as will be explicitly demonstrated in the next section) the $C$-metric is the
Petrov type~$D$ spacetime, i.e., it admits two double principal null directions.
These directions lie exactly in the section ${\tau\textdash\y}$ depicted
in Fig.~\ref{fig:ConfDiag}.

In this paper we are mainly interested in a behavior of fields near the
infinity. Therefore, we will concentrate mostly on the region I.
This region has similar properties for all possible values of the
parameters $\mass$, $\charge$, and $\acp$.
Its more precise diagrams are drawn in Fig.~\ref{fig:DiagDet}.
Observers in one of the regions~I (near the future infinity~$\scri^+$) will
consider themselves to live in an asymptotically de~Sitter-like universe \vague{containing}
two causally disconnected black holes (for ${\mass\ne0}$).
Here by \emph{two} black holes we understand those
black holes (i.e., regions III and IV)
immediately \vague{visible} from the given asymptotical
region~I, although the geodesically complete spacetime can, of course, contain
an infinite number of black holes.
As we have said, the conformal infinity $\scri$ is given by the condition
\eqref{xyscricond}, ${\om=0}$. Thanks to a timelike character of ${\grad\om}$ at
${\om=0}$ (cf.\  Eq.~\eqref{omsgmetric})
the infinity has indeed a spacelike character as for de~Sitter universe
(see Refs.~\cite{Penrose:1964,Penrose:1965,PenroseRindler:book}
for a general discussion of conformal infinity).
In Fig.~\ref{fig:xyDiag} the infinity corresponds to the diagonal line, in
Fig.~\ref{fig:ConfDiag}, however, it obtains a richer structure.
It comprises of two parts --- \defterm{future infinity} $\scri^+$ and
the \defterm{past infinity} $\scri^-$ --- both possibly consisting of several disjoint parts
(depending on the global topology) in different asymptotically de~Sitter domains I.
Because the conformal diagrams in Fig.~\ref{fig:ConfDiag} are slices with a
fixed coordinate $\x$, and the condition \eqref{xyscricond} depends on $\x$,
the conformal infinity $\scri$ would have a different
position in diagrams with different values of~$\x$.
We shall return to this fact at the end of this section. Note,
that for values of the coordinate $\y$ smaller than $\,{\x_\paxis\tanh\acp}$, the
hypersurface ${\y=\text{constant}}$ reaches  $\scri$.
Clearly, the coordinate $\y$ is not well
adapted to the region near the conformal infinity $\scri$.
Near the infinity it is more convenient to use the coordinates
${\om,\,\tau,\,\sg,\,\ph}$ defined by Eqs.~\eqref{omsgdef}
(cf.\  Eqs.~\eqref{ae:omdef}, \eqref{ae:sgdef}; see also Fig.~\ref{fig:DiagDet}).

The coordinate $\tau$ is a coordinate along the \vague{boost} Killing vector
$\cv\tau$, and in region~I it can be understood simply
as a translational spatial coordinate.
The coordinates ${\x,\,\ph}$ play roles of longitudinal and latitudinal coordinates
of a suitably defined hypersurface at an \vague{instant of time}.
For example, in region~I the spacelike
hypersurface ${\y=\text{constant}}$ has topology ${\realn\times S^2}$
(if it does not cross  infinity~$\scri$) with
the coordinate $\tau$ along the $\realn$~direction, and ${\x,\,\ph}$ on the sphere $S^2$.
To justify the \vague{longitudinal} character of the coordinate $\x$, we
introduce, instead of $\x$, an angular coordinate $\vartheta$ by
the relation ${\sin\vartheta=\sqrt{\G}}$ (cf.\  Eq.~\eqref{ae:Gval}).
This is a longitudinal angle measured by a circumference of the $\ph$-circle
(see the metric \eqref{xymetric}). Alternatively, we can introduce the angle
$\thacc$, defined by Eq.~\eqref{ae:xytoacc}, measured by the length of a \vague{meridian}.
At infinity $\scri$ or, in general, on any hypersurface ${\om=\text{constant}}$,
the coordinate lines ${\sg=\text{constant}}$ coincide with the lines of constant $\x$.
The coordinate $\sg$ thus also parametrizes the longitudinal direction near the
infinity, similarly to the coordinate~$\x$.

In Section~\ref{sc:Cmetric} we mentioned that the coordinate $\ph$ along
the second Killing vector $\cv\ph$ takes values
in the interval $(-\pi\conpar,\pi\conpar)$. Here ${\conpar>0}$ is the parameter
which allows us to change the conicity on the axis of the $\ph$~symmetry, i.e., it allows
us to choose a deficit (or excess) angle around the axis arbitrarily.
Such a change of the range of the coordinate $\ph$ is allowed for any axially symmetric
spacetime. The range is usually chosen in such a way that the axis of the
$\ph$~symmetry is regular. However, for the $C$-metric such a choice is not
globally possible. In this case the axis consists of two parts ${\x=\x_\maxis}$ and ${\x=\x_\paxis}$
--- one of them joins the \vague{north} poles of the black holes, the other one joins
the \vague{south} poles. The physical \defterm{conicity} (defined as a limiting
ratio of $\text{\vague{circumference}}$ and ${2\pi\!\times\!\text{\vague{radius}}}$ of a small circle
around the axis) calculated at the axes ${\x_\maxis}$ and ${\x_\paxis}$ is
\begin{equation}\label{conicity}
  \conicity_\maxis = {\textstyle\frac12}\,\conpar\,\G'\vert_{\x=\x_\maxis}\comma
  \conicity_\paxis =
  -{\textstyle\frac12}\,\conpar\,\G'\vert_{\x=\x_\paxis}\commae
\end{equation}
respectively, see, e.g., Ref.~\cite{PodolskyGriffiths:2001}.
In general, the values of ${\abs{\G'}}$ at ${\x_\maxis}$ and ${\x_\paxis}$
are not the same, cf.\  Eq.~\eqref{ConOrd} below.
Therefore we can set ${\conicity=1}$ (zero deficit of angle, i.e., a regular axis)
by a suitable choice of the parameter $\conpar$ only at \emph{one} part of the axis.

This fact has a clear physical interpretation.
The axis with nonregular conicity corresponds to a cosmic string
which causes the \vague{accelerated motion} of the black holes.
The cosmic string \cite{VilenkinShellard:book} is a one-dimensional object,
sort of a \vague{rod} or a \vague{spring},
which is characterized by its mass density equal to its linear tension.
These parameters are proportional to the deficit angle, namely, a string with a
deficit angle (${\conicity<1}$) has a positive mass density and it is stretched, a string with
an excess angle (${\conicity>1}$) has negative mass density and is squeezed.
In Appendix~\ref{apx:FGprop}, Eq.~\eqref{ae:ConOrd}, we prove for ${\mass\neq0}$,
${\accl\neq0}$ that
\begin{equation}\label{ConOrd}
  \conicity_\paxis< \conicity_\maxis\period
\end{equation}
Using this fact, we may conclude that by eliminating a nontrivial conicity at the axis ${\x=\x_\paxis}$
(so that ${\conicity_\paxis=1}$) we obtain ${\conicity_\maxis>1}$, i.e., a squeezed cosmic string at the axis
${\x=\x_\maxis}$. Alternatively, if we set the physical conicity ${\conicity_\maxis=1}$
at ${\x=\x_\maxis}$, we obtain ${\conicity_\paxis<1}$, i.e.,
a stretched cosmic string at the axis ${\x=\x_\paxis}$.
In both these cases, as well as in the general
cases of cosmic strings on both parts of the axis, the system
of black holes with string(s) between them is not in an equilibrium. The string(s)
acts on both black holes and cause what we  usually
call an \vague{accelerated motion} of black holes. However, the precise
interpretation of acceleration is not so straightforward.

The problem here is that we consider a fully self-gravitating system, not just a motion
of test particles on a fixed background. The motion of black holes is actually realized
through a nonstatic, nonspherical deformation of geometry of the spacetime in a
direction of motion, i.e., along the axis of $\ph$~symmetry. Moving black holes
together with the cosmic string(s) curve the spacetime in such a way
that, strictly speaking, it is not justified to use the term
\emph{acceleration} in a rigorous sense. This has several reasons.
First, black holes are nonlocal objects and one can hardly expect a uniquely defined
acceleration for such extended objects. Secondly, thanks to the equivalence
principle we cannot distinguish between acceleration of the black holes with respect to the
universe, and acceleration due to the gravitational field of each hole.
Finally, one has to expect a gravitational dragging of local inertial
frames by moving black holes, i.e., it is not obvious how to define an
acceleration of black holes with respect to these frames.
A plausible definition could be given if some privileged cosmological
coordinate system playing a role of \vague{nonmoving} background is available.
Unfortunately, we are not aware of such a system applicable in a general case. In the next
paragraph we shall demonstrate this approach just for a simple case of empty de~Sitter spacetime.
Summarizing, it is not straightforward to define the
acceleration of black holes in the general case. One usually identifies the
acceleration only in an appropriate limiting regime.
The usage of the term \emph{acceleration} for
the parameter $\accl$ in the \mbox{$C$-metric} (see Refs.~\cite{Bonnor:1982,BicakPravda:1999,LetelierOliveira:2001} for the
case ${\Lambda=0}$, and, e.g., Refs.~\cite{PodolskyGriffiths:2001} for the case ${\Lambda\neq0}$)
has been justified exactly in this way.

\begin{figure}
\includegraphics{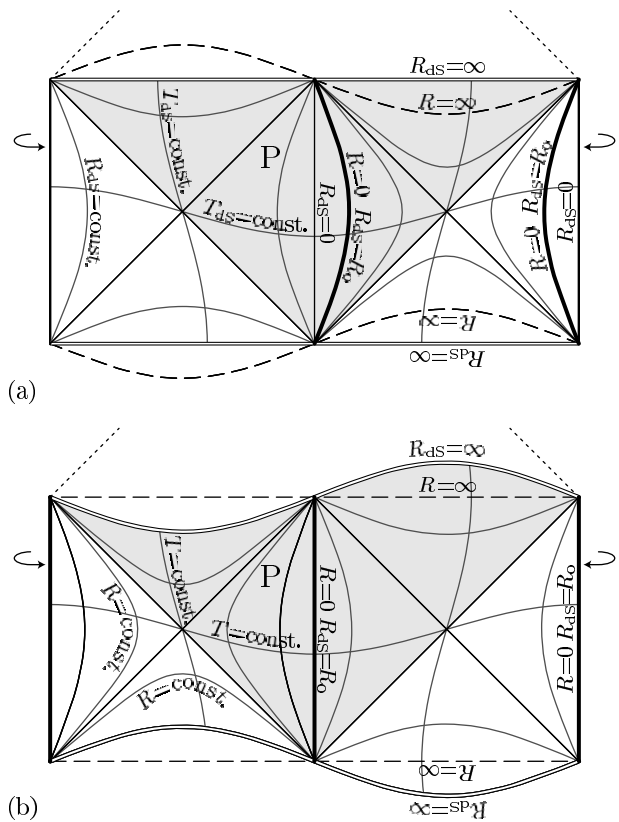}
\caption{\label{fig:deSitter}
Conformal diagrams of de~Sitter universe (a) in the standard cosmological
coordinates \eqref{ae:dSacctostat}, and (b) based on the accelerated coordinates
\eqref{xytoacc} (cf.\ Eq.~\eqref{ae:xytoacc}).
In  contrast to Fig.~\ref{fig:ConfDiag}(c), the diagram (b) depicts
two sections of constant $\x$, namely ${\x=\x_\maxis}$ (${\thacc=0}$) at the right half of the diagram,
and ${\x=\x_\paxis}$ (${\thacc=\pi}$) at the left half. We can see that the
position of infinity (double line) is different for these two values
of $\x$. For intermediate values of $\x$  infinity $\scri$ would attain an
intermediate position at $\,{\racc=\DSr/\x\,\coth\acp}\,$, according to Eq.~\eqref{xyscricond}.
The infinity has a simple shape in diagram (a), where it is indicated by
the horizontal lines ${\rdS=\infty}$.
In both diagrams the left and right boundaries are identified
--- they correspond to one of the two poles of the appropriate coordinates
(the other pole is located in the center of the diagram).
A horizontal line thus corresponds to the main
circle of a spatial $S^3$ section of de~Sitter universe.
Bold lines corresponds to the origins of the accelerated coordinates (${\racc=0}$)
which have been employed in the paper as \vague{remnants} of the sources.
In diagram (a) they move with respect to the cosmological frame.
Diagram (b) is adapted to their accelerated motion and therefore the sources are
located at origins.
Dashed line corresponds to value ${\racc=\infty}$, i.e., ${\y=0}$, where the
accelerated coordinates are not well defined. Relative position of the
hypersurface ${\racc=\infty}$ and of  infinity ${\rdS=\infty}$ can be
visualized with help of a conformally related Minkowski space (lower half indicated
by the shaded domain P, the upper half indicated by dotted line). In this space
the infinity corresponds to hypersurface ${\tM=0}$, the coordinate singularity
${\racc=\infty}$ corresponds to ${\tM'=0}$, where $\tM$ and $\tM'$ are Minkowski
time coordinates in inertial frames moving with relative velocity ${\tanh\acp}$
--- see Appendix~\ref{apx:Coor} for a related discussion.}%
\end{figure}

Of course, the situation extremely simplifies in the case of vanishing mass and
charge (${\mass=0}$, ${\charge=0}$). In this case there are no black
holes, and the spacetime reduces to de~Sitter universe. However, if we still keep
${\accl\neq0}$, a trace of (now vanished) \vague{accelerated} sources
remains in the metric \eqref{xymetric} through the parameter $\acp$. By a simple transformation
\eqref{ae:xytoacc},
\begin{equation}\label{xytoacc}
  \tacc = \DSr\,\tau\comma \racc=\frac\DSr\y\comma
  \cos\thacc=-\x\comma\phacc=\ph\commae
\end{equation}
we obtain the metric \eqref{ae:dSmetricacc} of the de~Sitter space in
\defterm{accelerated coordinates} ${\tacc,\,\racc,\,\thacc,\,\phacc}$
introduced in Ref.~\cite{PodolskyGriffiths:2001} and discussed in Ref.~\cite{BicakKrtous:BIS}.
These coordinates are analogue of the Rindler
coordinates in Minkowski space generalized to the case of the de~Sitter universe. They are
adapted to accelerated observers: the origins ${\racc=0}$
represent two uniformly accelerated observers which are decelerating from
antipodal poles of the spherical space section of the
de~Sitter universe toward each other until the moment of minimal contraction of
the universe, and then accelerate away back to the antipodal poles (see Fig.~\ref{fig:deSitter}).
In the standard de~Sitter static coordinates
${\tdS,\,\rdS,\,\thdS,\,\phdS}$ of the metric \eqref{ae:dSmetric},
related to Eq.~\eqref{xytoacc} by Eq.~\eqref{ae:dSacctostat},
these observers are characterized by ${\rdS=\Ro}$, ${\thdS=0}$.
Thus, they are static observers staying at constant
distance ${\Ro=\DSr\tanh\acp}$ (cf.\  Eq.~\eqref{ae:Rodef})
from the poles ${\rdS=0}$ of de~Sitter space, measured in their
instantaneous rest frame (or, equivalently, in the de~Sitter static frame).
They are uniformly
accelerated with acceleration ${\accl=\DSr^{-1}\sinh\acp}$ toward these poles ---
in fact, this acceleration exactly compensates the acceleration due to cosmological
contraction and subsequent expansion of de~Sitter universe.

We can consider the above accelerated observers as \vague{remnants} of accelerated
black holes of the full $C$-metric universe. Of course, in the oversimplified case of
de~Sitter space these \vague{sources} just move along the worldlines and we are able to measure
their acceleration explicitly. It is thus natural to draw the conformal diagram
(Fig.~\ref{fig:deSitter}(a)) of de~Sitter universe, based on the
standard global cosmological coordinates, in which the remnants of sources are
obviously depicted as moving \vague{objects}. On other hand, we can draw an alternative
conformal diagram based on the accelerated coordinates (Fig.~\ref{fig:deSitter}(b)),
in which the remnants of the sources are located at the \vague{fixed} poles of
the space sections of the universe. The diagram in Fig.~\ref{fig:deSitter}(a) is adapted to
global cosmological structure of the universe and explicitly visualizes the motion
of the sources, whereas the diagram in Fig.~\ref{fig:deSitter}(b) is
adapted to sources and thus \vague{hides} their motion.

This intuition can be carried on to the general case with  nontrivial sources. The coordinates
${\tau,\,\y,\,\x,\,\ph}$ (or alternatively the accelerated coordinates defined
in the general case by Eq.~\eqref{ae:xytoacc}) are adapted to sources and thus the
conformal diagrams in Fig.~\ref{fig:ConfDiag} \vague{hide} the motion of the black holes.
Therefore, it would be very useful to find an analogue of the coordinates of
Fig.~\ref{fig:deSitter}(a) for the general case ${\mass\neq0}$, ${\charge\neq0}$, to be able to explicitly identify the
accelerated motion of the black holes. However, as was  already mentioned, we are not
aware of such coordinates.

Using the insight obtained from the de~Sitter case, we also observe that the
\vague{changing of shape} of  infinity $\scri$ in the conformal diagrams for different values
of the coordinate $\x$, as discussed above, is actually an \emph{evidence} of nonvanishing
acceleration of the sources. In the case of pure de~Sitter space we have obtained
this \vague{changing of position} of $\scri$ when we have used the coordinates adapted
to the accelerated observers. We expect that the analogous \vague{changing of shape}
of the infinity in a general case also indicates accelerated motion of the sources.

\section{Privileged orthonormal and null tetrads near $\scri^+$}
\label{sc:tetrads}

We wish to investigate properties of null geodesics and
the character of fields near infinity $\scri$
(domain~I in Figs.~\ref{fig:xyDiag}, \ref{fig:ConfDiag}). Therefore, we will
assume ${\F<0}$, ${\G>0}$, and ${\E<0}$. Before we discuss the geodesics and behavior
of the fields
we first introduce some privileged tetrads which will be used for physical
interpretation. In the following, we will denote by $\ct_{\alpha}$ a normalized
vector tangent to the coordinate $x^\alpha$, i.e., the unit
vector proportional to the coordinate vector $\cvil{\alpha}$.

We will employ several types of orthonormal and null tetrads which will be
distinguished by specific labels in subscript.
We denote the vectors of an orthonormal tetrad  as
${\nG,\,\fG,\,\rG,\,\sG}$. Here $\nG$ is a unit timelike
vector and the  remaining three are spacelike.
With this normalized tetrad we associate a null tetrad
of null vectors ${\kG,\,\lG,\,\mG,\,\bG}$, such that
\begin{equation}\label{NormNullTetr}
\begin{aligned}
  \kG &= \textstyle{\frac1{\sqrt{2}}} (\nG+\fG)\comma&
  \lG &= \textstyle{\frac1{\sqrt{2}}} (\nG-\fG)\commae\\
  \mG &= \textstyle{\frac1{\sqrt{2}}} (\rG-i\,\sG)\comma&
  \bG &= \textstyle{\frac1{\sqrt{2}}} (\rG+i\,\sG)\period
\end{aligned}
\end{equation}
Using the associated tetrad of null 1-forms ${\KG,\,\LG,\,\MG,\,\BG}$
dual to the null tetrad ${\kG,\,\lG,\,\mG,\,\bG}$,
the metric can be written as
\begin{equation}\label{metrNullTetr}
  \mtrc = -\KG\stp\LG + \MG\stp\BG\commae
\end{equation}
which implies
\begin{equation}\label{NullTetrNorm}
   \kG\spr\lG = -1\comma
   \mG\spr\bG = 1\commae
\end{equation}
all other scalar products being zero.
From this it follows that
\begin{equation}\label{NullTetr}
\begin{aligned}
  \KG_\alpha &= -\mtrc_{\alpha\beta}\,\lG^\beta\comma&
  \LG_\alpha &= -\mtrc_{\alpha\beta}\,\kG^\beta\commae\\
  \MG_\alpha &= \mtrc_{\alpha\beta}\,\bG^\beta\comma&
  \BG_\alpha &= \mtrc_{\alpha\beta}\,\mG^\beta\period
\end{aligned}
\end{equation}

The Weyl tensor $\WT_{\alpha\beta\gamma\delta}$ has ten independent real components which can be parametrized
by five standard complex coefficients defined as its components
with respect to the above null tetrad
(see, e.g., Refs.~\cite{Krameretal:book,PenroseRindler:book}):
\begin{equation}\label{PsiDef}
\begin{aligned}
  \WTP{}{0} &= \spcm\WT_{\alpha\beta\gamma\delta}\,
    \kG^\alpha\,\mG^\beta\,\kG^\gamma\,\mG^\delta\commae\\
  \WTP{}{1} &= \spcm\WT_{\alpha\beta\gamma\delta}\,
    \kG^\alpha\,\lG^\beta\,\kG^\gamma\,\mG^\delta\commae\\
  \WTP{}{2} &= -\WT_{\alpha\beta\gamma\delta}\,
    \kG^\alpha\,\mG^\beta\,\lG^\gamma\,\bG^\delta\commae\\
  \WTP{}{3} &= \spcm\WT_{\alpha\beta\gamma\delta}\,
    \lG^\alpha\,\kG^\beta\,\lG^\gamma\,\bG^\delta\commae\\
  \WTP{}{4} &= \spcm\WT_{\alpha\beta\gamma\delta}\,
    \lG^\alpha\,\bG^\beta\,\lG^\gamma\,\bG^\delta\period
\end{aligned}
\end{equation}
The coefficients $\WTP{}{n}$ transform in a simple way under special Lorentz
transformations of the null tetrad ${\kG,\,\lG,\,\mG,\,\bG}$,
namely, under null rotation around null vectors $\kG$ or $\lG$, under a boost
in the ${\kG\textdash\lG}$ plane, and a spatial rotation in
the ${\mG\textdash\bG}$ plane \cite{Krameretal:book}.
These transformation are summarized in Appendix~\ref{apx:Transformations}.
The tensor of electromagnetic field $\EMF_{\alpha\beta}$ has six independent real
components which can be parametrized, similarly to the Weyl
tensor, as
\begin{equation}\label{PhiDef}
\begin{aligned}
  \EMP{}{0} &= \EMF_{\alpha\beta}\,
    \kG^\alpha\,\mG^\beta\commae\\
  \EMP{}{1} &= {\textstyle\frac12}\,\EMF_{\alpha\beta}\,
    \bigl(\kG^\alpha\,\lG^\beta-\mG^\alpha\,\bG^\beta\bigr)\commae\\
  \EMP{}{2} &= \EMF_{\alpha\beta}\,
    \bG^\alpha\,\lG^\beta\period
\end{aligned}
\end{equation}
The transformation properties of coefficients $\EMP{}{n}$ under the
null rotations, special boost, and spatial rotation can also be found in
Appendix~\ref{apx:Transformations}.

Now, we first introduce an \defterm{algebraically special tetrad} ${\nS,\,\fS,\,\rS,\,\sS}$ which
is associated with the \emph{principal null directions} of the $C$-metric spacetime. We define
\begin{equation}\label{AlgSpecTetr}
\begin{gathered}
  \nS = -\ct_{\y} = -\frac{\sqrt{-\F}}\rRT \,\cv{\y}\comma
  \rS = \ct_{\x} = \frac{\sqrt{\G}}\rRT  \,\cv{\x}\commae\\
  \fS = -\ct_{\tau} = - \frac1{\rRT\sqrt{-\F}} \,\cv{\tau}\comma
  \sS = \ct_{\ph} = \frac1{\rRT\sqrt{\G}} \,\cv{\ph}\commae
\end{gathered}
\end{equation}
and the corresponding null tetrad ${\kS,\,\lS,\,\mS,\,\bS}$ by
Eqs.~\eqref{NormNullTetr}. It is straightforward to check
that these null directions ${\kS,\,\lS}$ can be expressed as
\begin{equation}\label{AlgSpecNullTetr}
  \kS =   \frac{\sin\vGN}{\sqrt{2}\rRT\abs{\GNcoef}} \frac1{\sqrt{-\F}} \,\cv{\vGN}\comma
  \lS =   \frac{\sin\uGN}{\sqrt{2}\rRT\abs{\GNcoef}} \frac1{\sqrt{-\F}} \,\cv{\uGN}\commae
\end{equation}
where the global null coordinates ${\uGN,\,\vGN}$, parametrized by a constant $\GNcoef$,
are introduced in Eq.~\eqref{ae:GNcoor}.
It turns out that the Weyl tensor has the simplest form in this tetrad.
It can be expressed as
\begin{equation}\label{Weylxy}
\begin{split}
  \WT =& \frac1{12}(\F''+\G'')\,\rRT^2\;\times\\
       &\Bigl(\frac1{\F\G}\,\grad\y\wedge\grad\x\;\grad\y\wedge\grad\x
        -\F\G\,\grad\tau\wedge\grad\ph\;\grad\tau\wedge\grad\ph\\
       &\;\,+\frac\G\F\,\grad\y\wedge\grad\ph\;\grad\y\wedge\grad\ph
        -\frac\F\G\,\grad\tau\wedge\grad\x\;\grad\tau\wedge\grad\x\\
       &\;\,+2\,\grad\tau\wedge\grad\y\;\grad\tau\wedge\grad\y
        -2\,\grad\x\wedge\grad\ph\;\grad\x\wedge\grad\ph\Bigr)\period
\end{split}\raisetag{89pt}
\end{equation}
Transforming this into the null tetrad ${\KS,\,\LS,\,\MS,\,\BS}$
we find that the only nonvanishing component is
$\WTP{\spec}{2}$, namely,
\begin{equation}\label{WeylAlgSpecTetr}
\begin{aligned}
 \WTP{\spec}{2} &={\textstyle{\frac1{12}}}(\F''+\G'')\,\rRT^{-2}\\
 &=-\Bigl(\frac{\mass}{\DSr} -
  \frac{\charge^2}{\DSr^2}\,(\y\cosh\acp+\x\sinh\acp)\Bigr)\,\frac{\DSr}{\rRT^3}\\
 &=-\Bigl(\mass - 2\,\charge^2\accl\,\x - \frac{\charge^2}\rRT\Bigr)\,\frac1{\rRT^3}\commae\\
 \WTP{\spec}{0} &=\WTP{\spec}{1}=\WTP{\spec}{3}=\WTP{\spec}{4}=0\period
\end{aligned}
\end{equation}
This exhibits explicitly that $\kS$, $\lS$ are the double principal
null directions \cite{Krameretal:book}, which lie in the ${\tau\textdash\y}$
plane.

Similarly, the electromagnetic field tensor \eqref{KWEMF} in coordinates
${\tau,\,\y,\,\x,\,\ph}$ reads
\begin{equation}\label{xyEMF}
  \EMF = \charge\, \grad\y\wedge\grad\tau\period
\end{equation}
Using relations \eqref{NormNullTetr},
\eqref{AlgSpecTetr}, we find that the only nonvanishing
coefficient of electromagnetic field is $\EMP{\spec}{1}$,
\begin{equation}\label{EMFAlgSpecTetr}
  \EMP{\spec}{1}=-\frac{\charge}{2\,\rRT^2}\comma
  \EMP{\spec}{0} = \EMP{\spec}{2} = 0\period
\end{equation}

The special null tetrad defined above is appropriate
for discussion of algebraic properties of the fields.
However, near future infinity $\scri^+$ we will
also have to use a different tetrad ${\nK,\,\fK,\,\rK,\,\sK}$ and the
related null tetrad ${\kK,\,\lK,\,\mK,\,\bK}$.
These will serve as \defterm{reference tetrads} with respect to which
we will parametrize a general asymptotic direction. These tetrads are adapted to the Killing
vectors $\cvil{\tau}$, $\cvil{\ph}$ and to de~Sitter-like infinity $\scri^+$. Namely, the
timelike vector $\nK$ is asymptotically orthogonal to $\scri^+$,
and ${\fK,\,\rK,\,\sK}$ are tangent to $\scri^+$. We define
\begin{equation}\label{KillTetr}
\begin{gathered}
  \nK = \ct_{\om} = \frac{\sqrt{-\E}}\rRT  \,\cv{\om}\comma
  \rK = \ct_{\sg} = \frac1\rRT \sqrt{\frac{\E}{\F\G}} \,\cv{\sg}\commae\\
  \fK = -\ct_{\tau} = - \frac1{\rRT\sqrt{-\F}} \,\cv{\tau}\!\comma\!
  \sK = \ct_{\ph} = \frac1{\rRT\sqrt{\G}} \,\cv{\ph}\commae
\end{gathered}
\end{equation}
the corresponding null tetrad ${\kK,\,\lK,\,\mK,\,\bK}$ is given by Eqs.~\eqref{NormNullTetr}.

\begin{figure}
\includegraphics{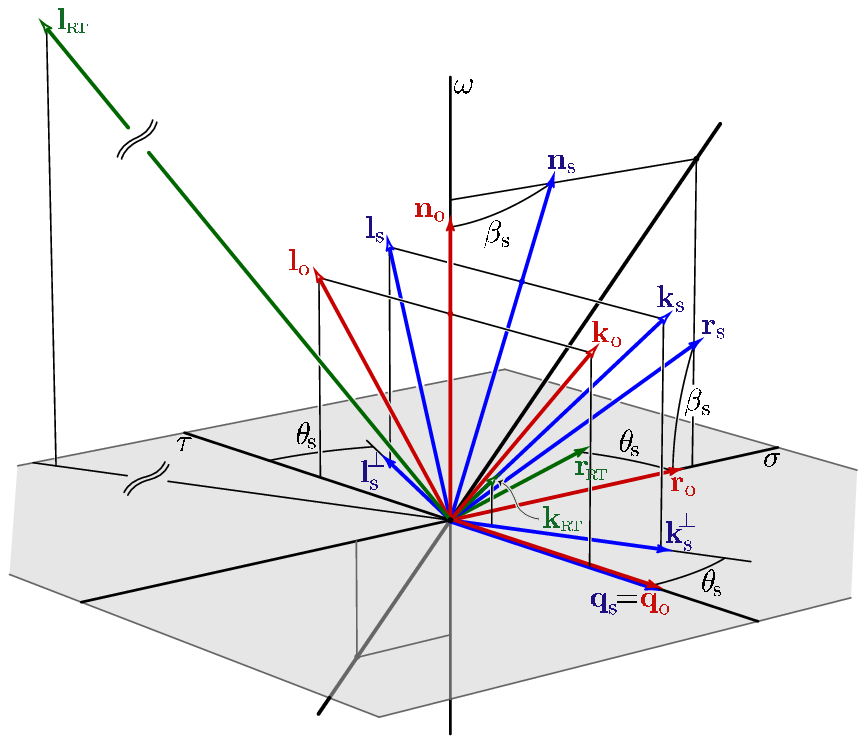}
\caption{\label{fig:tetrads}
A spacetime diagram ($\ph$ direction is suppressed) that depicts relations
between the reference tetrad ${\nK,\,\fK,\,\rK,\,\sK}$ (or ${\kK,\,\lK,\,\mK,\,\bK}$),
the algebraically special tetrad ${\nS,\,\fS,\,\rS,\,\sS}$ (or ${\kS,\,\lS,\,\mS,\,\bS}$),
and the Robinson-Trautman tetrad ${\kT,\,\lT,\,\mT,\,\bT}$. The reference tetrad is naturally
adapted to the infinity ($\nK$ is normal to $\scri^+$) and the Killing vectors
($\fK$ and $\sK$ are tangent to them), while
the algebraically special tetrad is adapted to both double principal null directions $\kS$ and $\lS$.
These two are related by a boost in the ${\nK\textdash\rK}$ plane, with
the boost parameter $\beta_\spec$ given by Eq.~\eqref{betasDef}. The vectors $\fK$
and $\fS$ are identical, similarly ${\sK=\sS}$. Orthogonal projections $\kS^\perp$, $\lS^\perp$
of the principal null directions onto ${\om=\text{constant}}$ hyperplane (shaded)
define the angle $\THT_\spec$ (Eq. \eqref{THTsDef}) that,
similarly to $\beta_\spec$, characterizes the relation between the reference and the special
tetrads. The vector $\kT$ of the Robinson-Trautman tetrad points into the
principal null direction $\kS$ with the coefficient of proportionality approaching zero
on $\scri^+$, cf.~Eq.~\eqref{BparRTdef}. The other null direction $\lT$ belongs to the ${\nK\textdash\kS}$ plane and
it becomes \vague{infinitely long} on~$\scri^+$.}%
\end{figure}

Relations between the tetrads ${\nS,\,\fS,\,\rS,\,\sS}$ and ${\nK,\,\fK,\,\rK,\,\sK}$ immediately follow
from the definitions \eqref{AlgSpecTetr}, \eqref{KillTetr}, and from relations
of  coordinates \eqref{omsgdef} (cf.\ Eqs.~\eqref{VecCoorFy}, \eqref{VecCoorFx}),
\begin{equation}\label{ASKtetrRel}
\begin{gathered}
  \nS = \sqrt{\frac\F\E}\cosh\acp\;\nK
         +\sqrt{\frac\G{-\E}}\sinh\acp\;\rK\commae\\
  \rS = \sqrt{\frac\G{-\E}}\sinh\acp\;\nK
         +\sqrt{\frac\F\E}\cosh\acp\;\rK\commae\\
  \fS = \fK \comma \sS = \sK\period
\end{gathered}
\end{equation}
A geometrical meaning of these transformations is seen in
Fig.~\ref{fig:tetrads}. Both tetrads are related by a simple boost in the ${\nK\textdash\rK}$
plane with a boost parameter $\beta_\spec$ given by
\begin{equation}\label{betasDef}
  \tanh\beta_\spec = \sqrt{\frac\G{-\F}}\tanh\acp\period
\end{equation}
This boost is described by relations similar to
Eq.~\eqref{ae:boostrotationON}, with the vectors $\fG$ and $\rG$ interchanged.

We obtain even a better visualization if we perform a projection of
the principal null directions ${\kS,\,\lS}$ to a \mbox{3-dimensional} hyperplane orthogonal to
the timelike vector $\nK$. We thus obtain \vague{spatial} directions
${\kS^\perp,\,\lS^\perp}$,
\begin{equation}\label{orthproj}
   \kS^\perp=\kS+(\kS\spr\nK)\,\nK\comma\text{etc.}\commae
\end{equation}
of the null vectors ${\kS,\,\lS}$ which lie in
the ${\fK\textdash\rK}$ plane, symmetrically with respect to the vector $\rK$
(see Fig.~\ref{fig:tetrads}). If we denote by $\THT_\spec$
the angle between $\fK$ and $\kS^\perp$, we can write
${\kS^\perp \propto \sin\THT_\spec\,\rK+\cos\THT_\spec\,\fK}$, and
taking into account the normalization \eqref{NullTetrNorm} we
obtain
\begin{equation}\label{SpecNullToKlg}
\begin{gathered}
  \kS = \frac1{\sqrt2\cos\THT_\spec}\,\bigl(
        \nK + \sin\THT_\spec\;\rK+\cos\THT_\spec\;\fK\bigr)\commae\\
  \lS = \frac1{\sqrt2\cos\THT_\spec}\,\bigl(
        \nK + \sin\THT_\spec\;\rK-\cos\THT_\spec\;\fK\bigr)\commae
\end{gathered}
\end{equation}
cf.\  also Eq.~\eqref{SpecTetrRel}.
Comparing this with  relations \eqref{ASKtetrRel} and using Eq.~\eqref{NormNullTetr}, we
find that the angle $\THT_\spec$ is given in terms of the metric functions
${\F,\,\G,\,\E}$ as
\begin{equation}\label{THTsDef}
  \sin\THT_\spec=\sqrt{\frac\G{-\F}}\,\tanh\acp\comma
  \cos\THT_\spec=\Bigl(\sqrt{\frac\F\E}\,\cosh\acp\Bigr)^{-1}\commae
\end{equation}
i.e., ${\tanh\beta_\spec=\sin\THT_\spec}$.

We will be interested mainly in the tetrads at the conformal
infinity $\scri^+$, i.e., for ${\om=0}$, where  ${\y = \x\tanh\acp}$ and
${\E=-1}$,
see Eqs.~\eqref{xyscricond}, \eqref{ae:Edef}.
From the definitions \eqref{betasDef}, \eqref{THTsDef} and using
Eqs.~\eqref{ae:Gval}, \eqref{ae:Fscrival}
we find that the boost parameter $\beta_\spec$ and the angle $\THT_\spec$
(which both characterize directions \vague{from the sources}) may on the $\scri^+$
have values in the ranges
\begin{equation}
  \beta_\spec \in \lclint0,\acp\rclint\comma
  \sin\THT_\spec \in \lclint 0,\tanh\acp \rclint\period
\end{equation}
The zero values occur on the axis of $\ph$~symmetry
(points \vague{between} the moving black holes; ${\x=\x_\maxis,\x_\paxis}$),
the maximal values occur on the \vague{equator} --- the $\ph$~circle of maximal circumference
(${\x=0}$, ${\y=0}$).

Transformation formulas \eqref{SpecTetrRel} allow us to find  components of the Weyl
tensor and tensor of electromagnetic field in the reference null tetrad
${\kK,\,\lK,\,\mK,\,\bK}$, namely,
\begin{gather}
\begin{gathered}
  \WTP{\klg}{2} = {\textstyle{\frac12}}\,\WTP{\spec}{2}\, \bigl(3\cos^{-2}\THT_\spec-1\bigr)\commae\\
  \WTP{\klg}{1} = \WTP{\klg}{3} =   - {\textstyle{\frac32}}\,\WTP{\spec}{2}\;{\sin\THT_\spec}\;{\cos^{-2}\THT_\spec}\commae\\
  \WTP{\klg}{0} = \WTP{\klg}{4} = {\textstyle{\frac32}}\,\WTP{\spec}{2}\;{\sin^2\THT_\spec}\;{\cos^{-2}\THT_\spec}\commae
\end{gathered}\label{WeylKillTetrTHT}\\
  \EMP{\klg}{0} = \EMP{\klg}{2} = -\tan\THT_\spec\,\EMP{\spec}{1}\comma
  \EMP{\klg}{1} = \cos^{-1}\THT_\spec\,\EMP{\spec}{1}\commae\label{EMFKillTetrTHT}
\end{gather}
or, more explicitly (using Eqs.~\eqref{WeylAlgSpecTetr}, \eqref{EMFAlgSpecTetr},
and \eqref{THTsDef})
\begin{gather}
\begin{gathered}
  \WTP{\klg}{2} = \frac{\F''+\G''}{8\,\E\,\rRT^2}\,\frac13\bigl(2\F\cosh^2\acp -\G\sinh^2\acp\bigr)\commae\\
  \WTP{\klg}{1} = \WTP{\klg}{3} = \frac{\F''+\G''}{8\,\E\,\rRT^2}\,\sqrt{-\F\G}\,\cosh\acp\,\sinh\acp\commae\\
  \WTP{\klg}{0} = \WTP{\klg}{4} =  -\frac{\F''+\G''}{8\,\E\,\rRT^2}\,\G\,\sinh^2\acp\commae
\end{gathered}\label{WeylKillTetr}\\
\begin{gathered}
  \EMP{\klg}{1} = -\frac{\charge}{2\,\rRT^2}\,\sqrt{\frac\F\E}\,\cosh\acp\commae\\
  \EMP{\klg}{0} = \EMP{\klg}{2} =
  \frac{\charge}{2\,\rRT^2}\,\sqrt{\frac\G{-\E}}\,\sinh\acp\period
\end{gathered}\label{EMFKillTetr}
\end{gather}

As we have already mentioned, the tetrad ${\nK,\,\fK,\,\rK,\,\sK}$ serves as the reference
tetrad with respect to which we characterize an arbitrarily \defterm{rotated tetrad} ${\nR,\,\fR,\,\rR,\,\sR}$.
The tetrad ${\nR,\,\fR,\,\rR,\,\sR}$ is obtained from the reference tetrad
by a spatial rotation given by angles ${\THT,\,\PHI}$,
\begin{equation}\label{GenKillTetr}
\begin{aligned}
  \nR &=  \nK\commae\\
  \fR &= \spcm\cos\THT\;\fK + \sin\THT\cos\PHI\;\rK + \sin\THT\sin\PHI\;\sK\commae\\
  \rR &= -\sin\THT\;\fK + \cos\THT\cos\PHI\;\rK + \cos\THT\sin\PHI\;\sK\commae\\
  \sR &= \mspace{120mu}-\sin\PHI\;\rK+\qquad\,\cos\PHI\;\sK\period
\end{aligned}
\end{equation}
Let us note that the angles ${\THT,\,\PHI}$, understood as standard spherical coordinates
spanned on the axes ${\fK,\,\rK,\,\sK}$, describe exactly the spatial direction
${\kR^\perp=\frac1{\sqrt2}\fR}$ of the null vector $\kR$, where the
\emph{spatial direction} means projection orthogonal to the vector $\nK$.
The relation between null tetrads following from Eq.~\eqref{GenKillTetr}
can be found in Eq.~\eqref{RotTetrRel}.
This transformation is obtained as a consecutive composition of null rotation with
fixed  $\kG$ (Eq.~\eqref{ae:kfixed}), null rotation with fixed $\lG$
(Eq.~\eqref{ae:lfixed}), and of  special boost and spatial rotation
\eqref{ae:boostrotation} with parameters%
\begin{equation}\label{GenKillCompTrans}
\begin{aligned}
  L &= -\tan\frac\THT2 \, \exp(-i\PHI)\commae\\
  K &= \sin\frac\THT2 \, \cos\frac\THT2 \, \exp(-i\PHI)\commae\\
  B &= \cos^{-2}\frac\THT2\comma \Phi=\PHI\period
\end{aligned}
\end{equation}

Finally, we also introduce the \defterm{Robinson-Trautman tetrad} ${\kT,\,\lT,\,\mT,\,\bT}$
(see, e.g., \cite{Krameretal:book})
naturally connected with the Robinson-Trautman coordinates ${\zRT,\,\bRT,\,\uRT,\,\rRT}$
(cf.\  Eqs.~\eqref{rRTdef} and \eqref{ae:rRTKW}, \eqref{ae:zetaRTKW})
\begin{equation}\label{RTtetrad}
\begin{aligned}
  \kT&=\cv{\rRT} \commae&&\\
  \lT&=-\textstyle{\frac{1}{2}}\HRT\,\cv{\rRT}+\cv{\uRT}\mspace{-12mu}
    &&=-\frac{r^2\,\E}{2\,\DSr^2}\,\cv{\rRT}+\cv{\uRT}\commae\\
  \mT&=\frac\PRT\rRT\,\cv{\bRT}&&=\frac1{\sqrt\G\,\rRT}\,\cv{\bRT} \commae\\
  \bT&=\frac\PRT\rRT\,\cv{\zRT}&&=\frac1{\sqrt\G\,\rRT}\,\cv{\zRT} \period
\end{aligned}
\end{equation}
Here we have written down equivalent expressions using both metric
functions $\HRT$, $\PRT$ commonly used in the Robinson-Trautman framework,
and the metric functions $\G$, $\E$ of the $C$-metric (cf.\  Eqs.~\eqref{RTmetric}, \eqref{PHRTdef}).
The vector $\kT$ of this tetrad is oriented along the principal null direction $\kS$,
and it will be demonstrated in Section~\ref{sc:AlgSpecDir}
that this tetrad is parallelly transported
along the geodesics tangent to principal null directions.

The tetrad \eqref{RTtetrad} is simply related to the
particularly rotated tetrad ${\kR,\,\lR,\,\mR,\,\bR}$
(Eq.~\eqref{RotTetrRel}) with ${\THT=\THT_\spec}$, ${\PHI=0}$,
${\THT_\spec}$ given by Eq.~\eqref{THTsDef}:
\begin{equation}\label{RTtoRS}
\begin{gathered}
  \kT=\exp(\beta_\RoTr)\kRS \comma
  \lT=\exp(-\beta_\RoTr)\lRS \commae\\
  \mT=\mRS \comma
  \bT=\bRS \commae
\end{gathered}
\end{equation}
i.e., the Robinson-Trautman tetrad can be obtained from the reference tetrad
${\kK,\,\lK,\,\mK,\,\bK}$ by the spatial rotation \eqref{RotTetrRel}
with ${\THT=\THT_\spec}$, ${\PHI=0}$,
followed by the boost \eqref{ae:boostrotation} with the parameter
\begin{equation}\label{BparRTdef}
   B = \exp\beta_\RoTr = -\frac{\sqrt2\,\om}{\sqrt{-\E}}
   =\sqrt{-\frac2\HRT}\period
\end{equation}
We also give the relation between the Robinson-Trautman and
the algebraically special tetrad. Because the vectors
$\kT$ and $\kS$ are proportional, the Robinson-Trautman
tetrad is obtained from the special tetrad by the null rotation
\eqref{ae:kfixed} followed by the boost \eqref{ae:boostrotation}
with parameters
\begin{equation}\label{RTtoASpar}
\begin{aligned}
  L &= -\sin\THT_\spec \mspace{46mu}=-\sqrt{\frac\G{-\F}}\,\tanh\acp\commae\\
  B &= \exp\beta_\RoTr\cos\THT_\spec =
      \frac{\sqrt2\,\DSr}{\rRT\sqrt{-\F}\,\cosh\acp}\period
\end{aligned}
\end{equation}
The explicit relation of both tetrads can be found in Eqs.~\eqref{SpecTetrRel} and  \eqref{RTTetrRel}.

Using the transformations \eqref{ae:kfixedWeyl},
\eqref{ae:boostrotationWeyl} and \eqref{ae:kfixedEM}, \eqref{ae:boostrotationEMF}
with these parameters $L$, $B$, we find
that the only nonvanishing components of the gravitational and
electromagnetic fields in the Robinson-Trautman tetrad are
\begin{gather}
\begin{gathered}
  \WTP{\RoTr}{2} = \WTP{\spec}{2}
  =-\Bigl(\mass - 2\,\charge^2\accl\,\x - \frac{\charge^2}\rRT\Bigr)\,\frac1{\rRT^3}
  \commae\\
  \WTP{\RoTr}{3} =
  -\frac3{\sqrt2}\,\frac{\accl\,\rRT}{\PRT}\,\WTP{\spec}{2}\comma
  \WTP{\RoTr}{4} =
  3\,\frac{\accl^2\,\rRT^2}{\PRT^2}\,\WTP{\spec}{2}\commae
\end{gathered}\label{WeylRTTetr}\\
  \EMP{\RoTr}{1} = \EMP{\spec}{1} =
  -\frac{\charge}{2\,\rRT^2}\comma
  \EMP{\RoTr}{2} =
  -\sqrt2\,\frac{\accl\,\rRT}{\PRT}\,\EMP{\spec}{1}\label{EMRTTetr}\commae
\end{gather}
with $\WTP{\spec}{2}$ and $\EMP{\spec}{1}$ also given by
Eqs.~\eqref{WeylAlgSpecTetr} and \eqref{EMFAlgSpecTetr},
cf.\  \cite{Krameretal:book}.

\section{Gravitational and electromagnetic fields near $\scri^+$}
\label{sc:RadChar}

Now we are prepared to discuss radiative properties of the $C$-metric fields
near the de~Sitter-like infinity $\scri^+$. As we have
already explained in Section~\ref{sc:intro}, by the \emph{radiative field} we
understand a field with a dominant component having the $1/\afp$ fall-off,
calculated in a tetrad parallelly transported
along a null geodesic $\geod{}(\afp)$.
We will in particular concentrate on investigation of a directional
dependence of the gravitational and electromagnetic radiation.

To study the dependence of the fields on the
directions along which the
spacelike infinity $\scri^+$ is approached, it is crucial to
find a parallelly  transported tetrad along all null geodesics.
However, it is difficult to find a general geodesic and the
corresponding tetrad in an explicit
form, except for the case of  very special geodesics along the
privileged principal null directions, which will be
discussed in Section~\ref{sc:AlgSpecDir}.
Fortunately, it is not, in fact, necessary to find an
explicit form of the geodesics and tetrads because we are
interested only in the dominant terms of the fields
close to $\scri^+$. It is fully sufficient to study only their
\emph{asymptotic forms}.

Near infinity $\scri^+$, null geodesics $\geod{}(\afp)$ can be expanded in
the inverse powers of the affine  parameter ${\afp\to\infty}$.
In particular, in coordinates ${\tau,\,\om,\,\sg,\,\ph}$
introduced in Eq.~\eqref{omsgdef}, the null geodesics $\geod{}(\afp)$ can
be expanded as
\begin{equation}\label{GeodExp}
\begin{aligned}
  \om(\afp)  &\lteq \mspace{40mu}\om_\dir\,\frac{\DSr}{\afp} + \dots\commae\\
  \tau(\afp) &\lteq \tau_\fix + \tau_\dir\,\frac{\DSr}{\afp} + \dots\commae\\
  \sg(\afp)  &\lteq \sg_\fix  + \sg_\dir\,\frac{\DSr}{\afp}  + \dots\commae\\
  \ph(\afp)  &\lteq \ph_\fix  + \ph_\dir\,\frac{\DSr}{\afp}  + \dots\commae
\end{aligned}
\end{equation}
where the affine parameter $\afp$ has the dimension of
length. There is no absolute term in  the expansion of the
coordinate $\om$ because ${\om=0}$ at $\scri^+$. The
constant  parameters ${\tau_\fix,\,\sg_\fix,\,\ph_\fix}$
(and the corresponding values $\y_\fix$ and $\x_\fix$ given by
Eq.~\eqref{omsgdef}) label the \emph{point} $N_\fix$ at
$\scri^+$ which is approached by the geodesic $\geod{}(\afp)$.
The parameters ${\tau_\dir,\,\sg_\dir,\,\ph_\dir}$
characterize the \emph{direction} along which this  point
$N_\fix$ is approached. The remaining coefficient $\om_\dir$ can
be determined from the normalization of the tangent vector
which must be null. The tangent vector has the form
\begin{equation}\label{ParTangVect}
\begin{split}
  \frac{D \geod{}}{d\afp}\lteq
    -\frac{\DSr}{\afp^2}\,
        \bigl(\om_\dir\,\cv{\om}+\tau_\dir\,\cv{\tau}+
        \sg_\dir\,\cv{\sg}+\ph_\dir\,\cv{\ph}\bigr)\period
\end{split}
\end{equation}
The asymptotic form of the metric \eqref{omsgmetric} along the null geodesic is
\begin{equation}\label{mtrcExpan}
  \mtrc \lteq \frac{\afp^2}{\om_\dir^2}\,\bigl(
    -\grad\om^2-\F_\fix\,\grad\tau^2-\F_\fix\G_\fix\,\grad\sg^2
    +\G_\fix\grad\ph^2\bigr)\commae
\end{equation}
where $\F_\fix$ and $\G_\fix$ are the functions $\F$ and $\G$ evaluated
at the point $N_\fix$ at infinity $\scri^+$,
and we used ${\E_\fix = -1}$. Therefore,
the condition that the tangent vector is a null vector implies
\begin{equation}\label{NormCond}
   \om_\dir^2 = -\F_\fix\tau_\dir^2 - \F_\fix\G_\fix\sg_\dir^2 +
   \G_\fix\ph_\dir^2\period
\end{equation}
Notice that ${\om_\dir<0}$ since ${\om<0}$, and
\begin{equation}\label{rRTafpRel}
  \rRT(\afp) \lteq \rafpc\,\afp + \dots
  \comma\text{where}\quad
  \rafpc=-\frac1{\om_\dir}\commae
\end{equation}
which follows from Eq.~\eqref{rRTomRel}.

We wish to compare
geodesics approaching the given point $N_\fix$
along different directions. We thus need to
ensure \vague{the same} universal  choice of the affine
parameter $\afp$ for all geodesics. It is natural to require
that the energy (or, equivalently, the frequency) of the ray
represented by the null geodesic
\begin{equation}\label{KlgEnergy}
  E_\klg = -\mom\spr\nK =
  -\DSr\,\frac{D\geod{}}{d\afp}\spr\nK
\end{equation}
(see \noteref{nt:PhysAffinePar}),
is \emph{the same} independently of the direction of the geodesic,
i.e., that the component of the tangent vector to the normal direction
$\nK$ is fixed. From Eqs.~\eqref{KlgEnergy}, \eqref{ParTangVect},
and \eqref{KillTetr} it immediately follows that
\begin{equation}\label{FreqIntrp}
  E_\klg \lteq \frac{\DSr^2}\afp
        = \rafpc\,\frac{\DSr^2}{\rRT}\period
\end{equation}
The value of the energy $E_\klg$ with respect to
any asymptotic observer characterized  by the 4-velocity
$\nK$ thus obviously approaches zero as ${\afp\to\infty}$.
This behavior is caused by the
de~Sitter-like character of $\scri^+$.
Therefore, we have  to
compare the values of $E_\klg$ at the same \vague{proximity} to
$\scri^+$, i.e., at some fixed large but \emph{finite}
value of the coordinate $\rRT$ (\noteref{nt:Proximity}).
We conclude from Eq.~\eqref{FreqIntrp} that
fixing the energy at a given prescribed value  of $\rRT$ is
equivalent to fixing the value of the constant parameter
$\rafpc$ independently of a direction of the geodesic.
Let us note that  this approach
is fully equivalent to fixing a finite value of
\emph{conformal energy}, i.e., of the energy defined with respect
to a vector normal to $\scri^+$ normalized using a
conformal metric ${\cmtrc = \om^2\mtrc}$.

Next, it is necessary to find an \defterm{interpretation tetrad}
${\kP,\,\lP,\,\mP,\,\bP}$ which is parallelly transported
along the geodesic $\geod{}(\afp)$. However, using only an
asymptotic expansion of the tetrad at infinity $\scri^+$,
we cannot determine unique initial conditions which define this tetrad
somewhere in a finite region of the spacetime.
But without specifying these initial conditions,
the parallelly transported tetrad at $\scri^+$ is given only up to
an arbitrary (finite) Lorentz transformation. It thus seems that we are losing all
information because of this nonuniqueness. However, it is not so.
It will be demonstrated that the crucial
information about the behavior of the fields at infinity $\scri^+$
is hidden in an \vague{infinite} Lorentz transformation corresponding to the parallel
transport from a finite region of the spacetime up to the infinity.
It will thus be sufficient to find only the leading term of this transformation.

To be more specific, we naturally choose the vector $\kP$
of the parallelly transported interpretation null
tetrad to be proportional to the (parallelly transported)
tangent vector of the geodesic.
This ensures that $\kP$ is finite in finite
regions of spacetime (see \noteref{nt:ArbFinCnd}).
However, we still have a freedom in the normalization of $\kP$ which can be
multiplied by an arbitrary finite
factor, constant along the geodesic.
Similarly to the choice of the \vague{universal} affine parameter for
different geodesics, we have to choose the parallelly
transported tetrads in some suitable \vague{comparable} way for various
geodesics approaching the same point $N_\fix$ at infinity from different
directions. Not having an explicit form of the geodesics
(except for those special ones discussed in Section~\ref{sc:AlgSpecDir}),
we have to eliminate the dependence on
initial conditions by fixing final conditions for the
tetrad at  infinity $\scri^+$. Namely, we will require
that the normalization of the vector $\kP$
is specified independently of the direction of the geodesics.
This is achieved, for example, by the condition
\begin{equation}\label{kPCond}
  \kP\ctr\grad\rRT = 1\period
\end{equation}
Thanks to Eq.~\eqref{rRTafpRel} we thus have
\begin{equation}\label{kPandTangVec}
  \kP=\frac1\rafpc\,\frac{D\geod{}}{d\afp}\period
\end{equation}

Concerning vectors ${\mP,\,\bP}$ of the
parallelly transported interpretation tetrad,
there is a priori no \vague{canonical} prescription how to choose
these in a universal way for different geodesics.
The only constraint is the correct normalization \eqref{NullTetrNorm}.
Therefore, we have to find such physical quantities which
are invariant under this freedom. It will be shown below
(see Eq.~\eqref{kFixIndep} and discussion therein) that
\emph{the magnitude of the leading term} of the fields at $\scri^+$
is, in fact, independent of the specific choice of the vectors
${\mP,\,\bP}$.

However, there is a natural possibility to
fix the null vector $\lP$ of the tetrad by the condition that the timelike unit
vector $\nK$, orthogonal to infinity $\scri^+$, lies in the ${\kP\textdash\lP}$
plane. In this case the parallelly transported tetrad can be obtained by a boost
in the ${\kP\textdash\lP}$ plane from the rotated tetrad ${\kR,\,\lR,\,\mR,\,\bR}$
(see Eqs.~\eqref{GenKillTetr} or \eqref{RotTetrRel}) with properly chosen
angles ${\THT,\,\PHI}$.
Clearly, the vector $\kR$ has to point exactly in the direction of the
geodesic, or equivalently, the spatial vector $\fR$ has to point in
the spatial direction of the geodesic
(here again by \emph{spatial vectors} we mean those orthogonal to ${\nK=\nR}$,
i.e., tangent to $\scri^+$).
Using Eqs.~\eqref{kPandTangVec}, \eqref{ParTangVect}, and \eqref{KillTetr} we obtain%
\begin{equation}\label{kParGenRel}
  \kP \lteq
  \frac\DSr{\rafpc\,\afp}\,\Bigl(
  \nK-\frac1{\afp}\,\bigl(
    \tau_\dir\,\cv{\tau}
    +\sg_\dir\,\cv{\sg}
    +\ph_\dir\,\cv{\ph}\bigr)\Bigr)\period
\end{equation}
The unit vector $\fR$ in the spatial direction of the geodesic is thus
\begin{equation}\label{GenForPAr}
\begin{aligned}
  \fR &\lteq -\frac1{\afp}\,\Bigl(
    \tau_\dir\,\cv{\tau}
    +\sg_\dir\,\cv{\sg}
    +\ph_\dir\,\cv{\ph}\Bigr)\\
  &=-\sqrt{-\F_\fix}\,\frac{\tau_\dir}{\om_\dir}\,\fK
    +\sqrt{-\F_\fix\G_\fix}\,\frac{\sg_\dir}{\om_\dir}\,\rK
    +\sqrt{\G_\fix}\,\frac{\ph_\dir}{\om_\dir}\,\sK\period
\end{aligned}
\end{equation}
The leading term of the expansion of the parallelly
transported tetrad near the infinity then can be written as%
\begin{equation}\label{ParGenRel}
\begin{aligned}
  \kP &\lteq \frac{\sqrt2\,\DSr}{\rafpc\,\afp}\,\kR
  = \frac{\DSr}{\rafpc\,\afp}\,(\nK+\fR)
  \commae   & \mP &\lteq \mR\commae\\
  \lP &\lteq \frac{\rafpc\,\afp}{\sqrt2\,\DSr}\;\lR\;
  = \frac{\rafpc\,\afp}{2\DSr}\,(\nK-\fR)
  \commae   & \bP &\lteq \bR\period
\end{aligned}
\end{equation}
Here, we have made a particular choice of the vectors ${\mP,\,\bP}$.
In general, $\mP$ could differ from $\mR$ by a phase factor
(a rotation in the ${\mP\textdash\bP}$ plane) which,
as we mentioned, cannot be fixed in a canonical way.
Our choice ${\mP\lteq\mR}$ is \vague{natural} for the approach presented here.
However, in the next section we will encounter another \vague{suitable} choice
of the vector $\mP$.

Now we have to identify the angles ${\THT,\,\PHI}$.
Let us recall that these angles are just
spherical coordinates of the spatial direction ${\fR\propto\kP^\perp}$
with respect to the reference frame ${\fK,\,\rK,\,\sK}$.
Comparing Eqs.~\eqref{GenForPAr} and \eqref{GenKillTetr}
we find that the parameters ${\tau_\dir,\,\sg_\dir,\,\ph_\dir}$,
characterizing the asymptotic spatial direction of the geodesic
\eqref{GeodExp}, fix the the angles ${\THT,\,\PHI}$ as
\begin{equation}\label{DirByTHTPHI}
\begin{aligned}
  \tau_\dir &= \spcm\frac1{\rafpc\sqrt{-\F_\fix}}\,\cos\THT\commae\\
  \sg_\dir  &=   -  \frac1{\rafpc\sqrt{-\F_\fix\G_\fix}}\,\sin\THT\cos\PHI\commae\\
  \ph_\dir  &=   -  \frac1{\rafpc\sqrt{\G_\fix}}\,\sin\THT\sin\PHI\period
\end{aligned}
\end{equation}
In the following we will use these angles ${\THT,\,\PHI}$
to parametrize the direction along which a null geodesic
approaches the point $N_\fix$ on $\scri^+$.

Now we are ready to calculate the leading terms of the
components $\WTP{\parl}{n}$ of the Weyl tensor in the
parallelly transported tetrad given above. First we find the components $\WTP{\rot}{n}$
in the rotated tetrad ${\kR,\,\lR,\,\mR,\,\bR}$.
These can easily be obtained from Eq.~\eqref{WeylKillTetr} using
 relations \eqref{ae:kfixedWeyl}, \eqref{ae:lfixedWeyl}, and
\eqref{ae:boostrotationWeyl} with the parameters \eqref{GenKillCompTrans}.
Notice that all these components are of the
same order in $\afp$, namely, ${\sim\afp^{-3}}$
(cf.\  also Eq.~\eqref{FplGddExp} below).
To obtain the components $\WTP{\parl}{n}$
in the parallelly transported tetrad we perform an
additional boost \eqref{ParGenRel} in the ${\kR\textdash\lR}$ plane
with the boost parameter given by
\begin{equation}\label{RotToParBoost}
  B=\frac{\sqrt2\,\DSr}{\rafpc\,\afp}\period
\end{equation}
Using relations~\eqref{ae:boostrotationWeyl}
we immediately observe that it
rescales $\WTP{\parl}{n}$ by different powers of $\afp$, namely,
\begin{equation}\label{PeelingWeyl}
  \WTP{\parl}{n} \sim \frac1{\afp^{5-n}}\comma
  n=0,\,1,\,2,\,3,\,4\period
\end{equation}
The field thus clearly exhibits the \defterm{peeling behavior}.
The leading term of the gravitational field representing radiation
near infinity $\scri^+$ is ${\WTP{\parl}{4}\sim 1/\afp}$.
Explicitly, this term asymptotically takes the form
\begin{equation}\label{DirCharKlg}
\begin{split}
  \WTP{\parl}{4} \lteq
  &\frac{1}{16\,\DSr^2\cos^2\THT_\spec}\,
  \bigl(\F''+\G''\bigr)\\
  &\!\!\times\bigl(\sin\THT+\sin\THT_\spec\cos\PHI
  -i \sin\THT_\spec\cos\THT\sin\PHI\bigr)^2\!\period
\end{split}
\end{equation}
Here we should note that (cf.\  \eqref{WeylAlgSpecTetr})
\begin{equation}\label{FplGddExp}
  {\textstyle\frac1{12}}\bigl(\F''+\G''\bigr)\lteq
  -\bigl(\mass-2\charge^2\accl\,\xî_\fix\bigr)\frac1{\rafpc\,\afp}\period
\end{equation}

The phase of the component $\WTP{\parl}{4}$
depends on the choice of the vector $\bP$ (cf.\  Eq.~\eqref{PsiDef}).
Because the vector $\bP$ was chosen arbitrarily,
only the modulus ${\abs{\WTP{\parl}{4}}}$ can have a physical meaning.
Using the peeling behavior \eqref{PeelingWeyl} we can even
justify that the magnitude ${\abs{\WTP{\parl}{4}}}$
does not depend on any change of the null vectors
${\lP,\,\mP,\,\bP}$ at infinity. Indeed, we may perform an arbitrary \emph{finite}
Lorentz transformation which leaves the vector $\kP$ fixed.
Such a transformation can be generated by a combination of
the discussed spatial rotation in the ${\mP\textdash\bP}$ plane \eqref{ae:boostrotation}
which change only a phase of $\WTP{\parl}{4}$,
and of a null rotation \eqref{ae:kfixed}.
Under this transformation, the component $\WTP{\parl}{4}$ transforms according to
Eq.~\eqref{ae:kfixedWeyl} as\pagebreak[1]
\begin{equation}\label{kFixIndep}
  \WTP{\hparl}{4}{} = \WTP{\parl}{4} + 4 {\bar L}\, \WTP{\parl}{3} +
    6 {\bar L}^2 \WTP{\parl}{2} + 4 {\bar L}^3 \WTP{\parl}{1} + {\bar L}^4 \WTP{\parl}{0}\period
\end{equation}
Since $L$ is finite and the components
${{\WTP{\parl}{n} \sim \afp^{n-5}}}$, ${{n=0,1,2,3}}$ are of the higher order in
${1/\afp}$ than ${{\WTP{\parl}{4}\sim \afp^{-1}}}$,
they do not change the leading term of the field, i.e.,
$\WTP{\parl}{4}$ remains invariant.
(Let us note that the same is obviously not true for
leading terms of other components of the Weyl tensor.)

\begin{figure}
\includegraphics{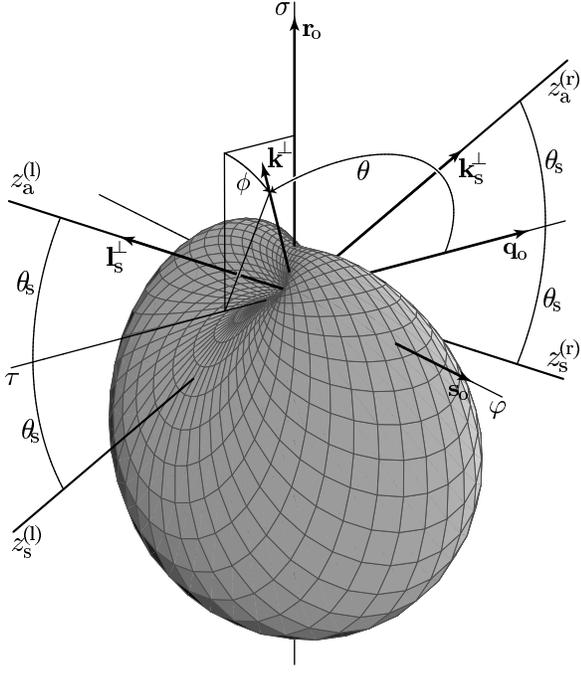}
\caption{\label{fig:RadChar3D}
The magnitude of the leading terms of gravitational and electromagnetic fields,
given by Eqs.~\eqref{DirChar} and \eqref{EMDirChar},
as a function of a direction from which the point $N_\fix$ at
infinity is approached --- the \defterm{directional pattern of radiation}.
The directions from the origin $N_\fix$ of the diagram correspond to spatial
directions in spacelike conformal infinity $\scri^+$.
The magnitude of the fields measured along a null geodesic with a tangent vector $\kG$
is drawn in the spatial direction ${-\kG^\perp}$
\emph{from} which the geodesic arrives (i.e., the geodesic points into
the spatial direction $\kG^\perp$). The angles ${\THT,\,\PHI}$ parametrizing the
spatial direction $\kG^\perp$ are measured from the axis $\fK$, and around the axis
$\fK$ starting from the ${\rK\textdash\fK}$ plane, respectively.
The special geodesics in principal null directions $\kS$ and $\lS$,
i.e., the null geodesics coming from the \vague{left} black hole and
the \vague{right} black hole (pointing \vague{from the sources}),
are denoted by $\geod[\lbh]{\spec}$ and $\geod[\rbh]{\spec}$.
They approach the point $N_\fix$ at infinity along
the spatial directions $\kS^\perp$ and $\lS^\perp$.
On the other hand, $\geod[\rbh]{\opp}$ and $\geod[\lbh]{\opp}$ are
\vague{antipodal} null geodesics approaching the infinity
along the spatial directions ${-\kS^\perp}$, ${-\lS^\perp}$, opposite to that of
$\geod[\lbh]{\spec}$ and $\geod[\rbh]{\spec}$, respectively.
The leading radiative term of the fields completely
vanishes along these antipodal geodesics.}%
\end{figure}
\begin{figure}
\includegraphics{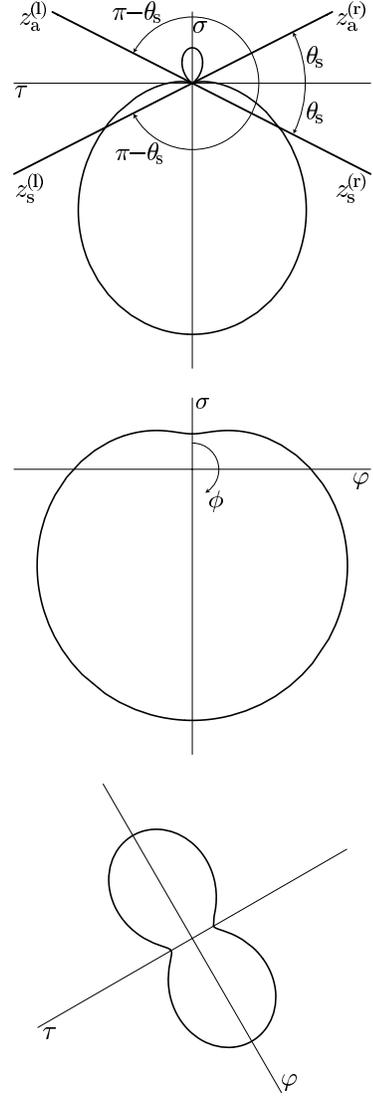}
\caption{\label{fig:RadChar2D}
The particular sections ${\tau\textdash\sg}$, ${\sg\textdash\ph}$
and ${\tau\textdash\ph}$ of the directional pattern of radiation
shown in Fig.~\ref{fig:RadChar3D}. The
\emph{oriented} angles $\THT$ of the spatial directions
of the geodesics from sources
(${\THT=\THT_\spec}$ and ${\THT=\pi\!-\THT_\spec}$, ${\PHI=0}$)
and of the antipodal geodesics
(${\THT=\THT_\spec}$ and ${\THT=\pi\!-\THT_\spec}$, ${\PHI=\pi}$)
are indicated.}%
\end{figure}
\begin{figure*}
\includegraphics{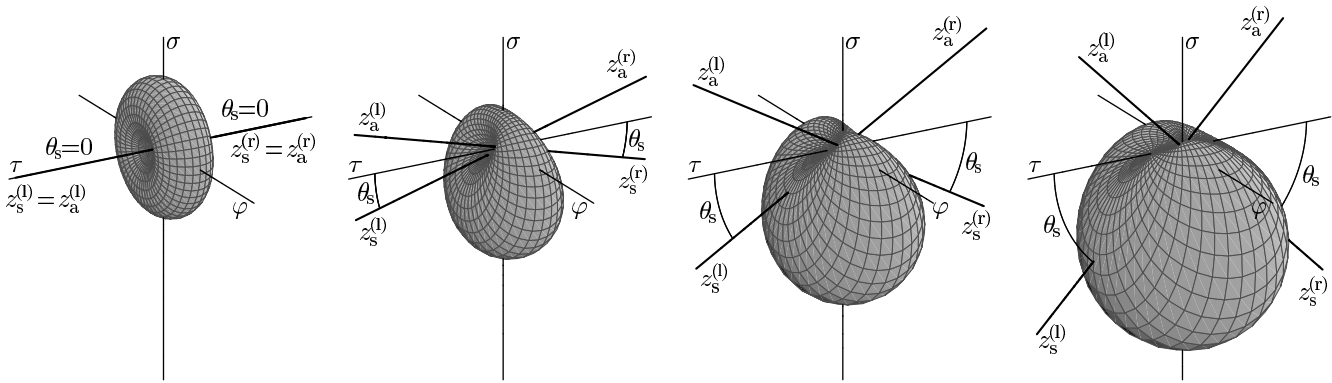}
\caption{\label{fig:chTHTs}
The directional pattern of radiation from Fig.~\ref{fig:RadChar3D}
for different values of the angular parameter $\THT_\spec$.
Because the directional dependence Eq.~\eqref{amplitude} of
the gravitational and electromagnetic radiation depends only on this
single parameter $\THT_\spec$ given by \eqref{THTsDef},
both changes of a position $N_\fix$ at infinity $\scri^+$ and
changes of the physical parameters $\mass$, $\charge$, $\accl$,
and $\Lambda$ manifest only through a change of the angle $\THT_\spec$.
The diagrams with different values of $\THT_\spec$ can thus be
interpreted either as the directional patterns at different points
of  infinity $\scri^+$, or as the directional characteristics
at \vague{the same} point (with fixed values of the metric functions
$\F$ and $\G$), but in spacetimes with, for example, different
acceleration of the black holes.}%
\end{figure*}

The invariant physical quantity ${\abs{\WTP{\parl}{4}}}$ is thus
\begin{equation}\label{DirChar}
\begin{split}
  \abs{\WTP{\parl}{4}} &\lteq
  \frac34\frac{(\mass-2\,\charge^2\accl\,\x_\fix)}{\rafpc\,\DSr^2\,\cos^2\THT_\spec}\,
  \frac1\afp\\
  &\;\;\times\Bigl((\sin\THT+\sin\THT_\spec\cos\PHI)^2
  +\sin^2\THT_\spec\cos^2\THT\sin^2\PHI\Bigr)\commae
\end{split}\raisetag{34pt}
\end{equation}
where the angle $\THT_\spec$ identifying the principal null
directions at infinity is, thanks to Eqs.~\eqref{THTsDef},
\eqref{acpdef}, \eqref{Edef} and ${\E_\fix=-1}$, given by
\begin{equation}\label{THTsDefonScri}
  \sin\THT_\spec=\sqrt{\frac{\G_\fix\DSr^2\accl^2}{1+\G_\fix\DSr^2\accl^2}}\comma
  \frac1{\cos^{2}\THT_\spec}=1+\G_\fix\DSr^2\accl^2\period
\end{equation}
Note that the term ${(\mass-2\,\charge^2\accl\,\x_\fix)}$ in Eq.~\eqref{DirChar} is positive,
which follows (although not immediately, see  Appendix~\ref{apx:FGprop})
from the conditions \eqref{assumtions}.

Analogously, we obtain the components $\EMP{\parl}{n}$ of the electromagnetic
field in the parallelly transported null tetrad in the form
\begin{equation}\label{PeelingEM}
  \EMP{\parl}{n} \sim \frac1{\afp^{3-n}}\comma
  n=0,\,1,\,2\commae
\end{equation}
which also exhibits the peeling behavior.
The leading term of the radiative component $\EMP{\parl}{2}$ is asymptotically
\begin{equation}\label{EMDirCharKlg}
\begin{split}
  \EMP{\parl}{2} \lteq&
  \frac{1}{2\sqrt2}\frac\charge{\rafpc\,\DSr\cos\THT_\spec}\frac1\afp\,\\
  &\,\times\bigl(\sin\THT+\sin\THT_\spec\cos\PHI
  -i \sin\THT_\spec\cos\THT\sin\PHI\bigr)\period
\end{split}
\end{equation}
Similarly to the $\WTP{\parl}{4}$ component, only the modulus of this
expression is independent of a choice of the interpretation tetrad.
Moreover, the square of modulus now has a clear physical meaning ---
it is exactly the leading term of the magnitude of the Poynting vector $\EMS_\parl$ in the
parallelly transported frame defined with respect to the timelike vector $\nP$.
Thus, we obtain
\begin{equation}\label{EMDirChar}
\begin{split}
  4\pi\abs{\EMS_\parl}&\lteq\abs{\EMP{\parl}{2}}^2\lteq
  \frac18\frac{\charge^2}{\rafpc^2\,\DSr^2\cos^2\THT_\spec}\frac1{\afp^2}\,\\
  &\;\;\times\bigl((\sin\THT+\sin\THT_\spec\cos\PHI)^2
  +\sin^2\THT_\spec\cos^2\THT\sin^2\PHI\bigr)\period
\end{split}\raisetag{34pt}
\end{equation}
The direction of the Poynting vector $\EMS_\parl$ is
asymptotically given by the vector $\fP$.
Interestingly, the dependence of ${\abs{\WTP{\parl}{4}}}$ and  ${\abs{\EMP{\parl}{2}}^2}$
on the direction along which a point $N_\fix$ at  infinity $\scri^+$ is
approached (i.e., the dependence on angles $\THT$ and $\PHI$)
is \emph{exactly the same}, namely,
\begin{equation}\label{amplitude}
  \dirampl(\THT,\PHI) =(\sin\THT+\sin\THT_\spec\cos\PHI)^2
  +\sin^2\THT_\spec\cos^2\THT\sin^2\PHI
  \period
\end{equation}
The angular dependence \eqref{amplitude} for a fixed value of $\THT_\spec$
which characterize the \emph{directional pattern of radiation}
at a given point of $\scri^+$ is shown in Figs.~\ref{fig:RadChar3D} and \ref{fig:RadChar2D}, and
for various $\THT_\spec$ in Fig.~\ref{fig:chTHTs}.

Let us now discuss the main results \eqref{DirChar} and \eqref{EMDirChar}.
These expressions can be understood as a more
detailed characterization of radiative fields near the spacelike conformal
infinity, supplementing thus the peeling behavior \eqref{PeelingWeyl},
\eqref{PeelingEM}.
It follows from \eqref{DirChar}, \eqref{EMDirCharKlg} that the dominant
components of both fields decay asymptotically near $\scri^+$,
corresponding to ${\rRT=\infty}$, as ${({\rafpc\,\afp})^{-1}=\rRT^{-1}}$.
The electromagnetic field is proportional to the charge parameter $\charge$
whereas the gravitational field is proportional to the mass parameter
$\mass$ modified, interestingly, by the term
${-2\,\charge^2\accl\,\x_\fix}$ which is a combination
of electric charge and acceleration parameters, and the constant
$\x_\fix$ denoting a specific point at infinity $\scri^+$. Both the
gravitational field ${\abs{\WTP{\parl}{4}}}$ and the
electromagnetic Poynting vector $
4\pi\,\abs{\EMS_\parl}\lteq\abs{\EMP{\parl}{2}}^2$ are
proportional to ${\DSr^{-2}=\frac13\Lambda}$ (but they also depend implicitly on
$\Lambda$ through the parameter $\THT_\spec$, cf.\  Eq.~\eqref{THTsDefonScri}).
The radiation at $\scri^+$ thus increases with a growing value of the
cosmological constant $\Lambda$.

The angular dependence of the magnitude of radiation ${\dirampl(\THT,\PHI)}$
is presented in Figs.~\ref{fig:RadChar3D} and \ref{fig:chTHTs}.
Their grid is given by the coordinate lines ${\THT=\text{constant}}$ and
${\PHI=\text{constant}}$, respectively.
It is straightforward to investigate
the behavior of the function ${\dirampl(\THT,\PHI)}$
for a fixed~$\THT$. The minimal value is
${\dirampl(\THT,\pi)=(\sin\THT-\sin\THT_\spec)^2}$,
and the maximum is
${\dirampl(\THT,0)=(\sin\THT+\sin\THT_\spec)^2}$.
The \emph{global maximum} ${\dirampl=(1+\sin\THT)^2}$ occurs for
${\THT=\frac{\pi}{2}}$, ${\PHI=0}$.
The greatest magnitude of radiation
thus arrives at infinity from the direction of $\rK$. On the other
hand, the minimal value ${\dirampl=0}$ is obtained for
${\THT=\THT_\spec}$, ${\PHI=\pi}$ and ${\THT=\pi-\THT_\spec}$,
${\PHI=\pi}$. These are exactly the spatial directions
${-\lS^\perp}$, ${-\kS^\perp}$ of \defterm{antipodal
null geodesics} $\geod[\lbh]{\opp}$ and $\geod[\rbh]{\opp}$, along
which the radiation completely vanishes. The value of $\dirampl$  along
the geodesics $\geod[\lbh]{\spec}$ and $\geod[\rbh]{\spec}$ coming
from the black holes in the directions $\kS^\perp$ (${\THT=\THT_\spec, \PHI=0}$)
and $\lS^\perp$ (${\THT=\pi-\THT_\spec}$, ${\PHI=0}$) is ${\dirampl=4\sin^2\THT_\spec}$.
The value along the direction  $\fK$ (corresponding to ${\THT=0}$) is
${\dirampl=\sin^2\THT_\spec}$, and along $\sK$ (${\THT=\frac{\pi}{2}}$,
${\PHI=\frac{\pi}{2}}$) is ${\dirampl=1}$.

Finally, it is interesting to observe that for a vanishing acceleration of the black
holes, i.e., for ${\accl=0}$ which implies ${\THT_\spec=0}$, we obtain
\begin{equation}\label{DirCharAzero}
\begin{aligned}
  \abs{\WTP{\parl}{4}} &\lteq
  \frac34\,\frac{\mass}{\rafpc\,\DSr^2}\,\frac1\afp\,\sin^2\THT \commae\\
  \abs{\EMP{\parl}{2}} &\lteq
  \frac{1}{2\sqrt2}\frac\charge{\rafpc\,\DSr}\frac1\afp\,\sin\THT
  \period
\end{aligned}
\end{equation}
The angular dependence  ${\dirampl=\sin^2\THT}$ is now independent of
$\PHI$ so that the directional pattern is axially symmetric
(see the diagram on the very left of Fig.~\ref{fig:chTHTs}).
Moreover, the gravitational and electromagnetic fields decay as
${1/\afp}$ even in this case of \emph{nonaccelerated} black holes if the fields are
measured along a nonradial null geodesic (${\THT\ne0,\pi}$).
A generic observer thus detects radiation. This effect
is intuitively caused by observer's asymptotic motion
relative to the \vague{static} black holes.
Only for special observers moving along null geodesics
radially from the black holes (${\THT=0,\pi}$)
the radiation vanishes as one would expect for \vague{static} sources.

Interestingly, the angular dependence ${\dirampl(\THT,\PHI)}$ is exactly the same
as that obtained in \cite{BicakKrtous:FUACS} for test electromagnetic field
of two accelerated charges in de~Sitter space (\noteref{nt:FUACS}).

\section{The radiation in the Robinson-Trautman framework}
\label{sc:RoTr}

In this part we rederive the above results
using the framework naturally adapted to the Robinson-Trautman
coordinate system \eqref{RTmetric}. This will not only provide us
with an independent way of deriving the characteristic
directional pattern of radiation generated by
accelerated charged black holes in the asymptotically de~Sitter
universe, but opens a possibility to investigate even more general
exact radiative solutions from  the large and important
Robinson-Trautman family.

We start again with investigation of  \emph{asymptotic null geodesics}
approaching  infinity $\scri^+$, i.e.,  those for which ${\rRT\to\infty}$.
Assuming a natural expansion of these geodesics in powers of
$\;{1/\rRT}\;$ (rather than in the affine parameter ${1/\afp}$ as was done in the previous
section),
\begin{equation}\label{expan}
\begin{gathered}
\begin{aligned}
  \zRT &\lteq\zRT_\fix+\frac{c}{\rRT}+\cdots\commae\\
  \uRT &\lteq\uRT_\fix+\frac{d}{\rRT}+\cdots \commae
\end{aligned}\\
  \rRT(\afp)\to\infty  \quad\text{as}\quad \afp \to \infty \commae
\end{gathered}
\end{equation}
where $\zRT_\fix$, $\uRT_\fix$, $c$, $d$
are constants, the derivatives with respect to the affine parameter $\afp$ are
\begin{equation}\label{expandot}
\begin{aligned}
  \dot\zRT  &\lteq-\frac{\dot\rRT}{\rRT^2}\,c+\cdots \commae\\
  \ddot\zRT &\lteq-\frac{\ddot\rRT}{\rRT^2}\Bigl(c+\cdots\Bigr)
         +\frac{{\dot\rRT}^2}{\rRT^3}\Bigl(2c+\cdots\Bigr) \period
\end{aligned}
\end{equation}
The expressions for $\dot\uRT$, $\ddot\uRT$ are obtained from \eqref{expandot} by
replacing $c$ with $d$.
Similarly, we may expand the metric functions and other quantities.
Using Eqs.~\eqref{expan} and \eqref{expandot} and the Christoffel symbols \eqref{ae:ChrisSymb},
the geodesic equations in the highest order read
\begin{equation}\label{geo}
  c\frac{\ddot\rRT}{\rRT}=0\comma
  d\frac{\ddot\rRT}{\rRT}=\mathcal{N}\,\frac{\dot\rRT^2}{\rRT^2}\comma
  \DSr^2\frac{\ddot\rRT}{\rRT} =-\mathcal{N}\,\frac{\dot\rRT^2}{\rRT^2}  \commae
\end{equation}
where
\begin{equation}\label{RTnfact}
  \mathcal{N} = 2\PRT_\fix^{-2}c\bar c+2d+\DSr^{-2}d^2\commae
\end{equation}
$P_\fix$ being the asymptotic value of $\PRT$ at the point $N_\fix$ at infinity.
However, a normalization of the tangent vector for \emph{null} geodesics  requires
\begin{equation}\label{norm}
\begin{aligned}
\mathcal{N}=0 \period
\end{aligned}
\end{equation}
Consequently, the asymptotic form of the null geodesics
approaching  $\scri^+$ is
\begin{equation}\label{asymp}
\begin{gathered}
  \rRT \lteq\rafpc\,\afp  \comma
  \zRT \lteq\zRT_\fix+\frac{c}{\rRT}  \comma
  \uRT \lteq\uRT_\fix+\frac{d}{\rRT}  \commae\\
  \frac{D\geod{}}{d\afp}=\rafpc\Bigl(\cv{\rRT}-\frac{c}{\rRT^2}\,\cv{\zRT}
            -\frac{\bar c}{\rRT^2}\,\cv{\bRT}-\frac{d}{\rRT^2}\,\cv{\uRT}\Bigr)\commae
\end{gathered}
\end{equation}
where the constant $\rafpc$ can be identified with that introduced in  Eq.~\eqref{rRTafpRel},
$\zRT_\fix$, $\uRT_\fix$ specify the \emph{point} $N_\fix$ on $\scri^+$
towards which the particular geodesic is approaching, and $c$, $d$
are parameters representing the \emph{direction}
along which $N_\fix$ is reached. In fact, this direction is
basically parametrized just by the complex constant $c$ since, using
 relations \eqref{norm}, \eqref{RTnfact}, $d$ is then given as
${d= -\DSr^2\bigl(\, 1 \pm \sqrt{1-2\DSr^{-2}\PRT_\fix^{-2}c\bar c}\;\bigr)}$.
For a particular  $c$, there are thus only two real values of $d$ which
represent two possible different \emph{orientations}
with which the null geodesics may approach $\scri^+$
in the given spatial direction.
In particular, for
the special choice ${c=0}$ we obtain ${d=0}$ and ${d=-2\DSr^2}$. The first
corresponds exactly to the privileged principal null direction along $\kS$
\vague{from the source} (i.e.,  the null geodesic $\geod[\lbh]{\spec}$
along the spatial direction $\kS^\perp$),
the second to an opposite orientation of this direction
\vague{away from the source}
(the \vague{antipodal} null geodesic $\geod[\rbh]{\opp}$ along  ${-\kS^\perp}$),
see Fig.~\ref{fig:RadChar3D}.

In order to find the behavior of  radiation near  $\scri^+$ we again have to set
up the interpretation tetrad transported parallelly along a general asymptotic null
geodesic, and project the Weyl tensor  and
the tensor of electromagnetic field  onto this tetrad.
We start with the Robinson-Trautman null tetrad
\eqref{RTtetrad}, naturally adapted to the Robinson-Trautman
coordinate system \eqref{RTmetric}.
We have seen in Section~\ref{sc:tetrads}
that the vector $\kT$ is oriented
along  one of the principal null direction, namely, $\kS$,
and (as we will see in Section~\ref{sc:AlgSpecDir})
the tetrad \eqref{RTtetrad} is parallelly transported
along the algebraically special geodesics.
In this standard tetrad the only nontrivial components
$\WTP{\RoTr}{n}$ and  $\EMP{\RoTr}{n}$,
which represent the gravitational and electromagnetic field,
are given by Eqs.~\eqref{WeylRTTetr} and \eqref{EMRTTetr}.
Let us now perform two subsequent null rotations and a boost of this
Robinson-Trautman null tetrad \eqref{RTtetrad}.
We first apply Eq.~\eqref{ae:lfixed}, then \eqref{ae:kfixed},
and finally \eqref{ae:boostrotation} with the parameters
\begin{equation}\label{para}
\begin{aligned}
  K&=-\frac{ c}{(1+\frac{1}{2}\DSr^{-2}d)\,\PRT\,\rRT}  \comma\\
  L&= \frac{c\,\rRT}{2\DSr^2 \PRT}  \comma\\
  B&=1+\textstyle{\frac{1}{2}}\DSr^{-2}d  \comma
  \Phi=0  \period\\
\end{aligned}
\end{equation}
The resulting null tetrad, using  relation
\eqref{norm}, then takes the following asymptotic
form as ${\rRT\to\infty}$:
\begin{equation}\label{genRTtetrad}
\begin{aligned}
  \kP&\lteq \left(\cv{\rRT}-\frac{c}{\rRT^2}\,\cv{\zRT}
            -\frac{\bar c}{\rRT^2}\,\cv{\bRT}-\frac{d}{\rRT^2}\,\cv{\uRT}\right) \commae\\
  \lP&\lteq\frac{\rRT^2}{2\DSr^2}
      \left(\cv{\rRT}+\frac{c}{\rRT^2}\,\cv{\zRT}
            +\frac{\bar c}{\rRT^2}\,\cv{\bRT}+\frac{d+2\DSr^2}{\rRT^2}\,\cv{\uRT}\right)   \commae\\
  \mP&\lteq\frac{\PRT}{\rRT} \left(\,\frac{c\,d}{2\DSr^2\bar c}\,\cv{\zRT}
    +(1+{\textstyle\frac{1}{2}}\DSr^{-2}d)\,\cv{\bRT}-\frac{c}{\PRT^2}\,\cv{\uRT}\right) \commae\\
  \bP&\lteq\frac{\PRT}{\rRT} \left(\,(1+{\textstyle\frac{1}{2}}\DSr^{-2}d)\,\cv{\zRT}
     +\frac{\bar c\,d}{2\DSr^2c}\,\cv{\bRT}-\frac{\bar c}{\PRT^2}\,\cv{\uRT}\right)
     \period
\end{aligned}
\end{equation}

Obviously, the above vector $\kP$ is tangent to a general
asymptotic null geodesics \eqref{asymp}. Moreover, the tetrad is chosen in such a way that
the timelike unit vector orthogonal to $\scri^+$
\begin{equation}\label{gennorm}
\nK=\frac{1}{\sqrt{-\HRT}}\,\left( -\HRT\,\cv{\rRT}+\cv{\uRT}\,\right)
\approx \frac{\rRT}{\DSr}\,\cv{\rRT}+\frac{\DSr}{\rRT}\,\cv{\uRT} \commae
\end{equation}
introduced in Eq.~\eqref{KillTetr},
belongs to the plane spanned  by the two null vectors $\kP$ and $\lP$.
Indeed,
\begin{equation}
  \nK\lteq\frac{1}{\sqrt2}\Bigl(
  \frac{\rRT}{{\sqrt2}\,\DSr }\,\kP
  + \frac{{\sqrt2}\,\DSr}{\rRT}\,\lP
  \Bigr)\period
\end{equation}
Note that this choice corresponds to a boost Eq.~\eqref{ae:boostrotation} which becomes
unbounded as ${\rRT\to\infty}$.

As discussed in the previous section,
in order to \emph{compare} the radiation for all null
geodesics approaching the given point at de~Sitter-like infinity $\scri^+$, it is
necessary to introduce a unique and universal normalization of the affine
parameter $\afp$ and of the vector $\kP$. We concluded that a natural and also the most
convenient choice is to keep the parameter $\rafpc$ fixed (see discussion near Eq.~\eqref{KlgEnergy})
and to require Eq.~\eqref{kPandTangVec}. These conditions are obviously satisfied by
Eq.~\eqref{genRTtetrad}, cf.\  Eq.~\eqref{asymp}.
Therefore, the tetrad \eqref{genRTtetrad} is exactly the interpretation
tetrad suitable for analysis of  behavior of fields on $\scri^+$.

Now we perform a projection of the above null tetrad onto the
spacelike infinity $\scri^+$.
These projections ${\kP^\perp,\,\lP^\perp,\,\mP^\perp}$
(cf. Eq.~\eqref{orthproj}) are
\begin{equation}\label{genRTproj}
\begin{gathered}
  \kP^\perp\lteq-\frac{1}{\rRT^2}\,\Bigl(\,c\,\cv{\zRT}
            +\bar c\,\cv{\bRT}+(d+\DSr^2)\,\cv{\uRT}\,\Bigr) \commae\\
  \lP^\perp\lteq-\frac{\rRT^2}{2\DSr^2}\,\kP^\perp \comma
  \mP^\perp=\mP \comma  \bP^\perp=\bP  \period
\end{gathered}
\end{equation}
The radiation approaching  $\scri^+$ along the null
vector $\kP$ propagates in the spatial direction ${\kP^\perp\propto\fR}$.
Imposing the normalization condition
${\fR\spr\fR=1}$, the unit vector of the radiation direction thus takes the form
\begin{equation}\label{directionf}
\begin{aligned}
  \fR\lteq-\frac{1}{\DSr\,\rRT}\,\Bigl(\,c\,\cv{\zRT}
      +\bar c\,\cv{\bRT}+(d+\DSr^2)\,\cv{\uRT}\,\Bigr) \period
\end{aligned}
\end{equation}
Of course, this vector is identical to the vector $\fR$ introduced
previously in Eq.~\eqref{GenKillTetr}.
Using Eqs.~\eqref{VecCoorFu}--\eqref{VecCoorFz},
\eqref{KillTetr}, and \eqref{THTsDef} we obtain
\begin{equation}\label{RTtransfo}
\begin{aligned}
  \cv{\zRT}&=\frac{1}{\sqrt2}\,\frac\rRT\PRT\,
   \left(-\sin\THT_\spec\,\fK+\cos\THT_\spec\,\rK+i\,\sK \,\right)\commae\\
 \cv{\uRT}&=-\sqrt{-\HRT}\,\left(\,\cos\THT_\spec\,\fK
       +\sin\THT_\spec\,\rK \,\right) \commae\\
 \cv{\rRT}&=\frac{1}{\sqrt{-\HRT}}\,\left(\,\nK+\cos\THT_\spec\,\fK
       +\sin\THT_\spec\,\rK \,\right) \period
\end{aligned}
\end{equation}
Substituting this into Eq.~\eqref{directionf}, using
${\E_\fix=-1}$, and comparing with the expression
\eqref{GenKillTetr}, we obtain the following relation between the
Robinson-Trautman parameters $\,c$, $d\,$ and the angles
$\,{\THT,\,\PHI}\,$
\begin{equation}\label{RTparam}
\begin{aligned}
   \frac{c+\bar c}{\sqrt2\,\DSr\PRT_\fix}&=\spcm\sin\THT_\spec\cos\THT-\cos\THT_\spec\sin\THT\cos\PHI \commae\\
  i\frac{c-\bar c}{\sqrt2\,\DSr\PRT_\fix}&=-\sin\THT\sin\PHI \commae\\
  1+\DSr^{-2}d&=\spcm \cos\THT_\spec\cos\THT+\sin\THT_\spec\sin\THT\cos\PHI \period
\end{aligned}
\end{equation}
Of course, this parametrization identically satisfies the
normalization condition \eqref{norm}. Moreover, it can now be
demonstrated that the above null tetrad \eqref{genRTtetrad} is in
fact identical to the parallelly transported tetrad
\eqref{ParGenRel}, except for the transverse vector $\mP$, which was previously defined as ${\mP\lteq\mR}$,
$\mR$ given by Eq.~\eqref{RotTetrRel} (cf.\  Eqs.~\eqref{GenKillTetr}, \eqref{NormNullTetr}).
Such a vector is related to the vector $\mP$ adapted to the
Robinson-Trautman framework \eqref{genRTtetrad} by the spatial rotation
\eqref{ae:boostrotation},
${\mP=\exp(-i\phi_\parl)\,\mR}$, where the rotation angle $\phi_\parl$ is given by
\begin{equation}\label{rotation}
\begin{gathered}
\begin{aligned}
   \sin\phi_\parl&=\frac{(\,\cos\THT_\spec+\cos\THT\,)\sin\PHI}{1+\cos\THT_\spec\cos\THT
       +\sin\THT_\spec\sin\THT\cos\PHI}  \commae\\
   \cos\phi_\parl&=\frac{\sin\THT_\spec\sin\THT+(1+\cos\THT_\spec\cos\THT\,)\cos\PHI}
       {1+\cos\THT_\spec\cos\THT+\sin\THT_\spec\sin\THT\cos\PHI}  \commae
\end{aligned}\\
   \exp(i\phi_\parl)=\frac
       {\exp(i\PHI)\cos\frac{\THT_\spec}2\cos\frac\THT2+\sin\frac{\THT_\spec}2\sin\frac\THT2}
       {\cos\frac{\THT_\spec}2\cos\frac\THT2+\exp(i\PHI)\sin\frac{\THT_\spec}2\sin\frac\THT2}
       \period
\end{gathered}
\end{equation}

Finally, we calculate the leading components of the gravitational
and electromagnetic fields in the  interpretation frame
\eqref{genRTtetrad} asymptotically close to  infinity $\scri^+$.
As we have said, the Lorentz transformation from
the tetrad \eqref{RTtetrad} to the tetrad \eqref{genRTtetrad} is given by two
subsequent null rotations and the boost with the parameters given by
Eq.~\eqref{para}.
Starting with the components \eqref{WeylRTTetr} in the
standard Robinson-Trautman frame, using
Eqs.~\eqref{ae:lfixedWeyl}, \eqref{ae:kfixedWeyl}, \eqref{ae:boostrotationWeyl}
and \eqref{ae:lfixedEM}, \eqref{ae:kfixedEM}, \eqref{ae:boostrotationEMF},
we obtain after somewhat lengthy calculation
\begin{align}
   \WTP{\parl}{4}&\lteq-\frac{3\accl^2(\mass-2\charge^2\accl\,\x_\fix)}{\rRT\,\PRT_\fix^2}\,
    \left(1-\frac{1}{\sqrt2\,\DSr^2\accl}\,\bar{c}+{\frac{1}{2\DSr^2}}\,d\right)^2    \commae\notag\\
   \EMP{\parl}{2}&\lteq\ \frac{\charge\,\accl}{\sqrt2\,\rRT\,\PRT_\fix}\,
    \left(1-\frac{1}{\sqrt2\,\,\DSr^2\accl}\,\bar{c}+{\frac{1}{2\DSr^2}}\,d\right) \period\label{RTfieldsScri}
\end{align}
Substituting from
Eq.~\eqref{RTparam} for the parameters $c$ and $d$,
and using Eqs.~\eqref{THTsDefonScri} and \eqref{PHRTdef} we get
\begin{equation}\label{RTfieldsScriF}
\begin{aligned}
   \WTP{\parl}{4}&\lteq-\frac{3}{4}\frac{(\mass-2\charge^2\accl\,\x_\fix)}{\DSr^2\,\rRT\,\cos^2\THT_\spec}\\
    &\;\times\!\left(\,\sin\THT_\spec+\sin\THT\cos\PHI+i\cos\THT_\spec\sin\THT\sin\PHI\,\right)^2    \commae\\
   \EMP{\parl}{2}&\lteq\frac{1}{2\sqrt2}
   \frac{\charge}{\DSr\,\rRT\,\cos\THT_\spec}\\
    &\;\times\!\left(\,\sin\THT_\spec+\sin\THT\cos\PHI+i\cos\THT_\spec\sin\THT\sin\PHI\,\right) \commae
\end{aligned}
\end{equation}

We should have recovered the previous results \eqref{DirCharKlg}
and \eqref{EMDirCharKlg}.
Comparing them we find that
the expressions differ in the angular part.
However, this is a consequence of the difference of interpretation tetrads
used in the previous and in this sections.
The results are, in fact, identical after performing
a spatial rotation \eqref{ae:boostrotation}
with the angular parameter $\phi_\parl$ given
by Eq.~\eqref{rotation}. This changes the phase of the
components according to Eqs.~\eqref{ae:boostrotationWeyl},
\eqref{ae:boostrotationEMF},
and we obtain
${\WTP{\parl}{4} = \exp(2i\phi_\parl)\, \WTP{\parl}{4}}$,
${\EMP{\parl}{2} = \exp(i\phi_\parl)\, \EMP{\parl}{2}}$, where the
left-hand side is given by Eq.~\eqref{RTfieldsScriF}, and the right hand side by
Eqs.~\eqref{DirCharKlg}, \eqref{EMDirCharKlg}. Both results are thus equivalent.

The tetrads \eqref{ParGenRel} and \eqref{genRTtetrad} have
been introduced in a way natural to each specific approach. The fact that they
differ in definitions of the vector $\mP$ documents what we have already discussed above:
there is no canonical way how to choose the interpretation tetrad. It also
means that the \emph{phase} of the results \eqref{DirCharKlg}, \eqref{EMDirCharKlg}, or
\eqref{RTfieldsScriF} is not physical. Invariant information,
independent of a choice of the interpretation tetrad,
is contained in the \emph{modulus} of the tetrad components of the fields.
Obviously, the magnitudes of the field
components \eqref{RTfieldsScriF} are the same as the results
\eqref{DirChar} and \eqref{EMDirChar}  derived previously.

\section{Radiation along the algebraically special null directions}
\label{sc:AlgSpecDir}

In the final section we concentrate on a family of special geodesics
${\geod[\lbh]{\spec}}$ approaching  infinity $\scri^+$ along principal
null direction $\kS$, and investigate the fields with respect to the
corresponding interpretation tetrad. Using Eqs.~\eqref{ae:ChrisSymb}
it is straightforward to observe that the coordinate lines
\begin{equation}\label{specgeodcoor}
  \uRT=\uRT_\fix=\text{constant}\comma\zRT=\zRT_\fix=\text{constant}
\end{equation}
(i.e., also ${\x=\text{constant}}$, ${\ph=\text{constant}}$) are null geodesics, $\rRT$
is their affine parameter, and the tangent vector is ${\kT=\cvil{\rRT}}$.
(For simplicity, in this section we use the affine parameter $\rRT$,
a general affine parameter $\afp$ can be introduced by a trivial rescaling
${\rRT=\rafpc\,\afp}$, cf.\  Eq.~\eqref{rRTafpRel}.)
The geodesics ${\geod[\lbh]{\spec}(\rRT)}$ emanate
\vague{radially} from the \vague{left} black hole up to the infinity
(similarly we could investigate analogous geodesics ${\geod[\rbh]{\spec}}$ along
$\lS$ from the \vague{right} black hole). As we have seen in
Section~\ref{sc:tetrads} (cf.\  Eq.~\eqref{RTtoRS} and the subsequent discussion),
the tangent vector $\kT$ is oriented along
the principal null direction $\kS$. These geodesics
thus approach the infinity from the specific spatial direction characterized by the angles
\begin{equation}\label{specgeod}
  \THT=\THT_\spec\comma\PHI=0\commae
\end{equation}
or by the parameters ${c=0}$, ${d=0}$ (see Eq.~\eqref{RTparam}).

Moreover, in such a case we can identify explicitly the parallelly transported interpretation
tetrad --- it can easily be shown using Eqs.~\eqref{ae:ChrisSymb} that the
Robinson-Trautman tetrad \eqref{RTtetrad} is parallelly transported along
${\geod[\lbh]{\spec}(\rRT)}$, i.e.,
\begin{equation}
  \kT\ctr\covd\kT=0\comma\kT\ctr\covd\lT=0\comma\kT\ctr\covd\mT=0\period
\end{equation}
We can thus set the interpretation tetrad
\begin{equation}\label{ITisRT}
  (\kP,\,\lP,\,\mP,\,\bP) \equiv (\kT,\,\lT,\,\mT,\,\bT)
\end{equation}
in the \emph{whole} spacetime, not only asymptotically near $\scri^+$,
as in Eq.~\eqref{genRTtetrad} for ${c=0}$, ${d=0}$.
As follows from Eqs.~\eqref{WeylRTTetr} and \eqref{EMRTTetr}, all components
of gravitational and electromagnetic fields are explicitly
\begin{equation}\label{WeylspTetr}
\begin{aligned}
  \WTP{\parl}{4} &=
  -3\,\Bigl(\mass - 2\,\charge^2\accl\,\x - \frac{\charge^2}\rRT\Bigr)\,
  \accl^2\G\,\frac1\rRT\commae\\
  \WTP{\parl}{3} &=
  \frac3{\sqrt2}\,\Bigl(\mass - 2\,\charge^2\accl\,\x - \frac{\charge^2}\rRT\Bigr)\,
  \accl\sqrt{\G}\,\frac1{\rRT^2}\commae\\
  \WTP{\parl}{2} &=
  -\Bigl(\mass - 2\,\charge^2\accl\,\x - \frac{\charge^2}\rRT\Bigr)\,
  \frac1{\rRT^3}\comma
  \WTP{\parl}{1} = \WTP{\parl}{0} = 0\commae
\end{aligned}
\end{equation}
and
\begin{equation}\label{EMspTetr}
  \EMP{\parl}{2} = \frac{\charge\,\accl\sqrt{\G}}{\sqrt2}\,\frac1\rRT\comma
  \EMP{\parl}{1} = -\frac{\charge}2\,\frac1{\rRT^2}\comma
  \EMP{\parl}{0} = 0\period
\end{equation}
Clearly, the leading terms in the ${1/\rRT}$ expansion give the previous general asymptotical
results \eqref{DirChar} and \eqref{EMDirChar} with ${\THT,\,\PHI}$ specified by
Eq.~\eqref{specgeod}, and ${\rRT=\rafpc\,\afp}$.
In the case of de~Sitter spacetime (${\mass=0}$, ${\charge=0}$) the field components
identically vanish, in the general case
the fields have a radiative character (${\sim 1/\rRT}$) except
for a vanishing acceleration~$\accl$ and/or for ${\G_\fix=0}$.
The \vague{static} case ${\accl=0}$ has been already discussed after Eq.~\eqref{DirCharAzero}.
The case ${\G_\fix=0}$ corresponds to observers located at the privileged position --- on the axes
${\x=\x_\maxis}$ and ${\x=\x_\paxis}$. This is analogous to the well-known situation of
an electromagnetic field of accelerated test charges in flat spacetime
which is also not radiative along the axis of symmetry.

Let us note that in this case the affine parameter $\rRT$ coincides, in fact, both with the
\defterm{luminosity distance} and the \defterm{parallax distance} for the congurence of the above null geodesics
--- as for any Robinson-Trautman spacetime described by the metric
\eqref{RTmetric}. Indeed, the luminosity distance $\lumdist$ is related to the affine
parameter $\rRT$ by the relation~\cite{Sachs:1962}
\begin{equation}
  \frac{d\lumdist}{d\rRT}=\frac12\,\lumdist\,\covd\ctr\kT\period
\end{equation}
Thanks to Eqs.~\eqref{ae:ChrisSymb} one obtains ${\frac12\covd\ctr\kT=1/\rRT}$, and thus
${\lumdist=\rRT}$.
This means that the radiative ${1/\rRT}$ fall-off of the fields is naturally measurable (even locally) by observers moving
radially to infinity, using both the parallax and the luminosity methods for determining the distance.

In the previous sections, when we studied the radiation along general geodesics,
we have been able to fix the interpretation tetrad only asymptotically, by specifying appropriate final conditions
at infinity (see Eq.~\eqref{kPCond}, \eqref{ParGenRel} and the discussion nearby).
For the special family of geodesics \eqref{specgeodcoor} discussed here we can specify
the interpretation tetrad by setting the initial conditions anywhere in the \emph{finite}
region inside the spacetime. Because any point at  infinity $\scri^+$ is only reached by
one algebraically special geodesic $\geod[\lbh]{\spec}$ from the \vague{left} black hole,
this does not allow us to study the \emph{directional} pattern of radiation
with respect to the interpretation tetrad fixed by these explicit initial conditions.
However, we can study
the standard \emph{positional} pattern of radiation along these special geodesics ---
the dependence of radiation on the position of asymptotic point $N_\fix$ in the infinity.

The initial conditions for interpretation tetrad inside a finite region of the spacetime can be chosen in many
different ways, e.g.,  using some natural tetrad on a spacelike hypersurface
(\vague{initial instant of time}, \noteref{nt:FUACSincond}),
on a \vague{surface of sources}, on a special null hypersurface, etc.
Obviously, geometrically privileged locations where
we can specify such initial conditions are \emph{horizons}, in particular the
cosmological horizon ${\y=\y_\chor}$, or the outer horizon ${\y=\y_\ohor}$ of the
\vague{left} black hole. The former one (its \vague{future} half) forms a
(past) boundary of the domain in which any observer has to reach the future
infinity $\scri^+$ (the domain I containing $\scri^+$ in Fig.~\ref{fig:ConfDiag}).
The latter one forms the \vague{surface} of black hole and can thus be
understood as a \vague{surface of sources} (the boundary between regions II and III).
Although we have in mind mainly these two cases,
the following discussion can be applied to any horizon ${\y=\y_h}$. The
special geodesics cross such horizon at null hypersurface ${\vGN=n\pi}$,
the global null coordinates ${\uGN,\,\vGN}$ being defined in
Eq.~\eqref{ae:GNcoor}, and the integer $n$ fixed by the horizon
under consideration (in particular ${n=0}$ and ${n=-1}$ in  Fig.~\ref{fig:ConfDiag}).

First, we observe that the choice \eqref{ITisRT} is the most natural one.
The Robinson-Trautman tetrads in the whole spacetime --- and thus the corresponding
initial conditions on any horizon ${\y=\y_h}$ --- are actually
invariant under a shift along the Killing vector $\cvil{\tau}$.
Indeed, expressing the Robinson-Trautman tetrad in terms of the coordinate vectors
${\cvil{\tau},\,\cvil{\om},\,\cvil{\sg},\,\cvil{\ph}}$ (using Eqs.~\eqref{RTtetrad},
and \mbox{\eqref{VecCoorFu}--\eqref{VecCoorFz}}) we find that the coefficients are
independent of $\tau$, i.e., the Lie derivatives vanish,
\begin{equation}
  \lied_{\cv{\tau}}\kT=0\comma
  \lied_{\cv{\tau}}\lT=0\comma
  \lied_{\cv{\tau}}\mT=0\period
\end{equation}
The definition of the interpretation tetrad \eqref{ITisRT} thus respects the symmetry of
spacetime.

There is also another possibility to fix the interpretation tetrad
${\kH,\,\lH,\,\mH,\,\bH}$ on the horizon ${\y=\y_h}$. We choose the null vector ${\kH\propto\kT}$
tangent to the geodesic, and the null vector $\lH$ tangent to the horizon.
Now we have to specify the length of one of these vectors, length of the other one
is then fixed by the normalization \eqref{NullTetrNorm}.
It will be achieved by requiring that the vector
$\lH$ is parallelly transported along the null geodesic generator of the
horizon (note, however, that this condition cannot be satisfied for the vector $\kH$).
Finally, we should fix the remaining vectors ${\mH,\,\bH}$. However, we
will be interested only in the magnitude of the leading terms of the field
components (as in the previous sections) and therefore a specific choice of the vectors
${\mH,\,\bH}$ is irrelevant --- see the discussion before Eq.~\eqref{kFixIndep}.
The interpretation tetrad defined in this way is a natural choice for
observers localized on the horizon --- its definition remains
\vague{the same} (is parallelly transported) along the generators of the horizon.

To follow explicitly the procedure described above, we use the global null coordinates
${\uGN,\,\vGN}$. The definition of these coordinates depends on a choice of parameter $\GNcoef$.
As explained in Appendix~\ref{apx:Coor}, the metric \eqref{ae:GNmetric}
is regular with respect of these coordinates on the horizon ${\y=\y_h}$ if we set
\begin{equation}
   \GNcoef = \GNcoef_h\commae
\end{equation}
where $\GNcoef_h$ is given by Eq.~\eqref{ae:yscoor}. In the following we assume
such a choice. We also introduce the sign of $\GNcoef_h$:
\begin{equation}\label{GNcoefSignDef}
  {\pm1=\sign\GNcoef_h}\commae
\end{equation}
then
\begin{equation}\label{GNcoefSign}
  {\cos\uGN\Big\vert_{\begin{subarray}{l}\y=\y_h\\\uGN=m\pi\end{subarray}}}
  ={\cos\vGN\Big\vert_{\begin{subarray}{l}\y=\y_h\\\vGN=n\pi\end{subarray}}}
  =\pm1\period
\end{equation}
For the cosmological horizon  ${\sign\GNcoef_\chor=1}$, whereas
for the outer horizon ${\sign\GNcoef_\ohor=-1}$.
With these definitions the metric coefficient $\mtrc_{\uGN\vGN}$
evaluated on the horizon ${\y=\y_h}$ reads
\begin{equation}\label{mtrcatH}
  \mtrc_{\uGN\vGN}\Big\vert_{\begin{subarray}{l}\y=\y_h\\\vGN=n\pi\end{subarray}}=
  -\rRT_h^2\frac{\abs{\GNcoef_h}}{\GNtcoef_h}(1\pm\cos\uGN)^{-1}\commae
\end{equation}
where $\GNtcoef_h$ is defined in Eq.~\eqref{ae:GNtcoef}, and
\begin{equation}\label{rRTatH}
   \rRT_h=\rRT\vert_{\y=\y_h}=\frac{\DSr}{\y_h\cosh\acp-\x\sinh\acp}\period
\end{equation}
Here and in the following we repeatedly use  relation \eqref{ae:tantan}.

Now, we fix the vectors ${\kH,\,\lH}$ at the bifurcation \mbox{2-surface} ${\uGN=m\pi}$, ${\vGN=n\pi}$
of the horizon in a \vague{symmetric way}, namely,
\begin{equation}\label{klatHcentr}
  \lH\Big\vert_{\begin{subarray}{l}\y=\y_h\\\uGN=m\pi\\\vGN=n\pi\end{subarray}}=
  \frac{\sqrt2}{\rRT_h}\sqrt{\frac{\GNtcoef_h}{\abs{\GNcoef_h}}}\;\cv{\uGN}\comma
  \kH\Big\vert_{\begin{subarray}{l}\y=\y_h\\\uGN=m\pi\\\vGN=n\pi\end{subarray}}=
  \frac{\sqrt2}{\rRT_h}\sqrt{\frac{\GNtcoef_h}{\abs{\GNcoef_h}}}\;\cv{\vGN}\period
\end{equation}
Using the fact that the only nonvanishing Christoffel coefficient
$\Gamma_{\uGN\uGN}^\alpha$ is
\begin{equation}
  \Gamma_{\uGN\uGN}^\uGN\Big\vert_{\y=\y_h}=
  \pm\Bigl(\tan\frac\uGN2\Bigr)^{\pm1}\commae
\end{equation}
we find that the vector $\lH$ defined by
\begin{equation}
  \lH\Big\vert_{\begin{subarray}{l}\y=\y_h\\\vGN=n\pi\end{subarray}}=
  \frac1{\sqrt2\,\rRT_h}\sqrt{\frac{\GNtcoef_h}{\abs{\GNcoef_h}}}\;(1\pm\cos\uGN)\,\cv{\uGN}\comma
\end{equation}
is parallelly transported along the geodesic null generators of the horizon
${\vGN=\text{constant}}$, ${\x=\text{constant}}$, ${\ph=\text{constant}}$,
with the initial condition \eqref{klatHcentr}.
Obviously, $\lH$ is tangent to the generator, and
${\lH\ctr\covd\lH \;\big\vert_{\begin{subarray}{l}{\scriptscriptstyle\y=\y_h}\\{\scriptscriptstyle\vGN=n\pi}\end{subarray}} = 0}$.
Taking into account the normalization \eqref{NullTetrNorm} and the metric coefficient \eqref{mtrcatH}
we find the normalization of the null vector $\kH$,
\begin{equation}\label{hparlGN}
  \kH\Big\vert_{\begin{subarray}{l}\y=\y_h\\\vGN=n\pi\end{subarray}}=
  \frac{\sqrt2}{\rRT_h}\sqrt{\frac{\GNtcoef_h}{\abs{\GNcoef_h}}}\;\cv{\vGN}\period
\end{equation}

The null vectors ${\kH,\,\lH}$ do \emph{not}
coincide on the horizon with the Robinson-Trautman null vectors ${\kP,\,\lP}$
given by Eq.~\eqref{ITisRT}. Expressing the tetrad \eqref{RTtetrad} in the ${\uGN,\,\vGN}$
coordinates (cf.\  Eqs.~\eqref{RTTetrRel}, \eqref{AlgSpecNullTetr}) we find
\begin{equation}\label{parlGN}
   \kP\Big\vert_{\begin{subarray}{l}\y=\y_h\\\vGN=n\pi\end{subarray}}
   = \frac{2\DSr\GNtcoef_h}{\rRT_h^2\cosh\acp}\,
   \Bigl(\cot\frac\uGN2\Bigr)^{\pm1}\;\cv{\vGN}\commae
\end{equation}
i.e., the vectors $\kH$ and $\kP$ are proportional.
The vectors $\lH$ and $\lP$ do not even point into the same null direction.
We could explicitly relate the interpretation tetrad ${\kH,\,\lH,\,\mH,\,\bH}$
to the tetrad \eqref{ITisRT} by a combination
of a boost in the ${\kG\textdash\lG}$ plane \eqref{ae:boostrotation}
followed by a transformation \eqref{ae:kfixed} leaving $\kG$ fixed. Of course, this relation
obtained on the horizon is propagated by a parallel transport up to infinity $\scri^+$.
The parameter $B$ of the boost transformation simply follows from  relation ${\kH=B\,\kP}$
between the vectors $\kH$ and $\kP$ (cf.\  Eqs.~\eqref{hparlGN}, \eqref{parlGN}),
\begin{equation}
   B = \frac{\rRT_h\cosh\acp}{\DSr{\sqrt{2\abs{\GNcoef_h}\GNtcoef_h}}}\,\Bigl(\tan\frac\uGN2\Bigr)^{\pm1}\period
\end{equation}
As we discussed in Section~\ref{sc:RadChar} (see Eq.~\eqref{kFixIndep})
the magnitude of the leading term of the fields is independent of the
transformation with $\kG$ fixed, so we do not
need to identify the second transformation \eqref{ae:kfixed} explicitly.

Using the transformation properties \eqref{ae:boostrotationWeyl} and
\eqref{ae:boostrotationEMF} of the fields we finally derive the magnitude of the
leading term of the fields with respect to the interpretation tetrad
${\kH,\,\lH,\,\mH,\,\bH}$ specified on the horizon ${\y=\y_h}$. We obtain
\begin{gather}
  \bigabs{\WTP{\hparl}{4}}\lteq B^{-2}\abs{\WTP{\parl}{4}}\comma
  \bigabs{\EMP{\hparl}{2}}^2\lteq B^{-2}\abs{\EMP{\parl}{2}}^2\commae\\
  B^{-2}=2\abs{\GNcoef_h}\GNtcoef_h\,(\y_h-\x_\fix\tanh\acp)^2\,
  \exp\Bigl(-\frac{\cosh\acp}{\DSr}\frac{\uRT_\fix}{\GNcoef_h}\Bigr)
  \commae\notag
\end{gather}
where $\x_\fix$ and $\uRT_\fix$ denote the coordinates of
the point $N_\fix$ on $\scri^+$, and we have used relations \eqref{rRTatH} and \eqref{ae:GNcoor}.

As expected, such a different choice of the interpretation tetrad does not change the
radiative character of the fields (the ${1/\rRT}$ fall-off),
it only modifies the field components by a finite factor.
Nevertheless, such modification can be substantial --- we have
obtained an additional factor which is exponential in the Robinson-Trautman coordinate $\uRT$,
namely, $\;{\exp\bigl(-\sqrt{(\Lambda/3)+\accl^2\,}\;\uRT_\fix/\GNcoef_h\,\bigr)}$.
This expresses the dependence of the magnitude of gravitational and
electromagnetic radiation on position of the asymptotic point $N_\fix$
at de~Sitter-like infinity $\scri^+$. Notice that the exponential \vague{damping}
of radiation depends not only on the cosmological constant $\Lambda$
but also on the acceleration $\accl$ of the black holes.
Interestingly, the factor ${\sqrt{(\Lambda/3)+\accl^2}}$ is
exactly the Hawking temperature ${2\pi T}$ recently discussed, e.g.,
in Ref.~\cite{DeserLevin:1999}.

\section{Summary}

In the present paper we have thoroughly investigated the
$C$-metric with a positive cosmological constant ${\Lambda>0}$. This
exact solution of the Einstein-Maxwell equations represents a
radiative spacetime in which the radiation is generated by a pair
of (charged) black holes uniformly accelerated in asymptotically
de~Sitter \nopagebreak universe. By introducing new convenient coordinates
and suitable interpretation tetrads near the
conformal light infinity $\scri^+$ we were able to analyze the
asymptotic behavior of gravitational and electromagnetic fields.
The peeling off  property has been demonstrated, the leading
components of the fields in the parallelly transported tetrad are
inversely proportional to the affine parameter of the
corresponding null geodesic.

In addition, as a main result of our investigation, an explicit
formula which describes the directional pattern of radiation has
been derived: it expresses the dependence of the fields on
spatial directions along which a given point $N_+$ at conformal
infinity $\scri^+$ is approached. This specific directional
characteristic supplements the peeling property, thus completing
the asymptotic behavior of gravitational and electromagnetic
fields near infinity $\scri^+$ with a spacelike character.

It was already observed  in the 1960s by Penrose \cite{Penrose:1964,Penrose:1967}
that radiation  is defined \vague{less invariantly}  when $\scri^+$
is spacelike than in the case when it is null
(asymptotically flat spacetimes in particular).
Our results can thus be understood
as an investigation of this \vague{nonuniqueness}.
In fact, the peeling off property supplemented by the directional pattern of
radiation \eqref{DirChar}, \eqref{EMDirChar} characterize \emph{fully}
the radiation near the de~Sitter-like infinity $\scri^+$.

The specific pattern of radiation has been obtained here by
analyzing  the exact model of uniformly accelerated black holes in
de Sitter universe. It is in agreement with the analogous recent
result for the test electromagnetic field generated by accelerated
charges in the de~Sitter background \cite{BicakKrtous:FUACS,BicakKrtous:BIS}.
We are convinced that the directional pattern of radiation derived
has a \vague{universal} validity and applies
to \emph{all} radiative fields of a given Petrov algebraic type
near the spacelike conformal infinity $\scri^+$. The proof of
this statement will be presented elsewhere~\cite{KrtousPodolskyBicak:inprep}.

\begin{acknowledgments}
We are very grateful to Ji\v r\'\i\  Bi\v c\'ak who brought our
attention to the problem of radiation in the de~Sitter-like universes,
the $C$-metric in particular. Thanks are also due to him for reading
the manuscript, for his comments and suggestions. The work has been supported
in part by the grants No.\  GA\v{C}R 202/02/0735 and GAUK 141/2000
of the Czech Republic and the Charles University in Prague.
\end{acknowledgments}

\appendix

\section{Various coordinates for the $C$-metric with $\Lambda$}
\label{apx:Coor}

The $C$-metric with possibly nonvanishing cosmological constant
$\,{\Lambda = 3/\DSr^2}\,$ can be written as
\begin{equation}\label{ae:KWmetric}
  \mtrc =
  \frac1{A^2(\xKW+\yKW)^2}\Bigl(
    -\FKW \,\grad\tKW^2
    +\frac1{\FKW} \,\grad\yKW^2
    +\frac1{\GKW} \,\grad\xKW^2
    +\GKW \,\grad\ph^2
    \Bigr)
\end{equation}
with
\begin{equation}\label{ae:KWFG}
\begin{aligned}
  \FKW &= -\frac1{\DSr^2\accl^2}-1+\yKW^2
  -2\mass\accl\,\yKW^3+\charge^2\accl^2\,\yKW^4\commae\\
  \GKW &= \mspace{81mu} 1-\xKW^2
  -2\mass\accl\,\xKW^3-\charge^2\accl^2\,\xKW^4\period
\end{aligned}
\end{equation}
The functions $\FKW$ and $\GKW$ are polynomials of the coordinates
$\yKW$ and $\xKW$, respectively, and are mutually related by
\begin{equation}\label{ae:KWFGQ}
  \FKW = -\QKW(\yKW)-\frac1{\DSr^2\accl^2}\comma
  \GKW = \QKW(-\xKW)\commae
\end{equation}
where ${\QKW(w)}$ denotes the polynomial
\begin{equation}\label{ae:KWQ}
  \QKW(w) = 1 - w^2 + 2\,\mass\accl\, w^3 - \charge^2\accl^2\, w^4
  \period
\end{equation}
The constants $\accl$, $\mass$,
$\charge$, and $\conpar$ (such that ${\ph\in(-\pi\conpar,\pi\conpar)}$)
parametrize acceleration, mass, charge of the black holes, and conicity
of the $\ph$~symmetry axis, respectively.

The metric \eqref{ae:KWmetric}, \eqref{ae:KWFG} is an ordinary
form of the \mbox{$C$-metric} in the case when the cosmological constant~$\Lambda$
vanishes, i.e., when ${\FKW=-\QKW(\yKW)}$. This has been
extensively used for investigation of uniformly accelerated (pair)
of black holes in asymptotically flat spacetime, see, e.g.,
Refs.~\cite{KinnersleyWalker:1970,LetelierOliveira:2001,Bonnor:1982,EhlersKundt:1962,Pravdovi:2000}.
However, for ${\Lambda\neq0}$ the form of
the generalization is not so obvious and unique. For example, in Ref.~\cite{MannRoss:1995} the
term with the cosmological constant was included in the metric
function $\GKW$ rather than in $\FKW$.
Also, the parametrization of the metric
\eqref{ae:KWmetric} is not unique. A  simple rescaling of
the coordinates can be performed which removes the acceleration
parameter $\accl$ from the conformal factor. These related metric
forms, which allow   an explicit limit ${\accl\to0}$, were
introduced e.g.\, in Refs.~\cite{PlebanskiDemianski:1976,Krameretal:book,PodolskyGriffiths:2001}.

Throughout this paper we use the particularly rescaled coordinates
${\tau,\,\y,\,\x,\,\ph}$ given by
\begin{equation}\label{ae:KWtoxy}
\begin{aligned}
  \tau &= \,\tKW\,\coth\acp \commae&
  \ph &= \ph\comma\\
  \y &= \yKW \,\tanh\acp \comma&
  \x &= -\xKW\commae
\end{aligned}
\end{equation}
where the dimensionless acceleration parameter $\acp$ is
introduced in Eq.~\eqref{acpdef}. In these convenient coordinates the
$C$-metric \eqref{ae:KWmetric}, \eqref{ae:KWFG} takes the form
\begin{equation}\label{ae:xymetric}
  \mtrc =
  \rRT^2\Bigl(
    -\F \,\grad\tau^2 +\frac1{\F} \,\grad\y^2
    +\frac1{\G} \,\grad\x^2 +\G \,\grad\ph^2
    \Bigr)\commae
\end{equation}
with the function $\,\rRT\,$ given by
\begin{equation}\label{ae:rRTdef}
  \rRT = \frac{1}{\accl(\xKW+\yKW)} =
   \frac{\DSr}{\y\cosh\acp-\x\sinh\acp}
\end{equation}
and
\begin{equation}\label{ae:xyFG}
\begin{aligned}
  -\F &= 1-\y^2+\cosh\acp\,\frac{2\mass}{\DSr}\;\y^3
        -\cosh^2\acp\,\frac{\charge^2}{\DSr^2}\;\y^4\commae\\
  \G &= 1-\x^2+\sinh\acp\,\frac{2\mass}{\DSr}\;\x^3
        -\sinh^2\acp\,\frac{\charge^2}{\DSr^2}\;\x^4\period
\end{aligned}
\end{equation}
These coordinates have the following ranges:
${\tau\in\realn}$,
${\ph\in(-\pi\conpar,\pi\conpar)}$,
${\x\in(\x_\maxis,\x_\paxis)}$, and
${\y\in(\x\tanh\acp,\infty)}$,
with ${\x_\maxis,\,\x_\paxis}$ being the two smallest roots of $\G$
--- see discussion in Section~\ref{sc:GlobStr}.

The metric functions ${\FKW,\,\GKW}$ and
${\F,\,\G}$ as functions on the spacetime manifold are related by
\begin{equation}\label{ae:FGFGrel}
  \F = \FKW\; \tanh^2\acp\comma
  \G = \GKW\comma
\end{equation}
but they are usually understood as functions of
different arguments, namely, ${\FKW(\yKW)}$, ${\GKW(\xKW)}$
and ${\F(\y)}$, ${\G(\x)}$. In this sense we will also use a
notation for differentiation of these functions
${\FKW'=d\FKW/d\yKW}$  and ${\F'=d\F/d\y}$ or
${\GKW'=d\GKW/d\xKW}$  and ${\G'=d\F/d\x}$.
The metric function $\G$ takes the values
\begin{equation}\label{ae:Gval}
  \G \in \lclint0,1\rclint\comma
\end{equation}
${\G=0}$ for ${\x=\x_\maxis,\x_\paxis}$ (axes of $\ph$~symmetry), and
${\G=1}$ for ${\x=0}$ (on \vague{equator}, i.e., a $\ph$~circle of maximum circumference).
At  infinity $\scri$ the metric function $\F$ takes the values
\begin{equation}\label{ae:Fscrival}
  -\F \in \lclint\cosh^{-2}\acp,1\rclint\comma
\end{equation}
with ${\F = -\cosh^{-2}\acp}$ on the axes of $\ph$~symmetry, and
${\F =-1}$ on the equator (${\x,\y=0}$).

The above coordinates ${\tau,\,\y,\,\x,\,\ph}$ are closely related
to the \defterm{accelerated coordinates} ${\tacc,\,\racc,\,\thacc,\,\phacc}$
introduced and discussed in Refs.~\cite{PodolskyGriffiths:2001,Podolsky:2002}. If we
define
\begin{equation}\label{ae:xytoacc}
  \tacc = \DSr\,\tau\comma \racc=\frac\DSr\y\comma
  \grad\thacc=\frac1{\sqrt\G}\,\grad\x\comma\phacc=\ph\commae
\end{equation}
the metric \eqref{ae:xymetric} takes the form
\begin{equation}\label{ae:accmetric}
   \mtrc = \frac{\rRT^2}{\racc^2}\Bigl(
   -\hacc\,\grad\tacc^2+\frac1\hacc\,\grad\racc^2 +
   \racc^2\bigl(\grad\thacc^2+\G\,\grad\phacc^2\bigr)
   \Bigr)\commae
\end{equation}
where
\begin{equation}\label{ae:accHdef}
   \hacc =  \frac1{\y^2}\,\F=1-\frac{\racc^2}{\DSr^2}
   -\cosh\acp\,\frac{2\mass}{\racc}
   +\cosh^2\acp\,\frac{\charge^2}{\racc^2}\period
\end{equation}
These coordinates have an obvious physical interpretation in two particular cases ---
in the case of a vanishing acceleration of the black holes (${\accl=0}$),
and for empty de~Sitter spacetime (${\mass=0}$, ${\charge=0}$).
In both these cases the metric function $\G$ reduces to
a simple form ${\G = 1-\x^2}$, so the definition \eqref{ae:xytoacc}
of the angle $\thacc$ gives
\begin{equation}
   \cos\thacc = -\x\comma \sin\thacc = \sqrt{1-\x^2}\period
\end{equation}
For vanishing acceleration ${\accl=0}$, i.e., by setting ${\acp=0}$,
we obtain ${\racc=\rRT}$, and the metric  \eqref{ae:accmetric} reduces to the well-known
metric for the Reissner-Nordstr\"{o}m black hole in Minkowski or
de~Sitter universe \cite{BrillHayward:1994,PodolskyGriffiths:2001},
\begin{equation}
  \mtrc\vert_{\acp=0}=-\hacc\,\grad\tacc^{\,2} + \frac1\hacc\,\grad\racc^2
  +\racc^2\,\bigl(\grad\thacc^2+\sin^2\thacc\,\grad\phacc^2\bigr)\commae
\end{equation}
with the metric function \eqref{ae:accHdef} simplified by ${\cosh\acp=1}$.

In the case of empty de~Sitter space (${\mass=0}$, ${\charge=0}$), but with
generally nonvanishing acceleration, the metric function $\F$ also simplifies to
${\F=\y^2-1}$. The de~Sitter metric in accelerated coordinates thus
takes the form (cf.\ Ref.~\cite{PodolskyGriffiths:2001})
\begin{equation}\label{ae:dSmetricacc}
\begin{split}\raisetag{48pt}
   \mtrc_\dS =&
   \frac{1-\DSr^{-2}{\Ro^2}}{\bigl(1+\DSr^{-2}\Ro\racc\cos\thacc\bigr)^2}\,
   \Biggl(\,-\Bigl(1-\frac{\racc^2}{\DSr^2}\Bigr)\,\grad\tacc^2
   \\&\quad
   +\Bigl(1-\frac{\racc^2}{\DSr^2}\Bigr)^{-1}\,\grad\racc^2 +
   \racc^2\bigl(\grad\thacc^2+\sin^2\thacc\,\grad\phacc^2\bigr)
   \Biggr)\commae
\end{split}
\end{equation}
where we introduced the constant
\begin{equation}\label{ae:Rodef}
   \Ro=\DSr\,\tanh\acp\period
\end{equation}
An explicit relation to the standard de~Sitter static coordinates
${\tdS,\,\rdS,\,\thdS,\,\phdS}$, in which
\begin{equation}\label{ae:dSmetric}
\begin{split}
   \mtrc_\dS =&
   -\Bigl(1-\frac{\rdS^2}{\DSr^2}\Bigr)\,\grad\tdS^2
   +\Bigl(1-\frac{\rdS^2}{\DSr^2}\Bigr)^{\!-1}\,\grad\rdS^2
   \\&\mspace{90mu}
   +\rdS^2\bigl(\grad\thdS^2+\sin^2\thdS\,\grad\phdS^2\bigr)
\end{split}
\end{equation}
is
\begin{equation}\label{ae:dSacctostat}
\begin{aligned}
   \rdS\cos\thdS&=\frac{\racc\cos\thacc+\Ro}{1+\DSr^{-2}\Ro\racc\cos\thacc}\commae&
   \tdS&=\tacc\commae\\
   \rdS\sin\thdS&=\frac{\racc\sin\thacc\sqrt{1-\DSr^{-2}{\Ro^2}}}{1+\DSr^{-2}\Ro\racc\cos\thacc}\commae&
   \phdS&=\phacc\commae
\end{aligned}
\end{equation}
The origin ${\racc=0}$ clearly corresponds to worldlines of two static observers ${\rdS=\Ro}$,
${\thdS=0}$ which move with a uniform acceleration $\accl$.
Further details concerning the interpretation of the accelerated coordinates in de~Sitter space
were discussed at the end of Section~\ref{sc:GlobStr} (see also Ref.~\cite{BicakKrtous:BIS}).

It is also instructive to elucidate a geometrical relation between these two coordinate
systems \eqref{ae:dSmetricacc} and \eqref{ae:dSmetric}. It is wellknown that
the de~Sitter spacetime is conformally related to Minkowski space
(see Refs.~\cite{Penrose:1963,Penrose:1965}, or recently Ref.~\cite{BicakKrtous:ASDS}).
Specifically, the (shaded) domain P of de~Sitter spacetime depicted
in Fig.~\ref{fig:deSitter} corresponds to the ${\tM<0}$ region of Minkowski spacetime
in standard spherical coordinates ${\tM,\,\rM,\,\thM,\,\phM}$, the metrics being
related by ${\mtrc_\dS=(\DSr/\tM)^2\,\mtrc_\Mink}$. The de~Sitter static
coordinates ${\tdS,\,\rdS,\,\thdS,\,\phdS}$ can be obtained from the spherical coordinates
of Minkowski space by a \vague{spherical Rindler} transformation, i.e., ${\rdS=\DSr\,\rM/\tM}$,
${\tdS/\DSr=\frac12\,\log\abs{(\tM^2-\rM^2)/\DSr^2}}$, ${\thdS=\thM}$, ${\phdS=\phM}$.
On the other hand, the accelerated coordinates ${\tacc,\,\racc,\,\thacc,\,\phacc}$ are also
obtained from conformally related Minkowski space
by the same construction, however, starting from a different spherical coordinates
${\tM',\,\rM',\,\thM',\,\phM'}$ which are defined in the inertial frame boosted
along the ${\thM=0}$ direction with the boost given exactly by the acceleration
parameter $\acp$ (i.e., with the relative velocity  ${\tanh\acp}$).
Using this insight we can easily visualize the relation
between the hypersurface ${\rdS=\infty}$ (the conformal infinity $\scri$ of de~Sitter
universe; the ${\tM=0}$ hypersurface of conformally related Minkowski space)
and the hypersurface ${\racc=\infty}$
(the coordinate singularity of accelerated coordinates in de~Sitter space, which is
easily removable, for example, by the coordinate ${\y=\DSr/\racc}$; the
hypersurface ${\tM'=0}$ of Minkowski space), as indicated in Fig.~\ref{fig:deSitter}.
For more details see Ref.~\cite{BicakKrtous:BIS}.

It is particularly  useful to introduce also new coordinates
${\tau,\,\om,\,\sg,\,\ph}$ for the $C$-metric naturally adapted both to the Killings
vectors $\cvil{\tau}$, $\cvil{\ph}$ and to   infinity $\scri$.
In terms of the new coordinate
\begin{equation}\label{ae:omdef}
  \om =  - \y\cosh\acp + \x\sinh\acp = -\frac\DSr\rRT \commae
\end{equation}
infinity $\scri$ is given by a simple condition
${\om = 0}$.
The coordinate $\sg$ is introduced by requiring an
orthogonality of the coordinates. Indeed, if we define $\sg$ by the
differential form
\begin{equation}\label{ae:sgdef}
\begin{gathered}
  \grad\sg = \frac{\sinh\acp}{\F}\,\grad\y +
  \frac{\cosh\acp}{\G}\,\grad\x\commae\\
  \sg=0\quad\text{for}\quad\x,\,\y=0\commae
\end{gathered}
\end{equation}
(which, thanks to Eq.~\eqref{ae:xyFG}, is integrable) the
$C$-metric takes the form
\begin{equation}\label{ae:omsgmetric}
  \mtrc =
  \frac{\DSr^2}{\om^2}\Bigl(
    -\F \,\grad\tau^2
    +\frac1\E \,\grad\om^2
    +\frac{\F\G}{\E} \,\grad\sg^2
    +\G \,\grad\ph^2
    \Bigr)\commae
\end{equation}
where
\begin{equation}\label{ae:Edef}
\begin{split}
  \E &= \F \cosh^2\acp + \G \sinh^2\acp \\
  &=-1 -\om \Bigl[\;\y\cosh\acp+\x\sinh\acp\\
  &\qquad-\frac{2\mass}{\DSr}\bigl(
  \y^2\cosh^2\acp+\y\x\cosh\acp\sinh\acp+
  \x^2\sinh^2\acp\bigr)\\
  &\qquad+\frac{\charge^2}{\DSr^2}\bigl(
  \y^3\cosh^3\acp+\y^2\x\cosh^2\acp\sinh\acp\\
  &\qquad\qquad\quad+\y\x^2\cosh\acp\sinh^2\acp+\x^3\sinh^3\acp\bigr)
  \Bigr]\period\raisetag{94pt}
\end{split}
\end{equation}
Obviously, on $\scri$, where ${\om=0}$, we obtain ${\E=-1}$.
Thanks to  relation ${\F<0}$ in region I of Fig.~\ref{fig:DiagDet}, we observe
from metric \eqref{ae:omsgmetric} that near infinity $\scri^+$,
the coordinate $\om$ plays the role of a time if ${\E<0}$.
It can be shown that for ${\E<0}$ the coordinate transformation
\eqref{ae:omdef}, \eqref{ae:sgdef} from ${\y,\,\x}$ to ${\om,\,\sg}$ is invertible.
We will use the coordinate $\sg$ only in this region. The hypersurface ${\E=0}$ is always located
above the cosmological horizon and it touches the horizon on the  axes
${\x=\x_\maxis,\,\x_\paxis}$, see the left part of Fig.~\ref{fig:DiagDet}.

The $C$-metric can also be put into the Robinson-Trautman
form (see Ref.~\cite{Podolsky:PhD}).
Introducing the coordinates $\rRT$ and $\uRT$,
\begin{equation}\label{ae:rRTKW}
\begin{aligned}
  \accl\,\rRT &= (\xKW+\yKW)^{-1}\commae\\
  \accl\,\grad\uRT &= \frac{\grad\yKW}{\FKW}+\grad\tKW\commae
\end{aligned}
\end{equation}
we obtain from Eq.~\eqref{ae:KWmetric} the metric
\begin{equation}\label{ae:KWmetricH}
\begin{aligned}
  \mtrc =
  &\,\rRT^2\,\Bigl(\frac1{\GKW} \,\grad\xKW^2
    +\GKW \,\grad\pKW^2    \Bigr)\\
  &\quad -\grad \uRT\stp\grad\rRT-\accl\,\rRT^2\grad\uRT\stp\grad\xKW
    -\accl^2\rRT^2\FKW\,\grad\uRT^2\commae
\end{aligned}
\end{equation}
where the function ${\accl^2\rRT^2\FKW}$, expressed in the
coordinates $\,\xKW$, $\rRT$ using $\,{\yKW=(\accl\rRT)^{-1}-\xKW}$,
reads
\begin{equation}\label{ae:KWH}
\begin{aligned}
 \accl^2\rRT^2\FKW &=-\frac{\rRT^2}{\DSr^2}-\accl^2\rRT^2\GKW
   +\accl\rRT\,\GKW'-\textstyle{\frac{1}{2}}\GKW''\\
 &\qquad\quad +\textstyle{\frac{1}{6}}(\accl\rRT)^{-1}\GKW'''
 -\textstyle{\frac{1}{24}}(\accl\rRT)^{-2}\GKW'''' \period
\end{aligned}
\end{equation}
This is the generalization of the Kinnersley-Walker coordinates
\cite{KinnersleyWalker:1970} to ${\Lambda\not=0}$.
Introducing the complex coordinates ${\zRT,\,\bRT}$
(or real coordinates $\nRT$, $\ph$, related by ${\zRT=\frac1{\sqrt 2}(\nRT-i\ph)}\,$)
instead of the coordinates ${\xKW,\,\ph}$,
\begin{equation}\label{ae:zetaRTKW}
\begin{aligned}
  \textstyle{\frac{1}{\sqrt2}}  (\grad\zRT+\grad\bRT)&= \grad\nRT \,= A\,\grad\uRT-\frac{\grad\xKW}{\GKW}\commae\\
  \textstyle{\frac{i}{\sqrt2}}(\grad\zRT-\grad\bRT)&= \grad\ph  \commae
\end{aligned}
\end{equation}
(notice that ${\nRT=\tau\tanh\acp+\sg\sech\acp}$),
we put the $C$-metric into the Robinson-Trautman form
\begin{equation}\label{ae:RTmetric}
  \mtrc =
  \frac{\rRT^2}{\PRT^2}\,\grad\zRT\stp\grad\bRT
    -\grad \uRT\stp\grad\rRT-\HRT\,\grad\uRT^2
\end{equation}
(or, alternatively, with ${\grad\zRT\stp\grad\bRT}$ replaced by ${\grad\nRT^2+\grad\ph^2}$),
where the metric functions are
\begin{equation}\label{ae:PandH}
\begin{aligned}
  \PRT^{-2}& = \GKW \qquad\qquad\quad=\G\commae\\
  \HRT & =\accl^2\rRT^2(\FKW+\GKW) =\frac{\rRT^2}{\DSr^2}\,\E \period
\end{aligned}
\end{equation}
Using Eqs.~\eqref{ae:KWH}, \eqref{ae:PandH}, and \eqref{ae:zetaRTKW},  which for
${\PRT=\GKW^{-\frac{1}{2}}}$ imply
${\accl\,\GKW'=-2\,(\ln\PRT)_{,u}}$ and
${\GKW''=-2\,\Delta\ln\PRT}$
with ${\Delta=2\PRT^2\partial_{\zRT}\partial_{\bRT}}$, we recover that
\begin{equation}\label{ae:cosi}
\begin{aligned}
 \HRT &=-\frac{\rRT^2}{\DSr^2}-2\,\rRT\,(\ln\PRT)_{,u}+\Delta\ln\PRT\\
 &\qquad\qquad\quad -\frac{2}{\rRT}(\mass-2\charge^2\accl\,\x)+
 \frac{\charge^2}{\rRT^2}\period
\end{aligned}
\end{equation}
This is the standard general expression for the metric function of the
Robinson-Trautman solution \cite{Krameretal:book}.
Let us finally note that the Christoffel coefficients for the metric \eqref{ae:RTmetric} are
\begin{equation}\label{ae:ChrisSymb}
\begin{gathered}
\Gamma^\zRT_{\rRT\zRT}=\frac{1}{\rRT}\commae\
\Gamma^\zRT_{\uRT\zRT}=-\frac{\PRT_{,\uRT}}{\PRT}\commae\
\Gamma^\zRT_{\zRT\zRT}=-\frac{2\PRT_{,\zRT}}{\PRT}\commae\
\Gamma^\zRT_{\uRT\uRT}=\frac{\PRT^2\HRT_{,\bRT}}{2\rRT^2} \commae
\\
\Gamma^\uRT_{\zRT\bRT}=\frac{\rRT}{\PRT^2}\comma\
\Gamma^\uRT_{\uRT\uRT}=-\textstyle{\frac{1}{2}}\HRT_{,\rRT}\comma
\Gamma^\rRT_{\uRT\uRT}=\textstyle{\frac{1}{2}}\HRT\HRT_{,\rRT}+\textstyle{\frac{1}{2}}\HRT_{,\uRT}\commae
\\
\Gamma^\rRT_{\zRT\bRT}=-\frac{\rRT\PRT\HRT+\rRT^2\PRT_{,\uRT}}{\PRT^3}\comma
\Gamma^\rRT_{\uRT\zRT}=\textstyle{\frac{1}{2}}\HRT_{,\zRT}\comma
\Gamma^\rRT_{\uRT\rRT}=\textstyle{\frac{1}{2}}\HRT_{,\rRT}\period
\end{gathered}
\end{equation}

Finally, for a discussion of the global structure of the spacetime it is necessary,
following the general approach \cite{HawkingEllis:book,Walker:1970},
to introduce global double null coordinates ${\uGN,\,\vGN,\,\x,\,\ph}$.
For this, we supplement the above defined null coordinate $\uRT$ with
the complementary null coordinate $\vRT$ (\noteref{nt:vupsilon}).
In terms of the coordinates ${\tau,\,\y}$
these are (cf.\  Eq.~\eqref{ae:rRTKW}, \eqref{ae:KWtoxy})
\begin{equation}\label{ae:DNcoor}
  \uRT=\frac\DSr{\cosh\acp}\bigl(\ys+\tau\bigr)\comma
  \vRT=\frac\DSr{\cosh\acp}\bigl(\ys-\tau\bigr)\commae
\end{equation}
where the tortoise coordinate $\ys$ is defined
by the differential relation
\begin{equation}\label{ae:ysder}
  \grad\ys=\frac1\F\,\grad\y\period
\end{equation}
Taking into account the  polynomial structure \eqref{ae:xyFG}
of the function $\F$,
\begin{equation}\label{ae:Fpolyn}
  \F=\prefact\prod_{h}(\y-\y_h)\commae
\end{equation}
where ${\prefact=\text{constant}}$, and $\y_h$ (${h=\ihor,\,\ohor,\,\chor,\,\mhor}$) are the
values of the coordinate $\y$ at the horizons (the roots of $\F$), we obtain
\begin{equation}\label{ae:yscoor}
  \ys=\sum_{h}\GNcoef_h \log\abs{\y-\y_h}\comma
  \GNcoef_h = (\F'\vert_{\y=\y_h})^{-1}\period
\end{equation}
In these coordinates the $C$-metric with ${\Lambda>0}$ takes the form
\begin{equation}\label{ae:DNmetric}
  \mtrc = \rRT^2\Bigl(\frac{\F\,\cosh^2{\acp}}{2\,\DSr^2}\,\grad\uRT\stp\grad\vRT
    +\frac1\G\,\grad\x^2+\G\,\grad\ph^2\Bigr)\period
\end{equation}

Now, we can define the global null coordinates ${\uGN,\,\vGN,\,\x,\,\ph}$ parametrized by
a constant coefficient $\GNcoef$, covering, for suitable values of $\GNcoef$, the
horizons smoothly,
\begin{equation}\label{ae:GNcoor}
\begin{gathered}
  \abs{\tan\frac{\uGN}{2}}=\exp\Bigl(\frac{\uRT\cosh\acp}{2\,\abs{\GNcoef}\,\DSr}\Bigr)
  \comma \sign\Bigl(\tan\frac{\uGN}{2}\Bigr)=(-1)^m\commae\\
  \abs{\tan\frac{\vGN}{2}}=\exp\Bigl(\frac{\vRT\cosh\acp}{2\,\abs{\GNcoef}\,\DSr}\Bigr)
  \comma \sign\Bigl(\tan\frac{\vGN}{2}\Bigr)=(-1)^n\period
\end{gathered}
\end{equation}
The $C$-metric in these coordinates then takes the form
\begin{equation}\label{ae:GNmetric}
  \mtrc = \rRT^2\Bigl(\frac{2\,\GNcoef^2\F}{\sin\uGN\sin\vGN}\,
    \grad\uGN\stp\grad\vGN
    +\frac1\G\,\grad\x^2+\G\,\grad\ph^2\Bigr)\period
\end{equation}
The horizons ${\y=\y_\ihor,\,\y_\ohor,\,\y_\chor}$ now correspond
to the values ${\uGN=m\pi}$ or ${\vGN=n\pi}$, with ${m,n\in\integern}$
(see Fig.~\ref{fig:ConfDiag}).

Notice, that it follows from Eqs.~\eqref{ae:GNcoor}, \eqref{ae:yscoor} that
\begin{equation}
  \abs{\tan\frac\uGN2\,\tan\frac\vGN2}^{\sign\GNcoef}=
  \prod_k\abs{\y-\y_k}^{\GNcoef_k/\GNcoef}\period
\end{equation}
Evaluating this expression on the particular horizon ${\y=\y_h}$ and comparing with Eq.~\eqref{ae:Fpolyn} we
find that with the choice ${\GNcoef=\GNcoef_h}$ the expression
\begin{equation}\label{ae:tantan}
\begin{split}
  &-\F\,\Bigl(\tan\frac\uGN2\,\tan\frac\vGN2\Bigr)^{\!-\sign\GNcoef}\,\Bigg\vert_{\y=\y_h}\\
  &\qquad\qquad=\prefact\prod_{k\ne h}\abs{\y_h-\y_k}^{1-\GNcoef_k/\GNcoef_h} =
  \frac1{\abs{\GNcoef_h}\GNtcoef_h}
\end{split}
\end{equation}
is finite and nonvanishing. Here we introduced the constant
\begin{equation}\label{ae:GNtcoef}
   \GNtcoef_h=\prod_{k\ne h}\abs{\y_h-\y_k}^{\GNcoef_k/\GNcoef_h}\period
\end{equation}
Using this fact it is possible to guarantee a regularity of the  metric
coefficient ${\GNcoef^2\F/(\sin\uGN\sin\vGN)}$ in Eq.~\eqref{ae:GNmetric}
(including smoothness and that it is finite and nonzero)
and smoothness of the coordinates $\rRT$ and $\y$ near the
horizon $\y_h$ by the choice ${\GNcoef=\GNcoef_h}$ of
the coefficient $\GNcoef$ in Eqs.~\eqref{ae:GNcoor},
assuming that ${\uGN,\,\vGN}$ forms a smooth
coordinate map in the neighborhood of this horizon.
However, such an appropriate factor $\GNcoef$
cannot be chosen for all horizons simultaneously --- a different
smooth map ${\uGN,\,\vGN}$ parametrized by different coefficients
$\GNcoef$ has to be used near the different horizons to
demonstrate the smoothness of the metric in the whole spacetime
(see, e.g., Refs.~\cite{Walker:1970,HawkingEllis:book} for a general discussion).

\section{Properties of the metric functions $\F$ and $\G$}
\label{apx:FGprop}

\begin{figure}
\includegraphics{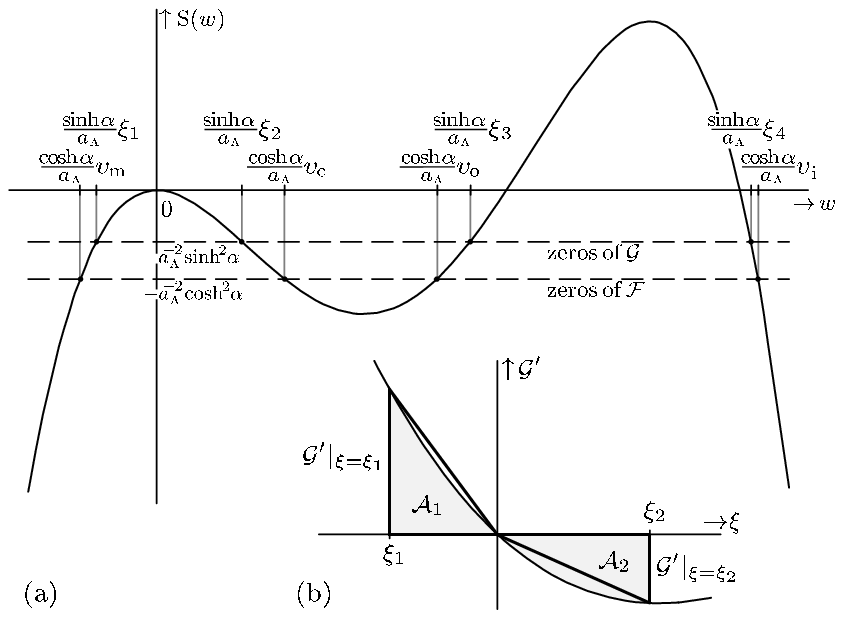}
\caption{\label{fig:PolS}
(a) A qualitative shape of the metric functions $\F$ and $\G$
(in the case ${\mass\ne0}$, ${\charge\ne0}$)
which are polynomials in $\y$ and $\x$, respectively. It follows
from the representation \eqref{ae:xyFGS} that both functions $\F$ and $\G$
are, up to the specific rescaling and the constant term, given by the same polynomial
${\SPD(w)}$, the graph of which is presented here. The zeros of $\F$ and $\G$
are thus given by intersections of the graph of $\SPD$ with the horizontal lines ${-\DSr^{-2}\cosh^2\acp}\,$
and ${-\DSr^{-2}\sinh^2\acp}$, respectively. Relations \eqref{zerosord} between the
zeros of $\F$ and $\G$ follows immediately from this fact.
(b) A graphical representation of the triangular estimate \eqref{ae:areasest}
for the areas ${\area_\maxis,\,\area_\paxis}$ under the graph of
$\G'$.}%
\end{figure}

First, let us note that $\F$ and $\G$ can be
represented in terms of polynomial ${\SPD(w)}$
\begin{equation}\label{ae:xyFGS}
\begin{aligned}
  -\frac{\cosh^2\acp}{\DSr^2}\,\F &=
     \SPD\Bigl(\frac{\cosh\acp}{\DSr}\,\y\Bigr) +
     \frac{\cosh^2\acp}{\DSr^2}\commae\\
  \frac{\sinh^2\acp}{\DSr^2}\,\G &=
     \SPD\Bigl(\frac{\sinh\acp}{\DSr}\,\x\Bigr) +
     \frac{\sinh^2\acp}{\DSr^2}\period
\end{aligned}
\end{equation}
where
\begin{equation}\label{ae:SPDdef}
  \SPD(w) = - w^2 (1-2\,\mass w + \charge^2 w^2)\period
\end{equation}
A typical graph of the polynomial ${\SPD(w)}$ is drawn in Fig.~\ref{fig:PolS}.
By inspecting the graph we obtain, e.g.,  relations \eqref{zerosord}
between the roots of the metric functions $\F$ and $\G$.

We may also prove some interesting properties, including Eq.~\eqref{ConOrd}, of the metric
function $\G$ in the case of charged accelerated black holes.
(Similar properties --- in particular the inequality \eqref{ae:ConOrd} --- can be also proved for
uncharged accelerated black holes, i.e., for ${\charge=0}$, ${\mass\neq0}$, ${\accl\neq0}$.)
In the case ${\charge\neq0}$, ${\accl\neq0}$, the metric function $\G$ is a polynomial
of the fourth order in $\x$ and its zeros have
been denoted in the ascending order as ${\x_\maxis,\,\x_\paxis,\,\x_3,\,\x_4}$.
The extremes of $\G$ (zeros of $\G'$) are ${\xt_0=0}$,
${\xt_\pm=(3\mass\pm\sqrt{9\mass^2-8\charge^2})/({4\accl\charge^2})}$.
The zero of $\,{\G'''=12(\accl\mass-2\charge^2\accl^2\x)}\,$
is $\,{\xttt=\mass/(2\accl\charge^2)}\,$, and ${\G'''>0}$ for ${\x<\xttt}$.
Using the conditions \eqref{assumtions}, a straightforward calculation
leads to an inequality ${\G\vert_{\xt_+}>0}$. The condition that $\G$ has four real roots
requires ${\G\vert_{\xt_-}<0}$. This confirms that the graph of $\G$ has always a
qualitative shape shown on Fig.~\ref{fig:PolS}.
The extremes of $\G$ are located between its zeros, so that ${\x_\maxis<0<\x_\paxis<\xt_-}$.
Expressing the vanishing linear coefficient in $\G$ in terms of
the roots we obtain
${(\x_\maxis+\x_\paxis)\x_3\x_4=-\x_\maxis\x_\paxis(\x_3+\x_4)}$, the right-hand
side is clearly positive, as well as ${\x_3\x_4}$, so we obtain
\begin{equation}\label{ae:xmlessxp}
-\x_\maxis<\x_\paxis\period
\end{equation}
From the conditions \eqref{assumtions} it
also follows that $\xt_-<\xttt$ and thus we have
\begin{equation}
  \x_\maxis<0<\x_\paxis<\xt_-<\xttt\period
\end{equation}
This means that $\G'$ is convex on the relevant interval ${(\x_\maxis,\x_\paxis)}$,
it is positive on the interval ${(\x_\maxis,0)}$ and negative on
${(0,\x_\paxis)}$. The positivity of $\G'''$ on the interval
${(\x_\maxis,\x_\paxis)}$
also implies ${\mass-2\charge^2\accl\,\x_\fix>0}$, which is the relation used in the
discussion following the result \eqref{DirChar}.
Clearly, ${\int_{\x_\maxis}^{\x_\paxis}\G'd\x=0}$, i.e., the areas
\begin{equation}\label{ae:areasdef}
  \area_\maxis = \int_{\x_\maxis}^{0}\G'd\x\comma
  \area_\paxis =-\int_{0}^{\x_\paxis}\G'd\x
\end{equation}
are the same. Thanks to the convexity of $\G'$, we can estimate $\area_\maxis$
and $\area_\paxis$ by  simpler triangular areas  (see Fig.~\ref{fig:PolS}(b)),
and we obtain
\begin{equation}\label{ae:areasest}
  -{\textstyle\frac12}\,\x_\paxis\,\G'\vert_{\x=\x_\paxis} < \area_\paxis =
  \area_\maxis < -{\textstyle\frac12}\,\x_\maxis\G'\vert_{\x=\x_\maxis}\period
\end{equation}
Using Eq.~\eqref{ae:xmlessxp} this implies
\begin{equation}\label{ae:ConOrd}
  -\G'\vert_{\x=\x_\paxis} < \G'\vert_{\x=\x_\maxis}\period
\end{equation}
Considering Eq.~\eqref{conicity} we obtain the important relation \eqref{ConOrd}
which is necessary for the discussion of  conicity in Section~\ref{sc:GlobStr}.

\section{Relations between the coordinate frames}
\label{apx:CoorFrames}

In this appendix we summarize for convenience the relations between
different coordinate 1-form and vector frames

\begin{widetext}
For coordinate 1-form frames
${(\grad{\tau},\,\grad{\y},\,\grad{\x},\,\grad{\ph})}$,
${(\grad{\tau},\,\grad{\om},\,\grad{\sg},\,\grad{\ph})}$, and
${(\grad{\zRT},\,\grad{\bRT},\,\grad{\rRT},\,\grad{\uRT})}$
we obtain
\begin{subequations}\label{FormCoorF}
\begin{alignat}{3}
  &\grad\tau                                                &
      & = \grad\tau                                         &
      & = \frac\E{\F\cosh\acp}\,\frac1\DSr\,\grad\uRT
        + \frac1{\F\cosh\acp}\frac\DSr{\rRT^2}\,\grad\rRT
        - \frac\G\F\tanh\acp\,\grad\nRT                     \commae\\
  &\grad\y                                                  &
      & = -\frac\F\E\,\cosh\acp\,\grad\om
        + \frac{\F\G}\E\,\sinh\acp\,\grad\sg                &
      & = -\G\,\frac{\sinh^2\acp}{\cosh\acp}\,\frac1\DSr\,\grad\uRT
        - \frac1{\cosh\acp}\frac\DSr{\rRT^2}\,\grad\rRT
        + \G\tanh\acp\,\grad\nRT                            \commae\\
  &\grad\x                                                  &
      & = \frac\G\E\,\sinh\acp\,\grad\om
        + \frac{\F\G}\E\,\cosh\acp\,\grad\sg                &
      & = -\G\sinh\acp\,\frac1\DSr\,\grad\uRT
        + \G\,\grad\nRT                                     \commae\\
  &\grad\om                                                 &
      & = -\cosh\acp\,\grad\y+\sinh\acp\,\grad\x            &
      & = \frac\DSr{\rRT^2}\,\grad\rRT                      \commae\\
  &\grad\sg                                                 &
      & = \frac{\sinh\acp}\F\,\grad\y
        + \frac{\cosh\acp}\G\,\grad\x                       &
      & = -\frac\E\F\,\tanh\acp\,\frac1\DSr\,\grad\uRT
        - \frac{\tanh\acp}\F\frac\DSr{\rRT^2}\,\grad\rRT
        + \frac\E{\F\cosh\acp}\,\grad\nRT                   \commae\displaybreak[0]\\
  \frac1\DSr\,&\grad\uRT                                    &
      & = \frac1{\cosh\acp}\,\grad\tau
        + \frac1{\F\cosh\acp}\,\grad\y                      &
      & = \frac1{\cosh\acp}\,\grad\tau
        - \frac1\E\,\grad\om
        + \frac\G\E\,\tanh\acp\,\grad\sg                    \commae\displaybreak[0]\\
  \frac1\DSr\,&\grad\rRT                                    &
      & = -\frac{\cosh\acp}{\om^2}\,\grad\y
        + \frac{\sinh\acp}{\om^2}\,\grad\x                  &
      & = \frac1{\om^2}\,\grad\om                           \commae\displaybreak[0]\\
  \sqrt2\,&\grad\zRT                                        &
      & = \tanh\acp\,\grad\tau
        + \frac{\tanh\acp}{\F}\,\grad\y
        + \frac1\G\,\grad\x
        - i\,\grad\ph                                       &
      & = \tanh\acp\,\grad\tau
        + \frac1{\cosh\acp}\,\grad\sg
        - i\,\grad\ph                                       \commae
\end{alignat}
where
\begin{equation}
\grad\nRT = \textstyle\frac1{\sqrt2}\bigl(\grad\zRT+\grad\bRT\bigr) \comma
\grad\ph  = \textstyle\frac{i}{\sqrt2}\bigl(\grad\zRT-\grad\bRT\bigr) \comma
\grad\zRT = \textstyle\frac1{\sqrt2}\bigl(\grad\nRT-i\,\grad\ph\bigr)\comma
\grad\bRT = \overline{\grad\zRT}\period
\end{equation}
\end{subequations}
Coordinate vector frames
${(\cvil{\tau},\,\cvil{\y},\,\cvil{\x},\,\cvil{\ph})}$,
${(\cvil{\tau},\,\cvil{\om},\,\cvil{\sg},\,\cvil{\ph})}$, and
${(\cvil{\zRT},\,\cvil{\bRT},\,\cvil{\rRT},\,\cvil{\uRT})}$
are related by
\begin{subequations}\label{VecCoorF}
\begin{alignat}{3}
  &\cv\tau                                                  &
      & = \cv\tau                                           &
      & = \frac1{\cosh\acp}\,\DSr\,\cv\uRT
        + \tanh\acp\,\cv\nRT                                \commae\\
  &\cv\y                                                    &
      & = -\cosh\acp\,\cv\om
        + \frac{\sinh\acp}\F\,\cv\sg                        &
      & = \frac1{\F\cosh\acp}\,\DSr\,\cv\uRT
        - \cosh\acp\,\frac{\rRT^2}{\DSr}\,\cv\rRT
        + \frac{\tanh\acp}{\F}\,\cv\nRT                     \label{VecCoorFy}\commae\\
  &\cv\x                                                    &
      & = \sinh\acp\,\cv\om
        + \frac{\cosh\acp}{\G}\,\cv\sg                      &
      & = \sinh\acp\,\frac{\rRT^2}{\DSr}\,\cv\rRT
        + \frac1\G\,\cv\nRT                                 \label{VecCoorFx}\commae\\
  &\cv\om                                                   &
      & = -\frac\F\E\,\cosh\acp\,\cv\y
        + \frac\G\E\,\sinh\acp\,\cv\x                       &
      & = -\frac1\E\,\DSr\,\cv\uRT
        +\frac{\rRT^2}\DSr\,\cv\rRT                         \commae\\
  &\cv\sg                                                   &
      & = \frac{\F\G}\E\,\sinh\acp\,\cv\y
        + \frac{\F\G}\E\,\cosh\acp\,\cv\x                   &
      & = \frac\G\E\,\tanh\acp\,\DSr\,\cv\uRT
        + \frac1{\cosh\acp}\,\cv\nRT                        \commae\\
  \DSr\,&\cv\uRT                                            &
      & = \frac\E{\F\cosh\acp}\,\cv\tau
        - \G\frac{\sinh^2\acp}{\cosh\acp}\,\cv\y
        - \G\sinh\acp\,\cv\x                                &
      & = \frac\E{\F\cosh\acp}\,\cv\tau
        - \frac\E\F\,\tanh\acp\,\cv\sg                      \label{VecCoorFu}\commae\\
  \DSr\,&\cv\rRT                                            &
      & = \frac{\om^2}{\F\cosh\acp}\,\cv\tau
        - \frac{\om^2}{\cosh\acp}\,\cv\y                    &
      & = \frac{\om^2}{\F\cosh\acp}\,\cv\tau
        + \om^2\,\cv\om
        - \frac{\tanh\acp}\F\,\om^2\,\cv\sg                 \commae\\
  \sqrt2\,&\cv\zRT                                          &
      & = -\frac\G\F\,\tanh\acp\,\cv\tau
        + \G\tanh\acp\,\cv\y
        + \G\cv\x
        + i\,\cv\ph                                         &
      & = -\frac\G\F\,\tanh\acp\,\cv\tau
        + \frac\E{\F\cosh\acp}\,\cv\sg
        + i\,\cv\ph                                         \label{VecCoorFz}\commae
\end{alignat}
where
\begin{equation}
\cv\nRT = \textstyle\frac1{\sqrt2}\bigl(\cv\zRT+\cv\bRT\bigr) \comma
\cv\ph  = -\textstyle\frac{i}{\sqrt2}\bigl(\cv\zRT-\cv\bRT\bigr) \comma
\cv\zRT = \textstyle\frac1{\sqrt2}\bigl(\cv\nRT+i\,\cv\ph\bigr)\comma
\cv\bRT = \overline{\cv\zRT}\period
\end{equation}
\end{subequations}

It is also useful to express the relations between different null tetrads introduced in the
paper. The special null tetrad (defined by Eq.~\eqref{AlgSpecTetr}, using Eq.~\eqref{NormNullTetr})
the reference null tetrad (see Eq.~\eqref{KillTetr}),
and the Robinson-Trautman tetrad \eqref{RTtetrad} are related by
\begin{align}
  \kS
    &= \frac12\tan\THT_\spec\,
         \Bigl(\cot\frac{\THT_\spec}2\;\kK+\tan\frac{\THT_\spec}2\;\lK+\mK+\bK\Bigr)
    &&\nquad= \exp(-\beta_\RoTr)\sec{\THT_\spec}\,\kT
    \commae\notag\\
  \lS
    &= \frac12\tan\THT_\spec\,
         \Bigl(\tan\frac{\THT_\spec}2\;\kK+\cot\frac{\THT_\spec}2\;\lK+\mK+\bK\Bigr)
    &&\nquad= \sin\THT_\spec\,
         \Bigl(\exp(-\beta_\RoTr)\tan\THT_\spec\,\kT
         +\exp(\beta_\RoTr)\cot\THT_\spec\,\lT
         +\mT+\bT\Bigr)
    \commae\notag\\
  \mS
    &= \frac12\tan\THT_\spec\,
         \Bigl(\kK+\lK+\cot\frac{\THT_\spec}2\;\mK+\tan\frac{\THT_\spec}2\;\bK\Bigr)
    &&\nquad= \mT + \exp(-\beta_\RoTr)\tan\THT_\spec\,\kT
    \commae\label{SpecTetrRel}\\
  \bS
    &= \frac12\tan\THT_\spec\,
         \Bigl(\kK+\lK+\tan\frac{\THT_\spec}2\;\mK+\cot\frac{\THT_\spec}2\;\bK\Bigr)
    &&\nquad= \bT + \exp(-\beta_\RoTr)\tan\THT_\spec\,\kT
    \commae\notag
\end{align}
\begin{align}
  \kT
     &= \frac12\exp(\spcm\beta_\RoTr)\sin\THT_\spec\,
         \Bigl(\cot\frac{\THT_\spec}2\,\kK + \tan\frac{\THT_\spec}2\,\lK+\mK+\bK\Bigr)
     &&\nquad= \exp(\beta_\RoTr)\cos\THT_\spec\,\kS
     \commae\notag\\
  \lT
     &= \frac12\exp(-\beta_\RoTr)\sin\THT_\spec\,
         \Bigl(\tan\frac{\THT_\spec}2\,\kK + \cot\frac{\THT_\spec}2\,\lK-\mK-\bK\Bigr)
     &&\nquad= \exp(-\beta_\RoTr)\tan\THT_\spec\,
         \Bigl(\sin\THT_\spec\,\kS+\csc\THT_\spec\,\lS-\mS-\bS\Bigr)
     \commae\notag\\
  \mT
     &= \frac12\sin\THT_\spec\,\Bigl(-\kK+\lK+\cot\frac{\THT_\spec}2\,\mK-\tan\frac{\THT_\spec}2\,\bK\Bigr)
     &&\nquad= \mS - \sin\THT_\spec\,\kS
     \commae\label{RTTetrRel}\\
  \bT
     &= \frac12\sin\THT_\spec\,\Bigl(-\kK+\lK-\tan\frac{\THT_\spec}2\,\mK+\cot\frac{\THT_\spec}2\,\bK\Bigr)
     &&\nquad= \bS - \sin\THT_\spec\,\kS
     \period\notag
\end{align}
The factor $\beta_\RoTr$ is defined in Eq.~\eqref{BparRTdef}.
The rotated null tetrad (cf.\  Eq.~\eqref{GenKillTetr}) is related to the
reference tetrad as
\begin{equation}\label{RotTetrRel}
\begin{aligned}
  \kR &= \frac12\sin\THT\,\Bigl(
       \cot{\frac\THT2}\;\kK + \tan{\frac\THT2}\;\lK
       + \exp(i\PHI)\,\mK + \exp(-i\PHI)\,\bK \Bigr)
    \commae\mspace{340mu}\\
  \lR &= \frac12\sin\THT\,\Bigl(
       \tan{\frac\THT2}\;\kK + \cot{\frac\THT2}\;\lK
       - \exp(i\PHI)\,\mK - \exp(-i\PHI)\,\bK \Bigr)
    \commae\\
  \mR &= \frac12\sin\THT\,\Bigl(-\kK+\lK
       +\cot{\frac\THT2}\,\exp(i\PHI)\,\mK
       -\tan{\frac\THT2}\,\exp(-i\PHI)\,\bK\Bigr)
    \commae\\
  \bR &= \frac12\sin\THT\,\Bigl(-\kK+\lK
       -\tan{\frac\THT2}\,\exp(i\PHI)\,\mK
       +\cot{\frac\THT2}\,\exp(-i\PHI)\,\bK\Bigr)
    \period
\end{aligned}
\end{equation}
Here, the angle $\THT_\spec$ is defined by Eq.~\eqref{THTsDef}.

\end{widetext}

\section{Transformations of the components $\WTP{}{n}$ and $\EMP{}{n}$}
\label{apx:Transformations}

The components $\WTP{}{n}$ of the Weyl tensor (see \noteref{nt:WTcomp}),
\begin{equation}
\begin{gathered}\label{WTPsi}
  \WTP{}{0}=\WT_{\kappa\mu\kappa\mu} \comma
  \WTP{}{4}=\WT_{\lambda\bmu\lambda\bmu}\commae\\
  \WTP{}{1}=-\WT_{\kappa\mu\mu\bmu}          = -\WT_{\mu\bmu\kappa\mu}
          = \WT_{\kappa\lambda\kappa\mu}     = \WT_{\kappa\mu\kappa\lambda}
          \commae\\
  \WTP{}{3}=-\WT_{\kappa\lambda\lambda\bmu}  = -\WT_{\lambda\bmu\kappa\lambda}
          = \WT_{\lambda\bmu\mu\bmu}         = \WT_{\mu\bmu\lambda\bmu}
          \commae\\
  \WTP{}{2}=-\WT_{\kappa\mu\lambda\bmu}      = -\WT_{\lambda\bmu\kappa\mu}\commae\\
  2\re\WTP{}{2}=\WT_{\kappa\lambda\kappa\lambda} = \WT_{\mu\bmu\mu\bmu}\commae\\
  2\im\WTP{}{2}=i\,\WT_{\kappa\lambda\mu\bmu} = i\,\WT_{\mu\bmu\kappa\lambda}\commae
\end{gathered}
\end{equation}
and the components $\EMP{}{n}$
of tensor of electromagnetic field
\begin{equation}\label{EMFPhi}
\begin{aligned}
  \EMP{}{0}&=\EMF_{\kappa\mu} \commae&
  2 \re\EMP{}{1}&=\EMF_{\kappa\lambda}\commae\\
  \EMP{}{2}&=\EMF_{\bmu\lambda}\commae&
  2 \im\EMP{}{1}&=i\,\EMF_{\mu\bmu}\commae
\end{aligned}
\end{equation}
transform in the well-known way under
special Lorentz transformations, see, e.g., Ref.~\cite{Krameretal:book}.

For a null rotation with $\kG$ fixed,
\begin{equation}\label{ae:kfixed}
\begin{aligned}
  \kG &= \kK\commae\\
  \lG &= \lK + \bar L \mK + L \bK + L\bar L \kK\commae\\
  \mG &= \mK + L\kK\commae\\
  \bG &= \bK + \bar L\kK\commae
\end{aligned}
\end{equation}
$L$ being a complex number which parametrize the rotation,
the components of the Weyl tensor transform as
\begin{equation}\label{ae:kfixedWeyl}
\begin{split}
  \WTP{}{0} &= \WTP{\klg}{0}\commae\\
  \WTP{}{1} &= {\bar L}\, \WTP{\klg}{0} + \WTP{\klg}{1} \commae\\
  \WTP{}{2} &= {\bar L}^2 \WTP{\klg}{0} + 2 {\bar L}\, \WTP{\klg}{1} + \WTP{\klg}{2}\commae\\
  \WTP{}{3} &= {\bar L}^3 \WTP{\klg}{0} + 3 {\bar L}^2 \WTP{\klg}{1} +
    3 {\bar L}\, \WTP{\klg}{2} + \WTP{\klg}{3}\commae\\
  \WTP{}{4} &= {\bar L}^4 \WTP{\klg}{0} + 4 {\bar L}^3 \WTP{\klg}{1} +
    6 {\bar L}^2 \WTP{\klg}{2} + 4 {\bar L}\, \WTP{\klg}{3} + \WTP{\klg}{4}\commae
  \raisetag{44.5pt}
\end{split}
\end{equation}
and the components of tensor of electromagnetic field transform
according to
\begin{equation}\label{ae:kfixedEM}
\begin{aligned}
  \EMP{}{0} &= \EMP{\klg}{0}\commae\\
  \EMP{}{1} &= {\bar L}\, \EMP{\klg}{0} + \EMP{\klg}{1} \commae\\
  \EMP{}{2} &= {\bar L}^2 \EMP{\klg}{0} + 2 {\bar L}\, \EMP{\klg}{1} + \EMP{\klg}{2}\period
\end{aligned}
\end{equation}

Under a null rotation with $\lG$ fixed,
\begin{equation}\label{ae:lfixed}
\begin{aligned}
  \kG &= \kK + \bar K \mK + K \bK + K\bar K \lK\commae\\
  \lG &= \lK \commae\\
  \mG &= \mK + K\lK\commae\\
  \bG &= \bK + \bar K\lK\commae
\end{aligned}
\end{equation}
$K$ being a complex number which parametrize the rotation,
the components of the Weyl tensor transform as
\begin{equation}\label{ae:lfixedWeyl}
\begin{split}
  \WTP{}{0} &= K^4 \WTP{\klg}{4} + 4 K^3 \WTP{\klg}{3} +
    6 K^2 \WTP{\klg}{2} + 4 K\, \WTP{\klg}{1} + \WTP{\klg}{0}\commae\\
  \WTP{}{1} &= K^3 \WTP{\klg}{4} + 3 K^2 \WTP{\klg}{3} +
    3 K\, \WTP{\klg}{2} + \WTP{\klg}{1}\commae\\
  \WTP{}{2} &= K^2 \WTP{\klg}{4} + 2 K\, \WTP{\klg}{3} + \WTP{\klg}{2}\commae\\
  \WTP{}{3} &= K\, \WTP{\klg}{4} + \WTP{\klg}{3} \commae\\
  \WTP{}{4} &= \WTP{\klg}{4}\commae
  \raisetag{40pt}
\end{split}
\end{equation}
and for electromagnetic field we have
\begin{equation}\label{ae:lfixedEM}
\begin{aligned}
  \EMP{}{0} &= K^2 \EMP{\klg}{2} + 2 K\, \EMP{\klg}{1} + \EMP{\klg}{0}\commae\\
  \EMP{}{1} &= K\, \EMP{\klg}{2} + \EMP{\klg}{1} \commae\\
  \EMP{}{2} &= \EMP{\klg}{2}\period
\end{aligned}
\end{equation}

A boost in the ${\nG\textdash\fG\equiv\kG\textdash\lG}$ plane
and a spatial rotation in the ${\rG\textdash\sG\equiv\mG\textdash\bG}$ plane
is given by
\begin{equation}\label{ae:boostrotation}
\begin{gathered}
  \kG = B\,\kK \comma   \lG = B^{-1}\, \lK \commae\\
  \mG = \exp(i\Phi)\,\mK \comma   \bG = \exp(-i\Phi)\,\bK \commae
\end{gathered}
\end{equation}
or, introducing ${B=\exp\beta}$,
\begin{equation}\label{ae:boostrotationON}
\begin{aligned}
  \nG &= \cosh\beta\,\nK+\sinh\beta\,\fK \commae\\
  \fG &= \sinh\beta\,\nK+\cosh\beta\,\fK \commae\\
  \rG &= \spcm \cos\Phi\,\rK+\sin\Phi\,\sK    \commae\\
  \sG &= -\sin\Phi\,\rK+\cos\Phi\,\sK    \commae
\end{aligned}
\end{equation}
$B$, $\beta$ being real numbers which parametrize
the boost, $\Phi$ parametrizing an angle of the rotation.
The components $\WTP{}{n}$ now transform
\begin{align}
  \WTP{}{0} &= B^2\,\exp(2i\Phi)\; \WTP{\klg}{0} \commae\notag\\
  \WTP{}{1} &= B\;\;\exp(i\Phi)\; \WTP{\klg}{1} \commae\notag\\
  \WTP{}{2} &= \WTP{\klg}{2} \commae\label{ae:boostrotationWeyl}\\
  \WTP{}{3} &= B^{-1}\;\exp(-i\Phi)\; \WTP{\klg}{3} \commae\notag\\
  \WTP{}{4} &= B^{-2}\,\exp(-2i\Phi)\; \WTP{\klg}{4}\commae\notag
\end{align}
and $\EMP{}{n}$ transform as
\begin{align}
  \EMP{}{0} &= B\;\;\exp(i\Phi)\; \EMP{\klg}{0} \commae\notag\\
  \EMP{}{1} &= \EMP{\klg}{1} \commae\label{ae:boostrotationEMF}\\
  \EMP{}{2} &= B^{-1}\,\exp(-i\Phi)\; \EMP{\klg}{2}\period\notag
\end{align}

\vfill


\begin{thebibliography}{56}
\expandafter\ifx\csname natexlab\endcsname\relax\def\natexlab#1{#1}\fi
\expandafter\ifx\csname bibnamefont\endcsname\relax
  \def\bibnamefont#1{#1}\fi
\expandafter\ifx\csname bibfnamefont\endcsname\relax
  \def\bibfnamefont#1{#1}\fi
\expandafter\ifx\csname citenamefont\endcsname\relax
  \def\citenamefont#1{#1}\fi
\expandafter\ifx\csname url\endcsname\relax
  \def\url#1{\texttt{#1}}\fi
\expandafter\ifx\csname urlprefix\endcsname\relax\def\urlprefix{URL }\fi
\providecommand{\bibinfo}[2]{#2}
\providecommand{\eprint}[2][]{\url{#2}}

\bibitem[{\citenamefont{Bondi}(1960)}]{Bondi:1960}
\bibinfo{author}{\bibfnamefont{H.}~\bibnamefont{Bondi}},
  \bibinfo{journal}{Nature} \textbf{\bibinfo{volume}{186}},
  \bibinfo{pages}{535} (\bibinfo{year}{1960}).

\bibitem[{\citenamefont{Bondi et~al.}(1962)\citenamefont{Bondi, van~der Burg,
  and Metzner}}]{BondiBurgMetzner:1962}
\bibinfo{author}{\bibfnamefont{H.}~\bibnamefont{Bondi}},
  \bibinfo{author}{\bibfnamefont{M.~G.~J.} \bibnamefont{van~der Burg}},
  \bibnamefont{and} \bibinfo{author}{\bibfnamefont{A.~W.~K.}
  \bibnamefont{Metzner}}, \bibinfo{journal}{Proc. R. Soc. Lond., Ser A}
  \textbf{\bibinfo{volume}{269}}, \bibinfo{pages}{21} (\bibinfo{year}{1962}).

\bibitem[{\citenamefont{Sachs}(1962)}]{Sachs:1962}
\bibinfo{author}{\bibfnamefont{R.~K.} \bibnamefont{Sachs}},
  \bibinfo{journal}{Proc. R. Soc. Lond., Ser A} \textbf{\bibinfo{volume}{270}},
  \bibinfo{pages}{103} (\bibinfo{year}{1962}).

\bibitem[{\citenamefont{van~der Burg}(1969)}]{Burg:1969}
\bibinfo{author}{\bibfnamefont{M.~G.~J.} \bibnamefont{van~der Burg}},
  \bibinfo{journal}{Proc. R. Soc. Lond., Ser A} \textbf{\bibinfo{volume}{310}},
  \bibinfo{pages}{221} (\bibinfo{year}{1969}).

\bibitem[{\citenamefont{Newman and Penrose}(1962)}]{NewmanPenrose:1962}
\bibinfo{author}{\bibfnamefont{E.}~\bibnamefont{Newman}} \bibnamefont{and}
  \bibinfo{author}{\bibfnamefont{R.}~\bibnamefont{Penrose}},
  \bibinfo{journal}{J. Math. Phys.} \textbf{\bibinfo{volume}{3}},
  \bibinfo{pages}{566} (\bibinfo{year}{1962}).

\bibitem[{\citenamefont{Newman and Unti}(1962)}]{NewmanUnti:1962}
\bibinfo{author}{\bibfnamefont{E.~T.} \bibnamefont{Newman}} \bibnamefont{and}
  \bibinfo{author}{\bibfnamefont{T.~W.~J.} \bibnamefont{Unti}},
  \bibinfo{journal}{J. Math. Phys.} \textbf{\bibinfo{volume}{3}},
  \bibinfo{pages}{891} (\bibinfo{year}{1962}).

\bibitem[{\citenamefont{Sachs}(1961)}]{Sachs:1961}
\bibinfo{author}{\bibfnamefont{R.}~\bibnamefont{Sachs}},
  \bibinfo{journal}{Proc. R. Soc. Lond., Ser A} \textbf{\bibinfo{volume}{264}},
  \bibinfo{pages}{309} (\bibinfo{year}{1961}).

\bibitem[{\citenamefont{Penrose}(1965)}]{Penrose:1965}
\bibinfo{author}{\bibfnamefont{R.}~\bibnamefont{Penrose}},
  \bibinfo{journal}{Proc. R. Soc. Lond., Ser A} \textbf{\bibinfo{volume}{284}},
  \bibinfo{pages}{159} (\bibinfo{year}{1965}).

\bibitem[{\citenamefont{Penrose}(1964)}]{Penrose:1964}
\bibinfo{author}{\bibfnamefont{R.}~\bibnamefont{Penrose}}, in
  \emph{\bibinfo{booktitle}{Relativity, Groups and Topology, Les Houches
  1963}}, edited by \bibinfo{editor}{\bibfnamefont{C.}~\bibnamefont{DeWitt}}
  \bibnamefont{and} \bibinfo{editor}{\bibfnamefont{B.}~\bibnamefont{DeWitt}}
  (\bibinfo{publisher}{Gordon and Breach}, \bibinfo{address}{New York},
  \bibinfo{year}{1964}).

\bibitem[{\citenamefont{Penrose}(1967)}]{Penrose:1967}
\bibinfo{author}{\bibfnamefont{R.}~\bibnamefont{Penrose}}, in
  \emph{\bibinfo{booktitle}{The Nature of Time}}, edited by
  \bibinfo{editor}{\bibfnamefont{T.}~\bibnamefont{Gold}}
  (\bibinfo{publisher}{Cornell University Press}, \bibinfo{address}{Ithaca, New
  York}, \bibinfo{year}{1967}), pp. \bibinfo{pages}{42--54}.

\bibitem[{\citenamefont{Penrose}(1963)}]{Penrose:1963}
\bibinfo{author}{\bibfnamefont{R.}~\bibnamefont{Penrose}},
  \bibinfo{journal}{Phys. Rev. Lett.} \textbf{\bibinfo{volume}{10}},
  \bibinfo{pages}{66} (\bibinfo{year}{1963}).

\bibitem[{\citenamefont{Bondi et~al.}(1959)\citenamefont{Bondi, Pirani, and
  Robinson}}]{BondiPiraniRobinson:1959}
\bibinfo{author}{\bibfnamefont{H.}~\bibnamefont{Bondi}},
  \bibinfo{author}{\bibfnamefont{F.~A.~E.} \bibnamefont{Pirani}},
  \bibnamefont{and} \bibinfo{author}{\bibfnamefont{I.}~\bibnamefont{Robinson}},
  \bibinfo{journal}{Proc. R. Soc. Lond., Ser A} \textbf{\bibinfo{volume}{251}},
  \bibinfo{pages}{519} (\bibinfo{year}{1959}).

\bibitem[{\citenamefont{Ehlers and Kundt}(1962)}]{EhlersKundt:1962}
\bibinfo{author}{\bibfnamefont{J.}~\bibnamefont{Ehlers}} \bibnamefont{and}
  \bibinfo{author}{\bibfnamefont{W.}~\bibnamefont{Kundt}}, in
  \emph{\bibinfo{booktitle}{Gravitation: an Introduction to Current Research}},
  edited by \bibinfo{editor}{\bibfnamefont{L.}~\bibnamefont{Witten}}
  (\bibinfo{publisher}{John Wiley}, \bibinfo{address}{New York},
  \bibinfo{year}{1962}), pp. \bibinfo{pages}{49--101}.

\bibitem[{\citenamefont{Robinson and Trautman}(1962)}]{RobinsonTrautman:1962}
\bibinfo{author}{\bibfnamefont{I.}~\bibnamefont{Robinson}} \bibnamefont{and}
  \bibinfo{author}{\bibfnamefont{A.}~\bibnamefont{Trautman}},
  \bibinfo{journal}{Proc. R. Soc. Lond., Ser A} \textbf{\bibinfo{volume}{265}},
  \bibinfo{pages}{463} (\bibinfo{year}{1962}).

\bibitem[{\citenamefont{Bonnor and
  Swaminarayan}(1964)}]{BonnorSwaminarayan:1964}
\bibinfo{author}{\bibfnamefont{W.~B.} \bibnamefont{Bonnor}} \bibnamefont{and}
  \bibinfo{author}{\bibfnamefont{N.~S.} \bibnamefont{Swaminarayan}},
  \bibinfo{journal}{Z. Phys.} \textbf{\bibinfo{volume}{177}},
  \bibinfo{pages}{240} (\bibinfo{year}{1964}).

\bibitem[{\citenamefont{Pirani}(1965)}]{Pirani:1965}
\bibinfo{author}{\bibfnamefont{F.~A.~E.} \bibnamefont{Pirani}}, in
  \emph{\bibinfo{booktitle}{Brandeis Lectures on General Relativity}}, edited
  by \bibinfo{editor}{\bibfnamefont{S.}~\bibnamefont{Deser}} \bibnamefont{and}
  \bibinfo{editor}{\bibfnamefont{K.~W.} \bibnamefont{Ford}}
  (\bibinfo{publisher}{Prentice-Hall}, \bibinfo{address}{New Jersey},
  \bibinfo{year}{1965}), pp. \bibinfo{pages}{249--372}.

\bibitem[{\citenamefont{Bonnor et~al.}(1994)\citenamefont{Bonnor, Griffiths,
  and MacCallum}}]{BonnorGriffithsMacCallum:1994}
\bibinfo{author}{\bibfnamefont{W.~B.} \bibnamefont{Bonnor}},
  \bibinfo{author}{\bibfnamefont{J.~B.} \bibnamefont{Griffiths}},
  \bibnamefont{and} \bibinfo{author}{\bibfnamefont{M.~A.~H.}
  \bibnamefont{MacCallum}}, \bibinfo{journal}{Gen. Rel. Grav.}
  \textbf{\bibinfo{volume}{26}}, \bibinfo{pages}{687} (\bibinfo{year}{1994}).

\bibitem[{\citenamefont{Bi\v{c}\'{a}k}(1985)}]{Bicak:Bonnor}
\bibinfo{author}{\bibfnamefont{J.}~\bibnamefont{Bi\v{c}\'{a}k}}, in
  \emph{\bibinfo{booktitle}{Galaxies, Axisymmetric Systems and Relativity}},
  edited by \bibinfo{editor}{\bibfnamefont{M.~A.~H.} \bibnamefont{MacCaluum}}
  (\bibinfo{publisher}{Cambridge University Press},
  \bibinfo{address}{Cambridge}, \bibinfo{year}{1985}), pp.
  \bibinfo{pages}{91--124}.

\bibitem[{\citenamefont{Bi\v{c}\'{a}k}(1997)}]{Bicak:1997}
\bibinfo{author}{\bibfnamefont{J.}~\bibnamefont{Bi\v{c}\'{a}k}}, in
  \emph{\bibinfo{booktitle}{Relativistic Gravitation and Gravitational
  Radiation, Les Louches 1995}}, edited by
  \bibinfo{editor}{\bibfnamefont{J.-A.} \bibnamefont{Marck}} \bibnamefont{and}
  \bibinfo{editor}{\bibfnamefont{J.-P.} \bibnamefont{Lasota}}
  (\bibinfo{publisher}{Cambridge University Press},
  \bibinfo{address}{Cambridge}, \bibinfo{year}{1997}), pp.
  \bibinfo{pages}{67--87}.

\bibitem[{\citenamefont{Bi\v{c}\'{a}k}(2000)}]{Bicak:Ehlers}
\bibinfo{author}{\bibfnamefont{J.}~\bibnamefont{Bi\v{c}\'{a}k}}, in
  \emph{\bibinfo{booktitle}{Einstein's Field Equations and Their Physical
  Implications}}, edited by \bibinfo{editor}{\bibfnamefont{B.~G.}
  \bibnamefont{Schmidt}} (\bibinfo{publisher}{Springer},
  \bibinfo{address}{Berlin}, \bibinfo{year}{2000}), vol. \bibinfo{volume}{540},
  pp. \bibinfo{pages}{1--126}, \eprint{gr-qc/0004016}.

\bibitem[{\citenamefont{Bi\v{c}\'{a}k and Krtou\v{s}}(2001)}]{BicakKrtous:ASDS}
\bibinfo{author}{\bibfnamefont{J.}~\bibnamefont{Bi\v{c}\'{a}k}}
  \bibnamefont{and}
  \bibinfo{author}{\bibfnamefont{P.}~\bibnamefont{Krtou\v{s}}},
  \bibinfo{journal}{Phys. Rev. D} \textbf{\bibinfo{volume}{64}},
  \bibinfo{pages}{124020} (\bibinfo{year}{2001}), \eprint{gr-qc/0107078}.

\bibitem[{\citenamefont{Bi\v{c}\'{a}k and
  Krtou\v{s}}(2002)}]{BicakKrtous:FUACS}
\bibinfo{author}{\bibfnamefont{J.}~\bibnamefont{Bi\v{c}\'{a}k}}
  \bibnamefont{and}
  \bibinfo{author}{\bibfnamefont{P.}~\bibnamefont{Krtou\v{s}}},
  \bibinfo{journal}{Phys. Rev. Lett.} \textbf{\bibinfo{volume}{88}},
  \bibinfo{pages}{211101} (\bibinfo{year}{2002}), \eprint{gr-qc/0207010}.

\bibitem[{\citenamefont{Bi\v{c}\'{a}k and Krtou\v{s}}()}]{BicakKrtous:BIS}
\bibinfo{author}{\bibfnamefont{J.}~\bibnamefont{Bi\v{c}\'{a}k}}
  \bibnamefont{and}
  \bibinfo{author}{\bibfnamefont{P.}~\bibnamefont{Krtou\v{s}}},
  \bibinfo{note}{unpublished}.

\bibitem[{\citenamefont{Bi\v{c}\'{a}k and Schmidt}(1989)}]{BicakSchmidt:1989}
\bibinfo{author}{\bibfnamefont{J.}~\bibnamefont{Bi\v{c}\'{a}k}}
  \bibnamefont{and} \bibinfo{author}{\bibfnamefont{B.~G.}
  \bibnamefont{Schmidt}}, \bibinfo{journal}{Phys. Rev. D}
  \textbf{\bibinfo{volume}{40}}, \bibinfo{pages}{1827} (\bibinfo{year}{1989}).

\bibitem[{\citenamefont{Levi-Civita}(1917)}]{LeviCivita:1917}
\bibinfo{author}{\bibfnamefont{T.}~\bibnamefont{Levi-Civita}},
  \bibinfo{journal}{Rend. Acc. Lincei} \textbf{\bibinfo{volume}{26}},
  \bibinfo{pages}{307} (\bibinfo{year}{1917}).

\bibitem[{\citenamefont{Weyl}(1918)}]{Weyl:1919}
\bibinfo{author}{\bibfnamefont{H.}~\bibnamefont{Weyl}}, \bibinfo{journal}{Ann.
  Phys. (Leipzig)} \textbf{\bibinfo{volume}{59}}, \bibinfo{pages}{185}
  (\bibinfo{year}{1918}).

\bibitem[{\citenamefont{Kinnersley and Walker}(1970)}]{KinnersleyWalker:1970}
\bibinfo{author}{\bibfnamefont{W.}~\bibnamefont{Kinnersley}} \bibnamefont{and}
  \bibinfo{author}{\bibfnamefont{M.}~\bibnamefont{Walker}},
  \bibinfo{journal}{Phys. Rev. D} \textbf{\bibinfo{volume}{2}},
  \bibinfo{pages}{1359} (\bibinfo{year}{1970}).

\bibitem[{\citenamefont{Bonnor}(1983)}]{Bonnor:1982}
\bibinfo{author}{\bibfnamefont{W.~B.} \bibnamefont{Bonnor}},
  \bibinfo{journal}{Gen. Rel. Grav.} \textbf{\bibinfo{volume}{15}},
  \bibinfo{pages}{535} (\bibinfo{year}{1983}).

\bibitem[{\citenamefont{Bi\v{c}\'{a}k and Pravda}(1999)}]{BicakPravda:1999}
\bibinfo{author}{\bibfnamefont{J.}~\bibnamefont{Bi\v{c}\'{a}k}}
  \bibnamefont{and} \bibinfo{author}{\bibfnamefont{V.}~\bibnamefont{Pravda}},
  \bibinfo{journal}{Phys. Rev. D} \textbf{\bibinfo{volume}{60}},
  \bibinfo{pages}{044004} (\bibinfo{year}{1999}).

\bibitem[{\citenamefont{Letelier and Oliveira}(2001)}]{LetelierOliveira:2001}
\bibinfo{author}{\bibfnamefont{P.~S.} \bibnamefont{Letelier}} \bibnamefont{and}
  \bibinfo{author}{\bibfnamefont{S.~R.} \bibnamefont{Oliveira}},
  \bibinfo{journal}{Phys. Rev. D} \textbf{\bibinfo{volume}{64}},
  \bibinfo{pages}{064005} (\bibinfo{year}{2001}), \eprint{gr-qc/9809089}.

\bibitem[{\citenamefont{Pravda and Pravdov\'{a}}(2000)}]{Pravdovi:2000}
\bibinfo{author}{\bibfnamefont{V.}~\bibnamefont{Pravda}} \bibnamefont{and}
  \bibinfo{author}{\bibfnamefont{A.}~\bibnamefont{Pravdov\'{a}}},
  \bibinfo{journal}{Czech. J. Phys.} \textbf{\bibinfo{volume}{50}},
  \bibinfo{pages}{333} (\bibinfo{year}{2000}).

\bibitem[{\citenamefont{Pleba\'{n}ski and
  Demia\'{n}ski}(1976)}]{PlebanskiDemianski:1976}
\bibinfo{author}{\bibfnamefont{J.}~\bibnamefont{Pleba\'{n}ski}}
  \bibnamefont{and}
  \bibinfo{author}{\bibfnamefont{M.}~\bibnamefont{Demia\'{n}ski}},
  \bibinfo{journal}{Ann. Phys. (N.Y.)} \textbf{\bibinfo{volume}{98}},
  \bibinfo{pages}{98} (\bibinfo{year}{1976}).

\bibitem[{\citenamefont{Carter}(1968)}]{Carter:1968}
\bibinfo{author}{\bibfnamefont{B.}~\bibnamefont{Carter}},
  \bibinfo{journal}{Commun. Math. Phys.} \textbf{\bibinfo{volume}{10}},
  \bibinfo{pages}{280} (\bibinfo{year}{1968}).

\bibitem[{\citenamefont{Debever}(1971)}]{Debever:1971}
\bibinfo{author}{\bibfnamefont{R.}~\bibnamefont{Debever}},
  \bibinfo{journal}{Bull. Soc. Math. Belg.} \textbf{\bibinfo{volume}{23}},
  \bibinfo{pages}{360} (\bibinfo{year}{1971}).

\bibitem[{\citenamefont{Mann and Ross}(1995)}]{MannRoss:1995}
\bibinfo{author}{\bibfnamefont{R.~B.} \bibnamefont{Mann}} \bibnamefont{and}
  \bibinfo{author}{\bibfnamefont{S.~F.} \bibnamefont{Ross}},
  \bibinfo{journal}{Phys. Rev. D} \textbf{\bibinfo{volume}{52}},
  \bibinfo{pages}{2254} (\bibinfo{year}{1995}).

\bibitem[{\citenamefont{Podolsk\'{y} and
  Griffiths}(2001)}]{PodolskyGriffiths:2001}
\bibinfo{author}{\bibfnamefont{J.}~\bibnamefont{Podolsk\'{y}}}
  \bibnamefont{and} \bibinfo{author}{\bibfnamefont{J.~B.}
  \bibnamefont{Griffiths}}, \bibinfo{journal}{Phys. Rev. D}
  \textbf{\bibinfo{volume}{63}}, \bibinfo{pages}{024006}
  (\bibinfo{year}{2001}), \eprint{gr-qc/0010109}.

\bibitem[{\citenamefont{Dias and Lemos}({\natexlab{a}})}]{DiasLemos:2003a}
\bibinfo{author}{\bibfnamefont{O.~J.~C.} \bibnamefont{Dias}} \bibnamefont{and}
  \bibinfo{author}{\bibfnamefont{J.~P.~S.} \bibnamefont{Lemos}}, \bibinfo{journal}{Phys. Rev. D}
  \textbf{\bibinfo{volume}{67}}, \bibinfo{pages}{064001}
  (\bibinfo{year}{2003}), \eprint{hep-th/0210065}.

\bibitem[{\citenamefont{Dias and Lemos}({\natexlab{b}})}]{DiasLemos:2003b}
\bibinfo{author}{\bibfnamefont{O.~J.~C.} \bibnamefont{Dias}} \bibnamefont{and}
  \bibinfo{author}{\bibfnamefont{J.~P.~S.} \bibnamefont{Lemos}}, \bibinfo{journal}{Phys. Rev. D}
  \textbf{\bibinfo{volume}{67}}, \bibinfo{pages}{084018}
  (\bibinfo{year}{2003}), \eprint{hep-th/0301046}.

\bibitem{nt:Units}
  We use here the unit conventions of Ref.~\cite{MTW}
  (which is somewhat different from Ref.~\cite{Krameretal:book}), namely,
  we assume \vague{geometrical} values of the gravitational constant
  ${\varkappa=8\pi}$, and of the electric permittivity ${\varepsilon=1/(4\pi)}$.
  These constants are fixed by the Einstein equation ${\tens{Ein}=\varkappa\,\tens{T}}$,
  and the Maxwell equation ${\nabla\spr\,\EMF=(1/\varepsilon)\,\EMJ}$.
  Both the mass and the charge thus have the dimension of length.
  The constants $\varkappa$, $\varepsilon$ can be explicitly reinserted into the expressions using
  a dimensional analysis with the help of  relations ${1\,\mathrm{m}=\frac\varkappa{8\pi}\times1\,\mathrm{kg}}$,
  and ${1\,\mathrm{m}=\sqrt{\frac{\varkappa}{32\pi^2\varepsilon}}\times1\,\mathrm{C}}$.
  For example, ${\GKW=1-\xKW^2-\frac{\varkappa\mass}{4\pi}\accl\,\xKW^3
  +\frac{\varkappa\charge^2}{32\pi^2\varepsilon}\accl^2\xKW^4}$, with the mass
  $\mass$ measured in $\mathrm{kg}$ and the charge $\charge$ in $\mathrm{C}$.
  Note, however, that because the tensor of electromagnetic field $\EMF_{\alpha\beta}$ is not dimensionless,
  we, in fact, obtain ${\EMF=\frac{\charge}{4\pi\varepsilon}\,\grad\yKW\wedge\grad\tKW}$.
  (Of course, ${c=1}$ is always assumed here.)

\bibitem[{\citenamefont{Misner et~al.}(1973)\citenamefont{Misner, Thorne, and
  Wheeler}}]{MTW}
\bibinfo{author}{\bibfnamefont{C.~W.} \bibnamefont{Misner}},
  \bibinfo{author}{\bibfnamefont{K.~S.} \bibnamefont{Thorne}},
  \bibnamefont{and} \bibinfo{author}{\bibfnamefont{J.~A.}
  \bibnamefont{Wheeler}}, \emph{\bibinfo{title}{Gravitation}}
  (\bibinfo{publisher}{Freeman}, \bibinfo{address}{San Francisco},
  \bibinfo{year}{1973}).

\bibitem{nt:SymTensProd}
  A \defterm{symmetric product} $\stp$ of two 1-forms is defined as
  \begin{equation*}
    \tens{a}\stp\tens{b}=\tens{a}\,\tens{b}+\tens{b}\,\tens{a}\period
  \end{equation*}
  For example, the metric \eqref{RTmetric} written in a more common (but less precise) notation would
  read
  \begin{equation*}
    \mtrc =
    2\frac{\rRT^2}{\PRT^{2}}\,
    \grad\zRT\,\grad\bRT
    -2\,\grad\uRT\,\grad\rRT
    -\HRT \,\grad\uRT^2\period
  \end{equation*}

\bibitem[{\citenamefont{Penrose and Rindler}(1984)}]{PenroseRindler:book}
\bibinfo{author}{\bibfnamefont{R.}~\bibnamefont{Penrose}} \bibnamefont{and}
  \bibinfo{author}{\bibfnamefont{W.}~\bibnamefont{Rindler}},
  \emph{\bibinfo{title}{Spinors and Space-time}} (\bibinfo{publisher}{Cambridge
  University Press}, \bibinfo{address}{Cambridge}, \bibinfo{year}{1984}).

\bibitem[{\citenamefont{Vilenkin and Shellard}(1994)}]{VilenkinShellard:book}
\bibinfo{author}{\bibfnamefont{A.}~\bibnamefont{Vilenkin}} \bibnamefont{and}
  \bibinfo{author}{\bibfnamefont{E.~P.~S.} \bibnamefont{Shellard}},
  \emph{\bibinfo{title}{Cosmic Strings and other Topological Defects}}
  (\bibinfo{publisher}{Cambridge University Press},
  \bibinfo{address}{Cambridge}, \bibinfo{year}{1994}).

\bibitem[{\citenamefont{Kramer et~al.}(1980)\citenamefont{Kramer, Stephani,
  Herlt, and MacCallum}}]{Krameretal:book}
\bibinfo{author}{\bibfnamefont{D.}~\bibnamefont{Kramer}},
  \bibinfo{author}{\bibfnamefont{H.}~\bibnamefont{Stephani}},
  \bibinfo{author}{\bibfnamefont{E.}~\bibnamefont{Herlt}}, \bibnamefont{and}
  \bibinfo{author}{\bibfnamefont{M.}~\bibnamefont{MacCallum}},
  \emph{\bibinfo{title}{Exact Solutions of Einstein's Field Equation}}
  (\bibinfo{publisher}{Cambridge University Press},
  \bibinfo{address}{Cambridge}, \bibinfo{year}{1980}).

\bibitem{nt:PhysAffinePar}
   The 4-momentum ${\mom=\frac{D\geod{}}{d\pafp}}$ of the ray is
   a tangent vector with respect to a physical dimensionless
   affine parameter $\pafp$. Since we are not interested
   in dynamical properties of the rays but rather in
   behavior of fields at \vague{large distances}, we are using the
   affine parameter $\afp$ which has the dimension of length. The
   relation between this \vague{distance} parameter $\afp$
   and the physical parameter $\pafp$ which determines energy
   can conventionally be chosen as ${\afp = \DSr\pafp}$.
   Therefore, ${\mom=\DSr\,\frac{D\geod{}}{d\afp}}$.

\bibitem{nt:Proximity}
   We cannot define here \vague{the same proximity to $\scri^+$}
   using some fixed value of the affine parameter $\afp$, since we wish to use
   this procedure to determine the specific normalization of the affine parameter.

\bibitem{nt:ArbFinCnd}
   A general choice of final conditions
   at infinity might cause a degenerated values of the vector $\kP$
   in finite domains of the spacetime.

\bibitem{nt:FUACS}
  In fact, Ref.~\cite{BicakKrtous:FUACS} use a different notation for
  quantities defined in the present paper. When comparing the angular
  dependence of Eq.~(15) of Ref.~\cite{BicakKrtous:FUACS} with
  Eq.~\eqref{EMDirChar} it is necessary to use the following dictionary:
  ${\alpha\rightarrow\DSr}$,\,
  ${a_{\mathrm{o}}\rightarrow\accl}$,\,
  ${\tht_*\rightarrow\THT}$,\,
  ${\ph_*\rightarrow\PHI}$,\, and
  \[
    \frac{a_{\mathrm{o}}\alpha}{\sqrt{1+a_{\mathrm{o}}^2\alpha^2}}\,\sin\tht_+
    \rightarrow\sin\THT_\spec\period
  \]
  In the case ${\mass=0}$, ${\charge=0}$ we can also identify
  ${\tht_+\rightarrow\thdS}$ and express $\chi_+$ of Ref.~\cite{BicakKrtous:FUACS} in
  terms of $\rdS$ and $\tdS$. The normalization of Eq.~\eqref{EMDirChar} and of
  the result of Ref.~\cite{BicakKrtous:FUACS} does not coincide due to different
  choice of initial conditions for the interpretation tetrads here and in
  Ref.~\cite{BicakKrtous:FUACS} (cf.\ Section~\ref{sc:AlgSpecDir}). Let us also
  note that Ref.~\cite{BicakKrtous:FUACS} uses the unit convention ${\varepsilon=1}$ in
  the sense of \noteref{nt:Units} instead of ${\varepsilon=\frac1{4\pi}}$ here.

\bibitem{nt:FUACSincond}
  This method has been used in Ref.~\cite{BicakKrtous:FUACS}.

\bibitem[{\citenamefont{Deser and Levin}(1999)}]{DeserLevin:1999}
\bibinfo{author}{\bibfnamefont{S.}~\bibnamefont{Deser}} \bibnamefont{and}
  \bibinfo{author}{\bibfnamefont{O.}~\bibnamefont{Levin}},
  \bibinfo{journal}{Phys. Rev. D} \textbf{\bibinfo{volume}{59}},
  \bibinfo{pages}{064004} (\bibinfo{year}{1999}).

\bibitem[{\citenamefont{Krtou\v{s} et~al.}()\citenamefont{Krtou\v{s},
  Podolsk\'{y}, and Bi\v{c}\'{a}k}}]{KrtousPodolskyBicak:inprep}
\bibinfo{author}{\bibfnamefont{P.}~\bibnamefont{Krtou\v{s}}},
  \bibinfo{author}{\bibfnamefont{J.}~\bibnamefont{Podolsk\'{y}}},
  \bibnamefont{and}
  \bibinfo{author}{\bibfnamefont{J.}~\bibnamefont{Bi\v{c}\'{a}k}},
  \bibinfo{note}{in preparation}.

\bibitem[{\citenamefont{Podolsk\'{y}}(2002)}]{Podolsky:2002}
\bibinfo{author}{\bibfnamefont{J.}~\bibnamefont{Podolsk\'{y}}},
  \bibinfo{journal}{Czech. J. Phys.} \textbf{\bibinfo{volume}{52}},
  \bibinfo{pages}{1} (\bibinfo{year}{2002}), \eprint{gr-qc/0202033}.

\bibitem[{\citenamefont{Brill and Hayward}(1994)}]{BrillHayward:1994}
\bibinfo{author}{\bibfnamefont{D.~R.} \bibnamefont{Brill}} \bibnamefont{and}
  \bibinfo{author}{\bibfnamefont{S.~A.} \bibnamefont{Hayward}},
  \bibinfo{journal}{Class. Quantum Grav.} \textbf{\bibinfo{volume}{11}},
  \bibinfo{pages}{359} (\bibinfo{year}{1994}).

\bibitem[{\citenamefont{Podolsk\'{y}}(1993)}]{Podolsky:PhD}
\bibinfo{author}{\bibfnamefont{J.}~\bibnamefont{Podolsk\'{y}}}, Ph.D. thesis,
  \bibinfo{school}{Charles University},
  \bibinfo{address}{Prague, Czech Republic} (\bibinfo{year}{1993}).

\bibitem[{\citenamefont{Hawking and Ellis}(1973)}]{HawkingEllis:book}
\bibinfo{author}{\bibfnamefont{S.~W.} \bibnamefont{Hawking}} \bibnamefont{and}
  \bibinfo{author}{\bibfnamefont{G.~F.~R.} \bibnamefont{Ellis}},
  \emph{\bibinfo{title}{The Large Scale Structure of Space-time}}
  (\bibinfo{publisher}{Cambridge University Press},
  \bibinfo{address}{Cambridge}, \bibinfo{year}{1973}).

\bibitem[{\citenamefont{Walker}(1970)}]{Walker:1970}
\bibinfo{author}{\bibfnamefont{M.}~\bibnamefont{Walker}}, \bibinfo{journal}{J.
  Math. Phys.} \textbf{\bibinfo{volume}{11}}, \bibinfo{pages}{2280}
  (\bibinfo{year}{1970}).

\bibitem{nt:vupsilon}
  Notice the difference between $\vRT$ (v) and
  $\y$ (upsilon). Fortunately, $\vRT$ is used only here, as an intermediate
  step, and in Fig.~\ref{fig:DiagDet} where the ranges of null coordinates are
  indicated.

\bibitem{nt:WTcomp}
  Here, e.g., $\WT_{\kappa\lambda\mu\bmu}=
  \WT_{\alpha\beta\gamma\delta}\;\kG^\alpha \lG^\beta \mG^\gamma \bG^\delta$.



\end{thebibliography}

\vfill

\end{document}